\documentstyle[epsf,eqsecnum,floats,preprint,aps]{revtex}

\tighten

\input epsf
\begin{document}

\newcommand{\be}{\begin{equation}}
\newcommand{\ee}{\end{equation}}
\newcommand{\bea}{\begin{eqnarray}}
\newcommand{\eea}{\end{eqnarray}}
\newcommand{\PSbox}[3]{\mbox{\rule{0in}{#3}\includegraphics{#1}\hspace{#2}}}
\newcommand{\modified}[1]{{\it #1}}

\def\5M{M^3_{(5)}}
\def\4M{M^2_{(4)}}
\def\mpsq{M_{\rm P}^2}
\def\om{\Omega_m}
\def\omt{\Omega_m^0}

\overfullrule=0pt
\def\Int{\int_{r_H}^\infty}
\def\d{\partial}
\def\e{\epsilon}
\def\M{{\cal M}}
\def\high{\vphantom{\Biggl(}\displaystyle}
\catcode`@=11
\def\@versim#1#2{\lower.7\p@\vbox{\baselineskip\z@skip\lineskip-.5\p@
    \ialign{$\m@th#1\hfil##\hfil$\crcr#2\crcr\sim\crcr}}}
\def\simge{\mathrel{\mathpalette\@versim>}} %
\def\simle{\mathrel{\mathpalette\@versim<}} %
\def\sun{\hbox{$\odot$}}
\catcode`@=12 
\ \\
\ \\
\vskip 0cm

\setcounter{footnote}{0}

\begin{center}
\Large{\bf \ \\ \ \\ \ \\THE PHENOMENOLOGY OF DVALI--GABADADZE--PORRATI COSMOLOGIES}
\ \\
\ \\
\large{Arthur Lue\footnote{E-mail: lue@lonestar.utsa.edu}}
\ \\
\normalsize{\em
Department of Physics and Astronomy\\
University of Texas at San Antonio\\
6900 North Loop 1604 West\\
San Antonio, TX 78249}
\end{center}
\begin{abstract}

\noindent
Cosmologists today are confronted with the perplexing reality that
the universe is currently accelerating in its expansion.  Nevertheless, the
nature of the fuel that drives today's cosmic acceleration is an open and
tantalizing mystery.  There exists the intriguing possibility that the acceleration
is not the manifestation of yet another mysterious ingredient in the cosmic gas
tank (dark energy), but rather our first real lack of understanding of
gravity itself, and even possibly a signal that there might exist
dimensions beyond that which we can currently observe.  The
braneworld model of Dvali, Gabadadze and Porrati (DGP) is a theory
where gravity is altered at immense distances by the excruciatingly slow leakage of
gravity off our three-dimensional Universe and, as a modified-gravity theory,
has pioneered this line of investigation.  I review the underlying structure of DGP
gravity and those phenomenological developments relevant to
cosmologists interested in a pedagogical treatment of this intriguing model.
\end{abstract}

\setcounter{page}{0}
\thispagestyle{empty}
\maketitle

\eject

\vfill

\baselineskip 18pt plus 2pt minus 2pt

\setcounter{page}{0}
\thispagestyle{empty}
\ \\
\newpage

\section{The Contemporary Universe}

Cosmology in this decade is said to be thriving in a golden age.  With the cornucopia of observational data from both satellite and ground-based surveys, an increasingly coherent  phenomenological and theoretical picture is now emerging.  And while our understanding of cosmic expansion, primordial nucleosynthesis, the microwave background and other phenomena allows particle physics and cosmology to use the very vastness of our Universe to probe the most incomprehensibly high energies, this golden age of cosmology is offering the first new data regarding the physics on immense scales in of themselves.  In other words, while modern cosmology is a ratification of the notion of the deep connection between the very small and the very large, it offers also the opportunity to challenge fundamental physics itself at the lowest of energies, an unexplored infrared domain.

A central example highlighting this theme is that physicists are currently faced with the perplexing reality that the Universe is accelerating in its expansion \cite{Perlmutter:1998np,Riess:1998cb}.  That startling reality is only driven home with the observation of the onset of this acceleration \cite{Riess}.  The acceleration represents, in essence, a new imbalance in the governing gravitational equations:  a universe filled only with ordinary matter and dark matter (ingredients for which we have independent corroboration) should decelerate in its expansion.  What drives the acceleration thus remains an open and tantalizing question.

Instructively, physics historically has addressed such imbalances in the governing gravitational equation in either one of two ways:  either by identifying sources that were previously unaccounted for (e.g., Neptune and dark matter) or by altering the governing equations (e.g., general relativity).  Standard cosmology has favored the first route to addressing the imbalance:  a missing energy-momentum component.  Indeed, a  ``conventional'' explanation exists for the cause of that acceleration --- in general relativity, vacuum energy provides the repulsive gravity necessary to drive accelerated cosmological expansion.  Variations on this vacuum-energy theme, such as quintessence, promote the energy density to the potential energy density of a dynamical field.  Such additions to the roster of cosmic sources of energy-momentum are collectively referred to as dark energy.  If it exists,
this mysterious dark energy would constitute the majority of the energy density of the
universe today.

However, one may also entertain the alternative viewpoint.  Can cosmic acceleration be the first new signal of a lack of understanding of gravitational interactions?  I.e., is the cosmic acceleration the result, not of the contents of the the cosmic gas tank, as it were, but a consequence of the engine itself.  This is the question that intrigues and excites us, and more importantly, {\em how} we can definitively answer that question.  How can one definitively differentiate this modification of the theory of gravity from dark energy?  Cosmology can offer a fresh opportunity to uncover new fundamental physics at the most immense of scales.\footnote{There is a bit of a semantic point about what one means by dark energy 		versus modified gravity,
i.e., altering the energy-momentum content of a theory versus altering the field equations
themselves.  For our qualitative discussion here, what I mean by dark energy is some (possibly
new) field or particle that is minimally coupled to the metric, meaning that its
constituents follow geodesics of the metric.  An alternative statement of this condition is that
the new field is covariantly conserved in the background of the metric.  I am presuming
that the metric {\em alone} mediates the gravitational interaction.  Thus, a modified-gravity
theory would still be a metric theory, minimally coupled to whatever energy-momentum exists
in that paradigm, but whose governing equations are not the Einstein equations.

We also wish to emphasize the point that we are not addressing the cosmological constant
problem here, i.e., why the vacuum energy is zero, or at least much smaller than the fundamental
Planck scale, $M_P^4$.  When we refer to either dark energy or a modified-gravity explanation of
cosmic acceleration, we do so with the understanding that a vanishing vacuum energy is
explained by some other means, and that dark energy refers to whatever residual vacuum energy or potential energy of a field may be driving today's acceleration,
and that modified-gravity assumes a strictly zero vacuum energy.}

Understanding cosmic acceleration and whether it indicates new fundamental physics serves as a first concrete step in the program of exploiting cosmology as a tool for understanding new infrared physics,  i.e., physics on these immense scales.  In 2000, Dvali, Gabadadze and Porrati (DGP) set forth a
braneworld model of gravity by which our observed four-dimensional Universe resides in a larger,
five-dimensional space.  However, unlike popular braneworld theories at the time, the extra dimension featured in this theory was astrophysically large and flat, rather than large compared to particle-physics scales but otherwise pathologically small compared to those we observe.  In DGP braneworlds, gravity is modified at large (rather than at short) distances through the excruciatingly slow evaporation of gravitational degrees of freedom off of the brane Universe.  It was soon shown by Deffayet that just such a model exhibits cosmological solutions that approach empty universes that nevertheless accelerate {\em themselves} at late times.

Having pioneered the paradigm of self-acceleration, DGP braneworld gravity remains a leading candidate for understanding gravity modified at ultralarge distances;   nevertheless, much work remains to be done to understand its far-reaching consequences.   This article is intended to be a coherent and instructive review of the material for those interested in carrying on the intriguing phenomenological work, rather than an exhaustive account of the rather dissonant, confusing and sometimes mistaken literature.  In particular, we focus on the simple cases and scenarios that best illuminate the pertinent properties of DGP gravity as well as those of highest observational relevance, rather than enumerate the many intriguing variations which may be played out in this theory.

We begin by setting out the governing equations and the environment in which this model exists,
while giving a broad picture of how its gross features manifest themselves.  We then provide a detailed view of how cosmology arises in this model, including the emergence of the celebrated self-accelerating phase.  At the same time, a geometric picture of how such cosmologies evolve is presented, from the
perspective of both observers in our Universe as well as hypothetical observers existing in the larger bulk space.  We touch on observational constraints for this specific cosmology.  We then address an important problem/subtlety regarding the recovery of four-dimensional Einstein gravity.  It is this peculiar story that leads to powerful and accessible observable consequences for this theory, and is the key to differentiating a modified-gravity scenario such as DGP gravity from dark-energy scenarios.  We then
illuminate the interplay of cosmology with the modification of the gravitational potentials and spend the next several sections discussing DGP gravity's astronomical and cosmological consequences.  Finally, we finish with the prospects of future work and some potential problems with DGP gravity.  We will see that DGP gravity provides a rich and unique environment for potentially shedding light on new cosmology and physics.

\section{Gravitational Leakage into Extra Dimensions}

\begin{figure} \begin{center}\PSbox{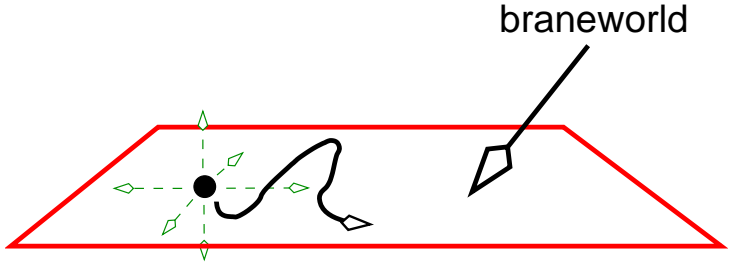
hscale=100 vscale=100 hoffset=0 voffset=0}{4in}{1.1in}\end{center}
\caption{
DGP gravity employs the braneworld scenario.  Matter and all standard model forces and particles
are pinned to a strictly four-dimensional braneworld.  Gravity, however, is free to explore the full
five-dimensional bulk.
}
\label{fig:world}
\end{figure}

We have set ourselves the task of determining whether there is more to gravitational physics than is revealed by general relativity.  Extra dimension theories in general, and braneworld models in particular, are an indispensable avenue with which to approach understand gravity, post--Einstein.  Extra dimensions provide an approach to modifying gravity with out abandoning Einstein's general relativity altogether as a basis for understanding the fundamental gravitational interaction.  Furthermore, the braneworld paradigm (Fig.~\ref{fig:world}) allows model builders a tool by which to avoid very real constraints on the number of extra dimensions coming from standard model observations.  By explicitly pinning matter and standard model forces onto a (3+1)--dimensional brane Universe while allowing
gravity to explore the larger, higher-dimensional space, all nongravitational physics follows the standard phenomenology.  Ultimately the game in braneworld theories is to find a means by which to hide the extra dimensions from gravity as well.  Gravity is altered in those regimes where the extra dimensions manifest themselves.  If we wish to explain today's cosmic acceleration as a manifestation of extra dimensions, it makes sense to devise a braneworld theory where the extra dimensions are revealed at only the largest of observable distance scales.

\subsection{The Formal Arena}

The braneworld theory \cite{Dvali:2000hr} of Dvali, Gabadadze, and Porrati (DGP) represents a leading model for understanding cosmic acceleration as a manifestation of new gravity.  The bulk is this model
is an empty five-dimensional Minkowski space; all energy-momentum is isolated on the
four-dimensional brane Universe.  The theory is described by the action \cite{Dvali:2000hr}:
\be
S_{(5)} = -\frac{1}{16\pi}M^3 \int d^5x
\sqrt{-g}~R +\int d^4x \sqrt{-g^{(4)}}~{\cal L}_m + S_{GH}\ .
\label{action}
\ee
$M$ is the fundamental five-dimensional Planck scale. The first term in $S_{(5)}$ is the  Einstein-Hilbert action in five dimensions for a five-dimensional metric $g_{AB}$ (bulk metric) with Ricci scalar $R$ and determinant $g$.  The metric $g^{(4)}_{\mu\nu}$ is the induced (four-dimensional) metric on the brane, and $g^{(4)}$ is its determinant.\footnote{
	Throughout this paper, we use $A,B,\dots = \{0,1,2,3,5\}$ as
	bulk indices, $\mu,\nu,\dots = \{0,1,2,3\}$ as brane spacetime
	indices, and $i,j,\dots = \{1,2,3\}$ as brane spatial indices.}
The contribution $S_{GH}$ to the action is a pure divergence necessary to ensure proper boundary conditions in the Euler-Lagrange equations.  An intrinsic curvature term is added to the brane action~\cite{Dvali:2000hr}:
\be
-\frac{1}{16\pi}M^2_P \int d^4x \sqrt{-g^{(4)}}\ R^{(4)}\ .
\label{action2}
\ee
Here, $M_P$ is the observed four-dimensional Planck scale.\footnote{
	Where would such a term come from?  The intrinsic curvature term may be viewed as coming from
	effective-action terms induced by quantum matter fluctuations that live exclusively on the brane
	 Universe  (see \cite{Dvali:2000hr,Dvali:2001gm,Dvali:2001gx} for details).  There is an ongoing
	 discussion as to whether this theory is unstable to further quantum
	 {\em gravity} corrections on the brane that reveal themselves at phenomenologically important scales \cite{Luty:2003vm,Rubakov:2003zb,Porrati:2004yi,Dvali:2004ph,Gabadadze:2004jk,Nicolis:2004qq,Barvinsky:2005db,Deffayet:2005ys}.  While several topics covered here are indeed relevant to that discussion
	 (particularly Sec.~\ref{sec:einstein}), rather than becoming embroiled in this technical issue,
	 we studiously avoid quantum gravity issues here and treat gravity as given by Eqs.~(\ref{action})
	 and~(\ref{action2}) for our discussion, and that for cosmological applications, classical gravity physics
	 is sufficient.}

The gravitational field equations resulting from the action Eqs.~(\ref{action}) and~(\ref{action2}) are
\be
	M^3G_{AB} + M_P^2~\delta\left(x^5 - z(x^\mu)\right) G^{(4)}_{AB}
		= 8\pi~\delta\left(x^5 - z(x^\mu)\right)T_{AB}(x^\mu)\ ,
\label{Einstein}
\ee
where $G_{AB}$ is the five-dimensional Einstein tensor, $G^{(4)}$ is the Einstein tensor of the induced metric on the brane $g^{(4)}_{\mu\nu}$, and
where $x^5$ is the extra spatial coordinate and $z(x^\mu)$ represents the location of the
brane as a function of the four-dimensional coordinates of our brane Universe, $\{x^\mu\}$.
Note that the energy-momentum tensor only resides on the brane surface, as we have
constructed.

While the braneworld paradigm has often been referred to as ``string-inspired," we are not
necessarily wedded to that premise.  One can imagine a more conventional scenario where
physics is still driven by field theory, where the brane is some sort of solitonic domain wall
and conventional particles and standard model forces are states bound to the domain wall
using usual quantum arguments.  This approach does require that DGP gravity still exhibits
the same properties as described in this review when the brane has a nonzero thickness \cite{Kiritsis:2001bc,Middleton:2002qa,Kolanovic:2003da,Middleton:2003yr,Porrati:2004yi}.  While the situation
seems to depend on the specifics of the brane's substrucutre, there exist specific scenarios
in which it is possible to enjoy the features of DGP gravity with a thick, soliton-like brane.

Unlike other braneworld theories, DGP gravity has a fanciful sort of appeal;  it uncannily resembles the
Flatland-like world one habitually envisions when extra dimensions are invoked.  The bulk is large and
relatively flat enjoying the properties usually associated with a Minkowski spacetime.  Bulk observers
may look down from on high upon the brane Universe, which may be perceived as being an imbedded
surface in this larger bulk.  It is important to note that the brane position remains fully dynamical and is
determined by the field equations, Eqs.~(\ref{Einstein}).  While a coordinate system may devised in which the brane appears flat, the brane's distortion and motion are, in that situation, registered
through metric variables that represent the brane's extrinsic curvature.  This is a technique used
repeatedly in order to ease the mathematical description of the brane dynamics.  Nevertheless, we
will often refer to a brane in this review as being warped or deformed or the like.  This terminology is
just shorthand for the brane exhibiting a nonzero extrinsic curvature while imagining a different coordinate system in which the brane is nontrivially imbedded.

\subsection{Preliminary Features}

In order to get a qualitative picture of how gravity works for DGP branewolrds, let us take small
metric fluctuations around flat, empty space and look at gravitational perturbations, $h_{AB}$, where
\be
	g_{AB} = \eta_{AB} + h_{AB}\ ,
\ee
where $\eta_{AB}$ is the five-dimensional Minkowski metric.  Choosing the harmonic gauge in
the bulk
\be
	\d^Ah_{AB} = {1\over 2}\d_Bh^A_A\ ,
\ee
where the $\mu 5$--components of this gauge condition leads to $h_{\mu 5} =0$ so that the
surviving components are $h_{\mu\nu}$ and $h_{55}$.  The latter component is solved by
the following:
\be
	\Box^{(5)}h_5^5 = \Box^{(5)}h_\mu^\mu\ ,
\ee
where $\Box^{(5)}$ is the five-dimensional  d'Alembertian.  The $\mu\nu$--component of the field
equations Eqs.~(\ref{Einstein}) become, after a little manipulation \cite{Dvali:2001gx},
\be
	M^3\Box^{(5)}h_{\mu\nu} + M_P^2\delta(x^5)(\Box^{(4)}h_{\mu\nu} - \d_\mu\d_\nu h_5^5)
				= 8\pi\left(T_{\mu\nu} - {1\over 3}\eta_{\mu\nu}T_\alpha^\alpha\right)\delta(x^5)\ ,
\label{prop-eqn}
\ee
where $\Box^{(4)}$ is the four-dimensional (brane) d'Alembertian, and where we take the
brane to be located at $x^5 = 0$.  Fourier transforming just the four-dimensional spacetime
$x^\mu$ to corresponding momentum coordinates $p^\mu$, and applying boundary conditions
that force gravitational fluctuations to vanish one approaches spatial infinity, then gravitational
fluctuations on the brane take the form \cite{Dvali:2001gx}
\be
	\tilde{h}_{\mu\nu}(p,x^5 = 0) = {8\pi\over M_P^2p^2 + 2M^3 p}
		\left[\tilde{T}_{\mu\nu}(p^\lambda)
		- {1\over 3}\eta_{\mu\nu}\tilde{T}_\alpha^\alpha(p^\lambda)\right]\ .
\label{prop}
\ee
We may recover the behavior of the gravitational potentials from this expression.

There exists a new physical scale, the crossover scale
\be
	r_0 = {\mpsq \over 2M^3}\ ,
\label{r0}
\ee
that governs the transition between four-dimensional behavior and five-dimensional
behavior.  Ignoring the tensor structure of Eq.~(\ref{prop}) until future sections, the
gravitational potential of a source of mass $m$ is
\be
	V_{\rm grav} \sim -{G_{\rm brane}m\over r}\ ,
\ee
when $r \ll r_0$.  When $r \gg r_0$
\be
	V_{\rm grav} \sim -{G_{\rm bulk}m\over r^2}\ ,
\ee
where the gravitational strengths are given by $G_{\rm bulk} = M^{-3}$ and $G_{\rm brane} = M_P^2$.
I.e., the potential exhibits four-dimensional behavior at short distances and
five-dimensional behavior (i.e., as if the brane were not there at all) at large distances.
For the crossover scale to be large, we need a substantial mismatch between $M_P$, the conventional
four-dimensional Planck scale (corresponding to the usual Newton's constant, $G_{\rm brane} = G$) and the fundamental, or bulk, Planck scale $M$.  The fundamental Planck scale $M$ has to be quite small\footnote{
	To have $r_0 \sim H_0^{-1}$, today's Hubble radius, one needs $M \sim 100\ {\rm MeV}$.  I.e.,
	bulk quantum gravity
	effects will come into play at this low an energy.  How does a low-energy quantum gravity
	not have intolerable effects on standard model processes on the brane?  Though one does
	not have a complete description of that quantum gravity, one may argue that there is a
	decoupling mechanism that
	protects brane physics from bulk quantum effects \cite{Dvali:2001gx}.}
 in order for the
energy of gravity fluctuations to be substantially smaller in the bulk versus on the brane, the energy of
the latter being controlled by $M_P$.  Note that when $M$ is small, the corresponding Newton's
constant in the bulk, $G_{\rm bulk}$, is large.  Paradoxically, for a given source mass $m$, gravity is much stronger in the bulk.

\begin{figure} \begin{center}\PSbox{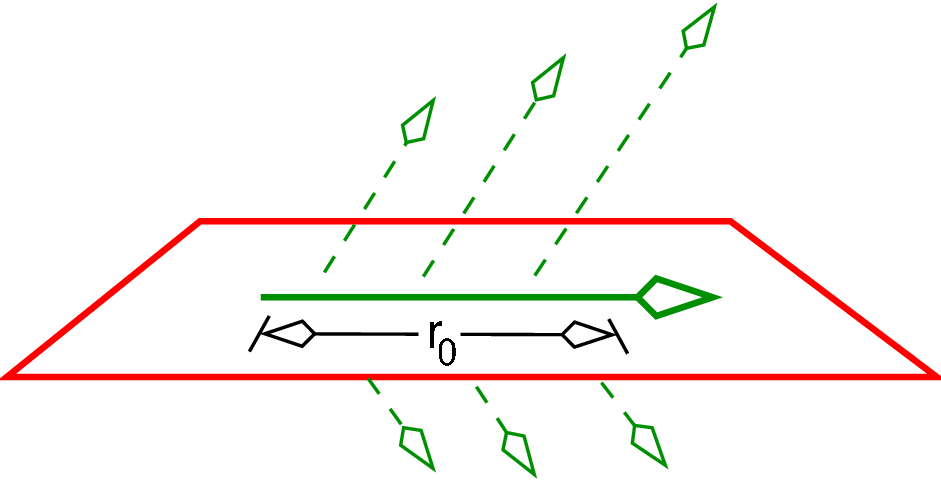
hscale=100 vscale=100 hoffset=0 voffset=0}{4in}{1.8in}\end{center}
\caption{At distances much smaller than the crossover scale $r_0$, gravity
appears four-dimensional.  As a graviton propagates, it's amplitude leaks
into the bulk.  On scales comparable to $r_0$ that amplitude is attenuated
significantly, thus, revealing the extra dimension.
}
\label{fig:propagator}
\end{figure}

There is a simple intuition for understanding how DGP gravity works and why the gravitational potential has its distinctive form.  When $M_P \gg M$, there is a large mismatch between the energy of a gravitational fluctuation on the brane versus that in the bulk.  I.e., imagine a gravitational field of a given amplitude (unitless) and size (measured in distance).  The corresponding energies are roughly
\bea
	E_{\rm brane} &\sim& M_P^2\int d^3x (\partial h)^2 \sim M_P^2 \times\ {\rm size}  \\
		E_{\rm brane} &\sim& M^3\int d^3x dx^5(\partial h)^2 \sim M^3 \times\ ({\rm size})^2
		\sim E_{\rm brane} \times {{\rm size}\over r_0}\ .
\eea
What happens then is that while gravitational fluctuations and field are free to explore the entire five-dimensional space unfettered, they are much less substantial energetically in the bulk.  Imagine an analogous situation.  Consider the brane as a metal sheet immersed in air.  The bulk modulus of the metal is much larger than that of air.  Now imagine that sound waves represent gravity.  If one strikes the metal plate, the sound wave can propagate freely along the metal sheet as well as into the air.  However, the energy of the waves in the air is so much lower than that in the sheet, the wave in the sheet attenuates very slowly, and the wave propagates in the metal sheet virtually as if there were
no bulk at all.  Only after the wave has propagated a long distance is there a substantial amount of attenuation and an observer on the sheet can tell an ``extra" dimension exists, i.e., that the sound
energy must have been lost into some unknown region (Fig.~\ref{fig:propagator}).  Thus at short
distances on the sheet, sound physics appears lower dimension, but at larger distances corresponding
to the distance at which substantial attenuation has occurred, sound physics appears higher dimensional.  In complete accord with this analogy, what results in DGP gravity is a model where
gravity deviates from conventional Einstein gravity at distances larger than $r_0$.

\begin{figure} \begin{center}\PSbox{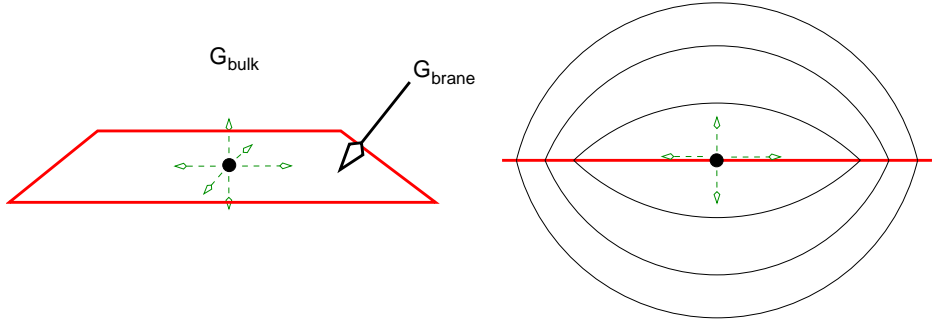
hscale=60 vscale=60 hoffset=0 voffset=0}{5in}{1.5in}\end{center}
\caption{A point source of mass $m$ is located on the brane.  The gravitational
constant on the brane is $G_{\rm brane} = M_P^{-2}$, whereas in the bulk $G_{\rm bulk} = M^{-3}$ (left diagram).
Gravitational equipotential surfaces, however, are not particularly pathological.  Near the matter
source, those surfaces are lens-shaped.  At distances farther from the matter source, the equipotential
surfaces become increasingly spherical, asymptoting to equipotentials of a free point source in
five-dimensions.  I.e., in that limit the brane has no influence (right diagram).
}
\label{fig:bias}
\end{figure}

While a brane observer is shielded from the presence of the extra dimensions at distance scales shorter
than the crossover scale, $r_0$.  But from the nature of the above analogy, it should be clear that
the bulk is not particularly shielded from the presence of brane gravitational fields.
In the bulk, the solution to Eq.~(\ref{prop-eqn}) for a point mass has equipotential surfaces as
depicted in Fig.~\ref{fig:bias}.  From the bulk perspective, the brane looks like a conductor which imperfectly repels gravitational potential lines of sources existing away from the brane, and one
that imperfectly screens the gravitational potential of sources located on the brane \cite{Kolanovic:2002uj}.

\section{Cosmology and Global Structure}

Just as gravity appears four-dimensional at short distances and five-dimensional at large distances,
we expect cosmology to be altered in a corresponding way.  Taking the qualitative features
developed in the last section, we can now construct cosmologies from the field equations
Eqs.~(\ref{Einstein}).  We will find that the cosmology of DGP gravity provides an intriguing
avenue by which one may explain the contemporary cosmic acceleration as the manifestation
of an extra dimension, revealed at distances the size of today's Hubble radius.

\subsection{The Modified Friedmann Equation}

The first work relevant to DGP cosmolgy appeared even before the article by Dvali, Gabadadze and Porrati, though these studies were in the context of older braneworld theories \cite{Collins:2000yb,Shtanov:2000vr,Nojiri:2000gv}.  We follow here the approach of Deffayet \cite{Deffayet}
who first noted how one recovers a self-accelerating solution from the DGP field equations
Eqs.~(\ref{Einstein}).  The general time-dependent line element with
the isometries under consideration is of the form
\begin{equation} \label{cosmback}
ds^{2} = N^{2}(\tau,z) d\tau^{2}
         -A^{2}(\tau,z)\delta_{ij}d\lambda^{i}d\lambda^{j}
         -B^{2}(\tau,z)dz^{2}\ ,
\label{metric1}
\end{equation}
where the coordinates we choose are $\tau$, the cosmological time; $\lambda^i$, the spatial
comoving coordinates of our observable Universe;  and $z$, the extra dimension into the bulk.
The three-dimensional spatial metric is the Kronecker delta, $\delta_{ij}$, because we
focus our attention on spatially-flat cosmologies.  While the analysis was originally done for
more general homogeneous cosmologies,
we restrict ourselves here to this observationally viable scenario.

Recall that all energy-momentum resides on the brane, so that the bulk is empty. The field
equations Eqs.~(\ref{Einstein}) reduce to
\bea
	G_0^0 &=& {3\over N^2}\left[{\dot{A}\over A}\left({\dot{A}\over A} + {\dot B\over B}\right)\right]
		- {3\over B^2}\left[{A''\over A} + {A'\over A}\left({A'\over A} - {B'\over B}\right)\right] = 0  \\
	G_j^i &=& {1\over N^2}\delta_j^i\left[{2\ddot A\over A} + {\ddot B\over B}
			- {\dot A\over A}\left({2\dot N\over N} - {\dot A\over A}\right)	\nonumber
			- {\dot B\over B}\left({\dot N\over N} - {2\dot A\over A}\right)\right]  \\
		&& \ \ \ - {1\over B^2}\delta_j^i\left[{N''\over N} +{2A''\over A}
			+ {A'\over A}\left({2N'\over N} + {A'\over A}\right)
			- {B'\over B}\left({N'\over N} +{2A'\over A}\right)\right] = 0  \\
	G_5^5 &=& {3\over N^2}\left[{\ddot A\over A}
		- {\dot{A}\over A}\left({\dot{N}\over N} - {\dot A\over A}\right)\right]
		- {3\over B^2}\left[{A'\over A}\left({N'\over N} + {A'\over A}\right)\right] = 0  \\
	G_{05} &=& 3\left[{\dot A'\over A} - {\dot A\over A}{N'\over N} - {\dot B\over B}{A'\over A}\right] = 0\ ,
\eea
in the bulk.  Prime denotes differentiation with respect to $z$ and dot denotes
differentiation with respect to $\tau$.  We take the brane to be located at $z = 0$.  This
prescription
does not restrict brane surface dynamics as it is tantamount to a residual coordinate
gauge choice.  Using a technique first developed in Refs.~\cite{Binetruy:1999ut,Shiromizu:1999wj,Binetruy:1999hy,Flanagan:1999cu}, the bulk equations, remarkably, may be
solved exactly given that the bulk energy-momentum content is so simple (indeed, it
is empty here).  Taking the bulk to be mirror (${\cal Z}_2$) symmetric across the brane,
we find the metric components in Eq.~(\ref{metric1}) are given by \cite{Deffayet}
\begin{eqnarray}
N &=& 1 \mp |z| {\ddot{a}\over \dot{a}} \nonumber \\
A &=& a \mp |z| \dot{a}	\label{bulkmet1} \\
B  &=& 1\ , \nonumber
\end{eqnarray}
where there remains a parameter $a(\tau)$ that is yet to be determined.  Note that
$a(\tau) = A(\tau, z=0)$ represents the usual scale factor of the four-dimensional
cosmology in our brane Universe.  Note that there are {\em two distinct possible cosmologies}
associated with the choice of sign.  They have very different global structures and
correspondingly different phenomenologies as we will illuminate further.
	
First, take the total energy-momentum tensor which includes matter and the cosmological
constant on the brane to be
\begin{equation}
T^A_B|_{\rm brane}= ~\delta (z)\ {\rm diag}
\left(-\rho,p,p,p,0 \right)\ .
\end{equation}
In order to determine the scale factor $a(\tau)$, we need to employ the proper boundary
condition at the brane.  This can be done by taking Eqs.~(\ref{Einstein}) and integrating those
equations just across the brane surface.  Then, the boundary conditions at the brane
requires
\bea
	\left.{N'\over N}\right|_{z=0} &=& - {8\pi G\over 3}\rho	\\
	\left.{A'\over A}\right|_{z=0} &=& {8\pi G\over 3}(3p + 2\rho)\ .
\eea
Comparing this condition to the bulk solutions, Eqs.~(\ref{bulkmet1}), these junction
conditions require a constraint on the evolution of $a(\tau)$.  Such an evolution is
tantamount to a new set of Friedmann equations \cite{Deffayet}:
\be
	H^2 \pm {H\over r_0} = {8\pi G\over 3}\rho(\tau)\ .
\label{Fried}
\ee
and
\be
\dot{\rho} + 3(\rho+p)H = 0\ ,
\label{friedmann2}
\ee
where we have used the usual Hubble parameter $H = \dot a/a$.  The second of these
equations is just the usual expression of energy-momentum conservation.  The first
equation, however, is indeed a new Friedmann equation that is a modification of the
convential four-dimensional Friedmann equation of the standard cosmological model.

Let us examine Eq.~(\ref{Fried}) more closely.  The new contribution from the DGP braneworld
scenario is the introduction of the term $\pm H/r_0$ on the left-hand side of the Friedmann equation.  The choice of sign represent two distinct cosmological phases.   Just as gravity is conventional four-dimensional gravity at short scales and appears five-dimensional at large distance scales, so too the Hubble scale, $H(\tau)$, evolves by the conventional Friedmann equation at high Hubble scales but is altered substantially as $H(\tau)$ approaches $r_0^{-1}$.

\begin{figure} \begin{center}\PSbox{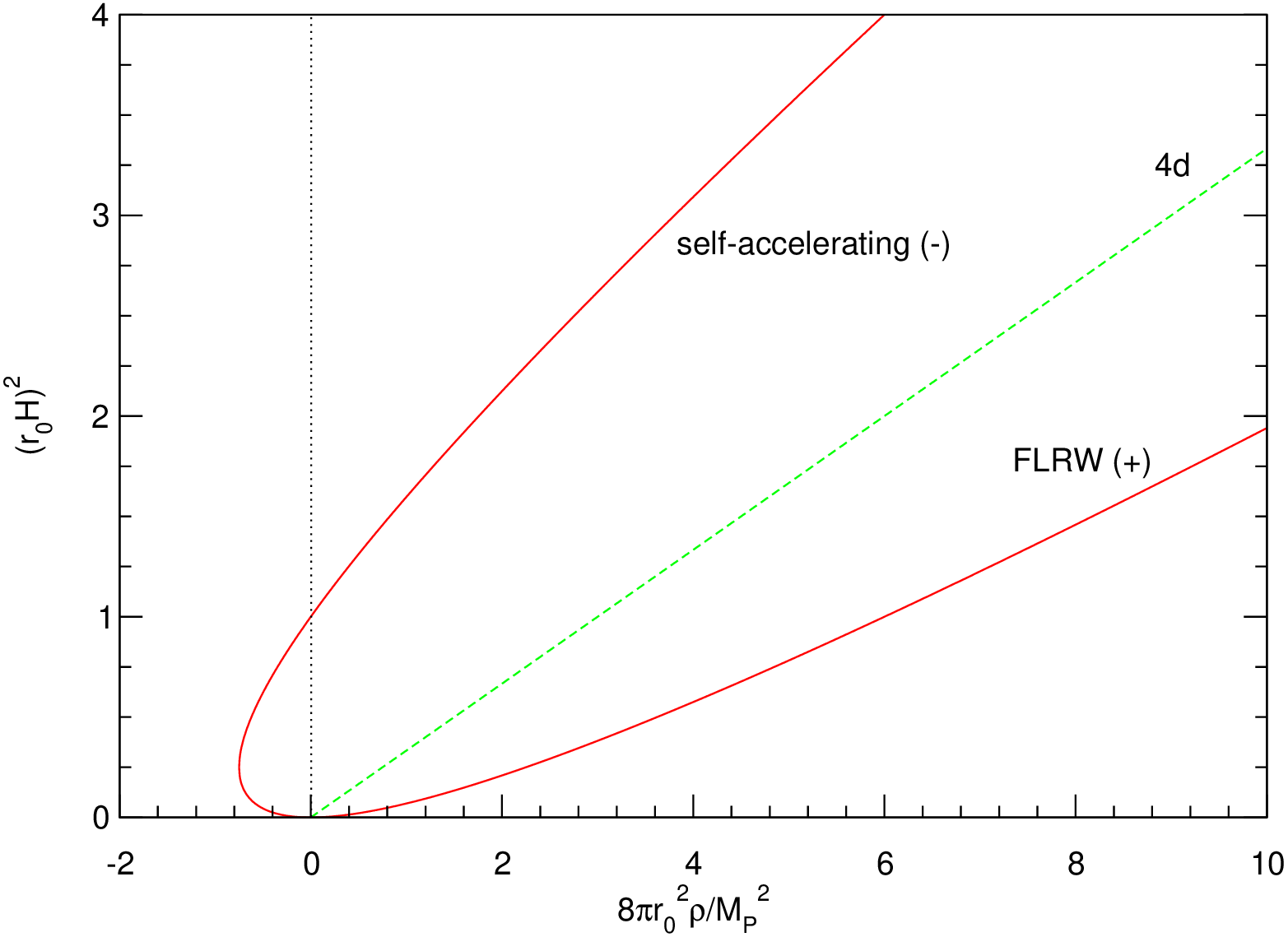
hscale=50 vscale=50 hoffset=-90 voffset=-20}{3in}{3.2in}\end{center}
\caption{
The solid curve depicts Eq.~(\ref{Fried}) while the dotted line represents the conventional four-dimensional Friedmann equation.  Two cosmological phases clearly emerge for any given spatially-homogeneous energy-momentum distribution.
}
\label{fig:cosmo}
\end{figure}

Figure~\ref{fig:cosmo} depicts the new Friedmann equation.  Deffayet \cite{Deffayet} first noted that
there are two distinct cosmological phases.  First, there exists the phase in Eq.~(\ref{Fried}) employing
the upper sign which had already been established in Refs.~\cite{Collins:2000yb,Shtanov:2000vr,Nojiri:2000gv}, which transitions between $H^2\sim \rho$ to $H^2\sim \rho^2$.  We refer to this phase as
the Friedmann--Lema\^{\i}tre--Robertson--Walker (FLRW) phase.
The other cosmological phase corresponds to the lower sign in Eq.~(\ref{Fried}).  Here, cosmology
at early times again behaves according to a conventional four-dimensional Friedmann equation,
but at late
times asymptotes to a brane self-inflationary phase (the asymptotic state was first noted by Shtanov
\cite{Shtanov:2000vr}).

In this latter self-accelerating phase, DGP gravity provides an alternative explanation for today's cosmic acceleration.  If one were to choose the cosmological phase associated with the lower sign in Eq.~(\ref{Fried}), and set the crossover distance scale to be on the order of $H_0^{-1}$, where $H_0$ is today's Hubble scale, DGP could account for today's cosmic acceleration in terms of the existence of extra dimensions and a modifications of the laws of gravity.  The Hubble scale, $H(\tau)$, evolves by the conventional Friedmann equation when the universe is young and $H(\tau)$ is large.  As the universe expands and energy density is swept away in this expansion, $H(\tau)$ decreases.  When the Hubble scale approaches a critical threshold, $H(\tau)$ stops decreasing and saturates at a constant value, even when the Universe is devoid of energy-momentum content.  The Universe asymptotes to a deSitter accelerating expansion.

\subsection{The Brane Worldsheet in a Minkowski Bulk}
\label{sec:global}

It is instructive to understand the meaning of the two distinct cosmological phases in the new
Friedmann equation Eq.~(\ref{Fried}).  In order to do that, one must acquire a firmer grasp on the
global structure of the bulk geometry Eqs.~({\ref{bulkmet1}) and its physical meaning \cite{Deffayet,Lue:2002fe}.  Starting with Eqs.~(\ref{bulkmet1}) and using the technique developed by Deruelle and Dolezel \cite{Deruelle:2000ge} in a more general context, an explicit change of coordinates may be obtained to go to a canonical five-dimensional Minkowskian metric
\begin{equation}
ds^2 = dT^2 - (dX^1)^2 - (dX^2)^2 - (dX^3)^2 - (dY^5)^2.
 \label{canon}
\end{equation}
The bulk metric in this cosmological environment is strictly flat, and whatever nontrivial spacetime
dynamics we experience on the brane Universe is derived from the particular imbedding of a nontrivial brane surface evolving in this trivial bulk geometry.  Rather like the popular physics picture of our
standard cosmology, we may think of our Universe as a balloon or some deformable surface expanding and evolving in a larger-dimensional space.  Here in DGP gravity, that picture is literally true.

The coordinate transformation from Eqs.~(\ref{bulkmet1}) to the explicitly flat metric Eq.~(\ref{canon})
is given by
\begin{eqnarray}
T &=& A(z,\tau) \left( \frac{\lambda^2}{4}+1-\frac{1}{4\dot{a}^2} \right)
-\frac{1}{2}\int d\tau \frac{a^2}{\dot{a}^3}~
{d\ \over d\tau}\left(\frac{\dot{a}}{a}\right)\ ,	\nonumber	\\
Y^5 &=& A(z,\tau) \left( \frac{\lambda^2}{4}-1-\frac{1}{4\dot{a}^2} \right)
-\frac{1}{2} \int d\tau \frac{a^2}{\dot{a}^3}~
{d\ \over d\tau}\left(\frac{\dot{a}}{a}\right)\ ,  \label{changun} \\
X^i &=&  A(z,\tau) \lambda^i\ ,	\nonumber
\end{eqnarray}
where $\lambda^2= \delta_{ij}\lambda^i\lambda^j$.

For clarity, we can focus on the early universe of cosmologies of DGP
braneworlds to get a picture of the global structure, and in particular of the
four-dimensional big bang evolution at early times.  Here, we restrict ourselves
to radiation domination (i.e., $p = {1\over 3}\rho$) such that, using
Eqs.~(\ref{Fried}) and (\ref{friedmann2}), $a(\tau) = \tau^{1/2}$ when $H
\gg r_0^{-1}$ using appropriately normalized time units.  A more
general equation of state does not alter the qualitative picture.  Equation~(\ref{Fried})
shows that the early cosmological evolution on the brane is independent of
the sign choice, though the bulk structure will be very different for the two different
cosmological phases through the persistence of the sign in Eqs.~(\ref{bulkmet1}).
The global configuration of the brane worldsheet is determined by
setting $z = 0$ in the coordinate transformation Eq.~(\ref{changun}).
We get
\bea
	T &=& \tau^{1/2}\left(\frac{\lambda^2}{4}+1-\tau\right)
					- {4\over 3}\tau^{3/2} \nonumber	\\
	Y^5 &=& \tau^{1/2}\left(\frac{\lambda^2}{4}-1-\tau\right)
					- {4\over 3}\tau^{3/2}
\label{Ybrane}		\\
	X^i &=&  \tau^{1/2} \lambda^i\ .		\nonumber
\eea
The locus of points defined by these equations, for all $(\tau,\lambda^i)$,
satisfies the relationship
\be
	Y_+ = {1\over 4Y_-}\sum_{i=1}^3(X^i)^2 + {1\over 3}Y_-^3\ ,
\label{branesurf}
\ee
where we have defined $Y_\pm = {1\over 2}(T\pm Y^5)$.  Note that if
one keeps only the first term, the surface defined by Eq.~(\ref{branesurf})
would simply be the light cone emerging from the origin at $(T,X^i,Y^5) = 0$.
However, the second term ensures that the brane worldsheet is
timelike except along the $Y_+$--axis.  Moreover, from
Eqs.~(\ref{Ybrane}), we see that
\be
	Y_- = \tau^{1/2}\ ,
\ee
implying that $Y_-$ acts as an effective cosmological time coordinate
on the brane.  The $Y_+$--axis is a singular locus corresponding to
$\tau=0$, or the big bang.\footnote{
The big bang singularity when $r < \infty$ is just the origin $Y_- =
Y_+ = X^i = 0$ and is strictly pointlike.  The rest of the big bang
singularity (i.e., when $Y_+ > 0$) corresponds to the pathological
case when $r = \infty$.
}

\begin{figure} \begin{center}\PSbox{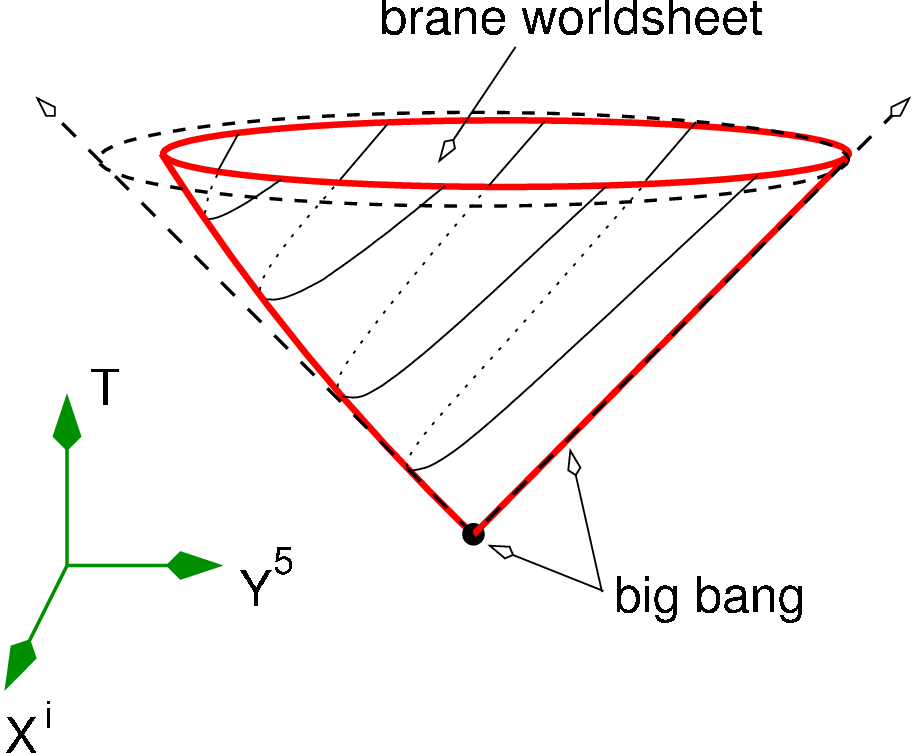
hscale=100 vscale=100 hoffset=-50 voffset=0}{3in}{2.8in}\end{center}
\caption{
A schematic representation of the brane worldsheet from an inertial
bulk reference frame.  The bulk time coordinate, $T$, is the
vertical direction, while the other directions represent all four spatial bulk coordinates,
$X^i$ and $Y^5$.  The big bang is located along the locus $Y^5 = T$, while the dotted surface is the future lightcone of the event located at the origin denoted by the solid dot.  The curves on the brane worldsheet are examples of equal cosmological time, $\tau$, curves and each is in a plane of constant $Y^5 + T$.  Figure from Ref.~[28].
}
\label{fig:brane}
\end{figure}

This picture is summarized in Figs.~\ref{fig:brane} and \ref{fig:exp}.  Taking $Y^0$ as
its time coordinate, a bulk observer perceives the braneworld as a
compact, hyperspherical surface expanding relativistically from an
initial big bang singularity.  Figure~\ref{fig:brane} shows a spacetime diagram
representing this picture where the three dimensional hypersurface of the brane
Universe is depicted as a circle with each time slice.  Note that a bulk
observer views the braneworld as spatially compact, even while a
cosmological brane observer does not.  Simultaneously, a bulk observer
sees a spatially varying energy density on the brane, whereas a
brane observer perceives each time slice as spatially homogeneous.
Figure~\ref{fig:exp} depicts the same picture as Fig.~\ref{fig:brane}, but with each
successive image in the sequence representing
a single time slice (as seen by a bulk observer).  The big bang starts as a strictly
pointlike singularity, and the brane surface looks like a relativistic shock originating
from that point.  The expansion factor evolves by Eq.~(\ref{Fried}) implying that at
early times near the big bang, its expansion is indistinguishable from a four-dimensional
FLRW big bang and the expansion of the brane bubble decelerates.  However, as
the size of the bubble becomes comparable to $r_0$, the expansion of the
bubble starts to deviate significantly from four-dimensional FLRW.

\begin{figure} \begin{center}\PSbox{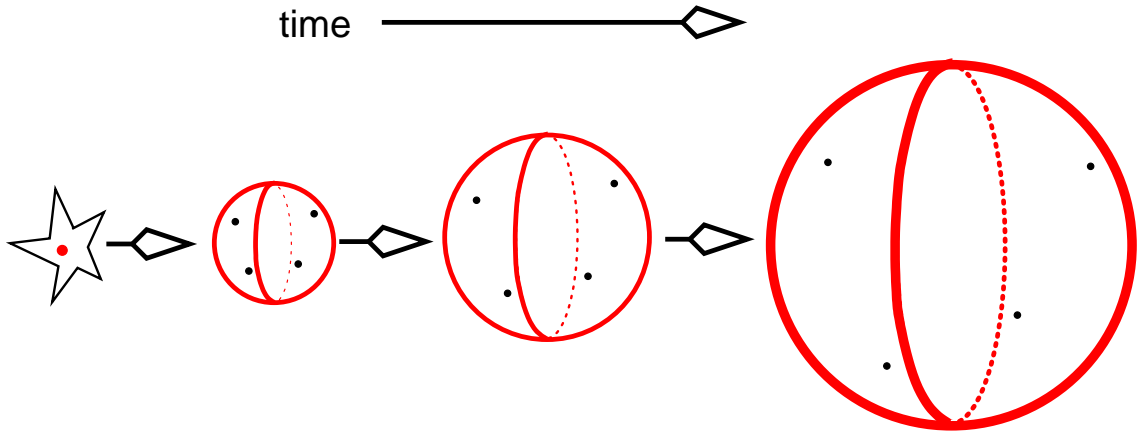
hscale=100 vscale=100 hoffset=0 voffset=0}{6in}{2in}\end{center}
\caption{
Taking time-slicings of the spacetime diagram shown in Fig.~\ref{fig:brane} (and now only suppressing
one spatial variable, rather than two) we see that the big bang starts as a pointlike singularity and
the brane universe expands as a relativistic shockwave from that origin.  Resulting from the
peculiarities of the coordinate transformation, the big-bang persists eternally as a lightlike singularity
(see Fig.~\ref{fig:brane}) so that for any given time slice, one point on the brane surface is singular
and is moving at exactly the speed of light relative to a bulk observer.
}
\label{fig:exp}
\end{figure}
  
Though the brane cosmological evolution between the FLRW phase and the
self-accelerating phase is indistinguishable at early times, the bulk
metric Eqs.~(\ref{bulkmet1}) for each phase is quite
distinct.  That distinction has a clear geometric interpretation: The
FLRW phase (upper sign) corresponds to that part of the bulk
interior to the brane worldsheet, whereas the self-accelerating phase
(lower sign) corresponds to bulk exterior to the brane
worldsheet  (see Fig.~\ref{fig:shock}).  The full bulk space is two copies of either the interior
to the brane worldsheet (the FLRW phase) or the exterior (the self-accerating
phase), as imposed by ${\cal Z}_2$--symmetry.  Those two copies are
then spliced across the brane, so it is clear that the full bulk space cannot really be
represented as imbedded in a flat space.

\begin{figure} \begin{center}\PSbox{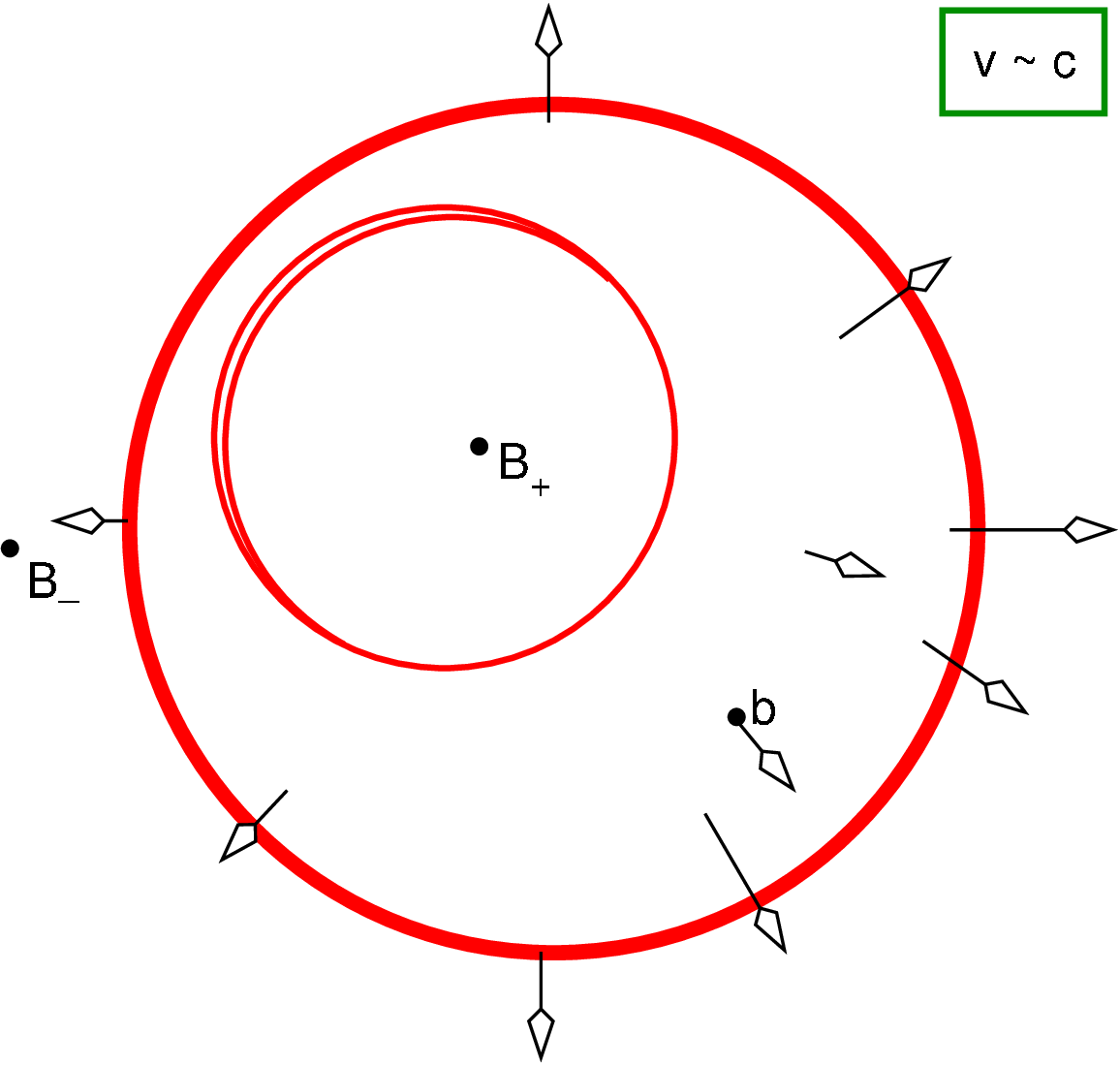
hscale=50 vscale=50 hoffset=-20 voffset=0}{2in}{2in}\end{center}
\caption{
The brane surface at a given time instant as seen from a inertial bulk observer.
While from a brane observer's point of view (observer $b$), a constant-time slice of
the universe is infinite in spatial extent, from a bulk observer's point of view, the
brane surface is always compact and spheroidal (imagine taking a time slice in
Fig.~\ref{fig:brane}).  That spheroidal brane surface expands at near the speed
of light from a bulk observer's point of view.  In the FLRW phase, a bulk observer
exists only inside the expanding brane surface, watching that surface expand
away from him/her (observer $B_+$).  In the self-accelerating phase, a bulk
observer only exists {\em outside} the expanding brane surface, watch that
surface expand towards him/her (observer $B_-$).
}
\label{fig:shock}
\end{figure}

It is clear that the two cosmological phases really are quite distinct, particularly at
early times when the universe is small.  In the FLRW phase, the bulk is the tiny
interior of a small brane bubble.  From a bulk observer's point of view,
space is of finite volume, and he/she witnesses that bulk space grow as the brane bubble
expands away from him/her.  The intriguing property of this space is that there
are shortcuts through the bulk one can take from any point on the brane Universe to
any other point.  Those shortcuts are faster than the speed of light on the brane
itself, i.e., the speed of a photon stuck on the brane surface \cite{Lue:2002fe}.
In the self-accelerating phase, the bulk is two copies of the infinite volume exterior,
spliced across the tiny brane bubble.  Here a bulk observer witnesses the brane
bubble rapidly expanding towards him/her, and eventually when the bubble size is
comparable to the crossover scale $r_0$, the bubble will begin to accelerate towards
the observer approaching the speed of light.  Because of the nature of the bulk space,
one {\em cannot} take shortcuts through the bulk.  The fastest way from point A to B
in the brane Universe is disappointingly within the brane itself.

\subsection{Luminosity Distances and Other Observational Constraints}

How do we connect our new understanding of cosmology in DGP gravity to the real
(3+1)--dimensional world we actually see?  Let us focus
our attention on the expansion history governed by Eq.~(\ref{Fried}) and ask how one
can understand this with an eye toward comparison with existing data.
In the dark energy paradigm, one assumes that general relativity is still valid and that today's cosmic
acceleration is driven by a new smooth, cosmological component of sufficient negative
pressure (referred to as dark energy) whose energy density is given by $\rho_{DE}$ and
so that the expansion history of the universe is driven by the usual Friedmann equation
\be
	H^2 = {8\pi G\over 3}\left(\rho_M + \rho_{DE}\right)\ ,
\label{oldFried}
\ee
the dark energy has an equation of state $w = p_{DE}/\rho_{DE}$, so that
\be
      \rho_{DE}(\tau) = \rho^0_{DE}a^{-3(1+w)}\ ,
\label{eos}
\ee
if $w$ is constant;  whereas $\rho_{DE}$ has more complex time dependence if $w$ varies
with redshift.  Dark energy composed of just a cosmological constant ($w = -1$) is fully consistent with existing observational data;  however, both $w > -1$ and $w < -1$ remain observationally
viable possibilities \cite{Riess}.

We wish to get a more agile feel for how the modified Friedmann equation of DGP gravity,
Eq.~(\ref{Fried}),
\bea
	H^2 \pm {H\over r_0} = {8\pi G\over 3}\rho(\tau)\ ,
\nonumber
\eea
behaves in of itself as an evolution equation.  We are concerned with the situation where
the Universe is only populated with pressureless energy-momentum constituents, $\rho_M$,
while still accelerating in its expansion at late time.  We must then focus on the self-accelerating
phase (the lower sign) so that
\be
     H^2 - {H\over r_0} = {8\pi G\over 3}\rho_M(\tau)\ .
\label{Fried2}
\ee
Then the effective dark energy density of the modified Friedmann equation is then
\be
     {8\pi G\over 3}\rho^{\rm eff}_{DE} = {H\over r_0}\ .
\label{DEeff}
\ee
The expansion history of this model and its corresponding luminosity distance redshift
relationship was first studied in Ref.~\cite{Deffayet:2001pu,Deffayet:2002sp}.

By comparing this expression to Eq.~(\ref{eos}), one can mimic a $w$--model,
albeit with a time-varying $w$.  One sees immediately that the effective dark energy
density attenuates with time in the self-accelerating phase.  Employing the intuition
devised from Eq.~(\ref{eos}), this implies that the effective $w$ associated with this
effective dark energy must always be greater than negative one.\footnote{
	It must be noted that if one were to go into the FLRW phase, rather than self-accelerating
	phase, and relax the presumption that the cosmological constant be zero (i.e.,
	abandon the notion of completely replacing dark energy), then there exists the intriguing
	possibility of gracefully achieving $w_{\rm eff} < -1$ without violating the null-energy
	condition, without ghost instabilities and without a big rip \cite{Sahni:2002dx,Alam:2002dv,Lue:2004za}.}

What are the parameters in this model of cosmic expansion?  We assume that the
universe is spatially flat, as we have done consistently throughout this review.  Moreover,
we assume that $H_0$ is given.  Then one may define the parameter $\Omega_M$ in
the conventional manner such that
\be
	\Omega_M = \Omega_M^0(1+z)^3\ ,
\ee
where
\be
	\Omega^0_M = {8\pi G\rho^0_M\over 3H_0^2}\ .
\ee
It is imperative to remember that while $\rho_M$ is the sole energy-momentum component
in this Universe, {\em spatial flatness does not imply} $\Omega_M = 1$.  This identity
crucially depends on the conventional four-dimensional Friedmann equation.
One may introduce a new effective dark energy component, $\Omega_{r_0}$, where
\be
	\Omega_{r_0} = {1 \over r_0H}\ ,
\ee
to resort to anologous identity:
\be
	1 = \Omega_M + \Omega_{r_0}\ .
\label{identity}
\ee
This tactic, or something similar, is often used in the literature.  Indeed, one may even
introduce an effective time-dependent $w_{\rm eff}(z) \equiv p^{\rm eff}_{DE}/\rho^{\rm eff}_{DE}$.
Using Eq.~(\ref{friedmann2}) and the time derivative of Eq.~(\ref{Fried2}),
\be
	w_{\rm eff}(z) = -{1\over 1+\Omega_M}\ .
\label{weff}
\ee
Taking Eq.~(\ref{Fried2}), we can write the redshift dependence of $H(z)$ in terms of the
parameters of the model:
\be
     {H(z)\over H_0} = {1\over 2}\left[{1\over r_0H_0}
       + \sqrt{{1\over r_0^2H_0^2} + 4\Omega_M^0(1+z)^3}\right]\ .
\label{hubble}
\ee
While it seems that Eq.~(\ref{hubble}) exhibits two independent parameters, $\Omega_M^0$ and
$r_0H_0$, Eq.~(\ref{identity}) implies that
\be
	r_0H_0 = {1\over 1-\Omega_M^0}\ ,
\ee
 yielding only a single free parameter in Eq.~(\ref{hubble}).
 
The luminosity-distance takes the standard form in spatially flat cosmologies:
\be
      d^{DGP}_L(z) = (1+z)\int_0^z {dz\over H(z)}\ ,
\label{dDGP}
\ee
using Eq.~(\ref{hubble}).  We can compare this distance
with the luminosity distance for a constant--$w$ dark-energy model
\be
     d^w_L(z) = (1+z)\int_0^z {H_0^{-1}dz\over \sqrt{\Omega^w_M(1+z)^3
	 + (1-\Omega^w_M)(1+z)^{3(1+w)}}}\ ,
\label{dQ}
\ee
and in Fig.~\ref{fig:d} we compare these two luminosity distances, normalized
using the best-fit $\Lambda$CDM model
for a variety of $\Omega_M^0$ values.

\begin{figure}
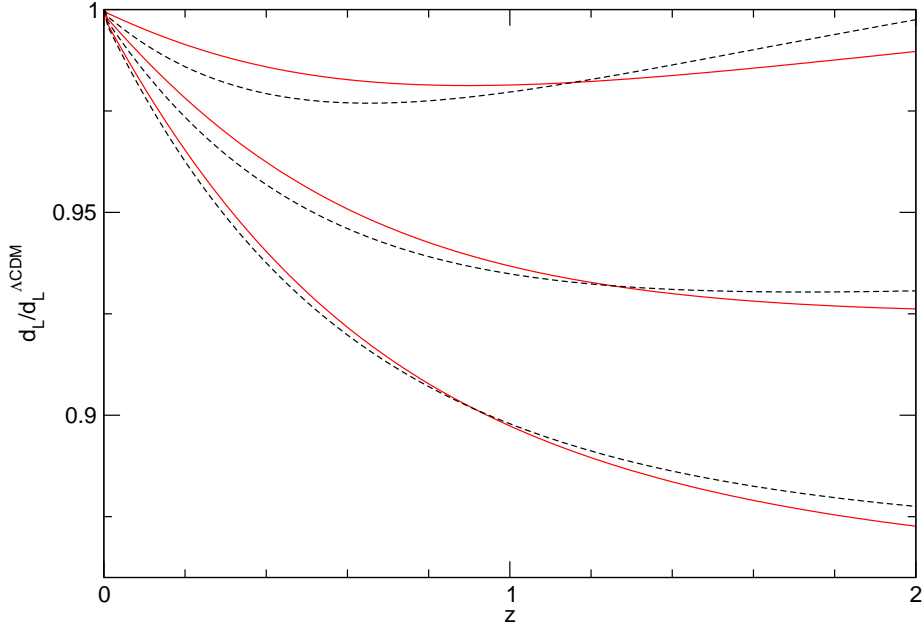
 \begin{center}\PSbox{d.eps
hscale=50 vscale=50 hoffset=-20 voffset=-20}{5in}{3.2in}\end{center}
\caption{
$d^w_L(z)/d^{\Lambda CDM}_L(z)$ and $d^{DGP}_L(z)/d^{\Lambda
CDM}_L(z)$ for a variety of models.  The reference model is the
best-fit flat $\Lambda$CDM, with $\Omega^0_M =0.27$.  The dashed
curves are for the constant--$w$ models with $(\Omega^0_M,w) =
(0.2,-0.765)$, $(0.27,-0.72)$, and $(0.35,-0.69)$ from top to bottom.
The solid curves are for the DGP models with the same $\Omega^0_M$
as the constant--$w$ curves from top to bottom.}
\label{fig:d}
\end{figure}

What is clear from Fig.~\ref{fig:d} is that for all practical purposes, the
expansion history of DGP self-accelerating cosmologies are indistinguishable
from constant--$w$ dark-energy cosmologies.  They would in fact be
identical except for the fact that $w_{\rm eff}(z)$ has a clear and specific
redshift dependence given by Eq.~(\ref{weff}).  The original analysis done
in Ref.~\cite{Deffayet:2001pu,Deffayet:2002sp} suggests that SNIA data favors an
$\Omega_M^0$ low compared to other independent measurements.
Such a tendency is typical of models resembling $w > -1$ dark energy.  Supernova
data from that period implies that the best fit $Omega_M^0$ is
\be
	\Omega_M^0 = 0.18^{+0.07}_{-0.06}\ ,
\ee
at the one-sigma level resulting from chi-squared minimization.
Equation~(\ref{identity}) implies that the corresponding best-fit estimation for the
crossover scale is
\be
	r_0 = 1.21^{+0.09}_{-0.09}\ H_0^{-1}\ .
\ee
Subsequent work using supernova data
\cite{Avelino:2001qh,Deffayet:2001xs,Sahni:2002dx,Linder:2002et,Alam:2002dv,Elgaroy:2004ne,Sahni:2005mc,Alcaniz:2004kq}  refined and generalized these results.
The most recent supernova results \cite{Riess} are able to probe the deceleration/acceleration
transition epoch \cite{Alcaniz:2004kq}.  Assuming a flat universe, this data suggests a
best fit $\Omega^0_M$
\be
	\Omega_M^0 = 0.21\ ,
\ee
corresponding to a best-fit crossover scale
\be
	r_0 = 1.26\ H_0^{-1}\ .
\ee
Similar results were obtained when relaxing flatness or while applying a gaussian prior on
the matter density parameter.

Another pair of interesting possibilities for probing the expansion history of
the universe is using the angular size of high-z compact radio-sources
\cite{Alcaniz:2002qm} and using the estimated age of high-z objects \cite{Alcaniz:2002qh}.
Both these constraints are predicated on the premise that the only meaningful effect of
DGP gravity is the alteration of the expansion history.
However, if the objects are at high enough redshift, this may be a plausible
scenario (see Sec.~\ref{sec:modforce}).

Finally, one can combine supernova data with data from the cosmic microwave
background (CMB).  Again, we are presuming the only effect of DGP gravity is to alter
the expansion history of the universe.  While that is mostly likely a safe assumption
at the last scattering surface (again see Sec.~\ref{sec:modforce}), there are ${\cal O}(1)$--redshift
effects in the CMB, such as the late-time integrate Sachs--Wolfe effect, that may be
very sensitive to alterations of gravity at scales comparable to today's Hubble
radius.  We pursue such issues later in this review.  For now, however, we may
summarize the findings on the simpler presumption \cite{Deffayet:2002sp}.  Supernova
data favors slightly lower values of $\Omega_M^0$ compared to CMB data for a flat
universe.  However, a concordance model with $\Omega_M^0 = 0.3$ provided a good
fit to both sets (pre-WMAP CMB data) with $\chi^2 \approx 140$ for the full data set (135
data points) with a best fit crossover scale $r_0 \sim 1.4 H_0^{-1}$.

\section{Recovery of Einstein Gravity}
\label{sec:einstein}

Until now, we have ignored the crucial question of whether adopting DGP gravity
 yields anomalous phenomenology beyond the alteration of cosmic expansion history.
If we then imagine that today's cosmic
acceleration were a manifestation of DGP self-acceleration, the naive expectation would
be that all anomalous gravitational effects of this theory would be safely hidden at distances
substantially smaller than today's Hubble radius, $H_0^{-1}$, the distance at which the
extra dimension is revealed. We will see in this section
that this appraisal of the observational situation in DGP gravity is too naive.

DGP gravity represents an infrared modification of general relativity.  Such theories often
have pathologies that render them phenomenologically not viable.  These pathologies are
directly related to van Dam--Veltman--Zakharov discontinuity found in massive gravity
\cite{Iwasaki:1971uz,vanDam:1970vg,Z}.  DGP does not evade such concerns:
although gravity in DGP is four-dimensional at distances shorter than $r_0$,
it is not four-dimensional Einstein gravity -- it is augmented by the presence of an ultra-light
gravitational scalar, roughly corresponding to the unfettered fluctuations of the braneworld
on which we live.  This extra scalar gravitational interaction persists even in the limit where
$r_0^{-1} \rightarrow 0$.  This is a phenomenological disaster which is only averted in a
nontrivial and subtle matter
\cite{Deffayet:2001uk,Lue:2001gc,Gruzinov:2001hp,Porrati:2002cp}.  Let us first describe
the problem in detail and then proceed to understanding its resolution.

\subsection{The van Dan--Veltman--Zakharov Discontinuity}

General relativity is a theory of gravitation that supports a massless
graviton with two degrees of freedom, i.e., two polarizations.  However, if one were to
describe gravity with a massive tensor field, general covariance is
lost and the graviton would possess five degrees of freedom.
The gravitational potential (represented by the quanitity
$h_{\mu\nu} = g_{\mu\nu} - \eta_{\mu\nu}$) generated by a static source
$T_{\mu\nu}$ is then given by (in three-dimensional momentum space, $q^i$)
\be
	h^{massive}_{\mu\nu}(q^2) = - {8\pi\over M_P^2}{1\over q^2+m^2}
	\left(T_{\mu\nu} - {1\over 3}\eta_{\mu\nu}T_\alpha^\alpha\right)
\label{potential-massive}
\ee
for a massive graviton of mass $m$ around a Minkowski-flat background.
While similar in form to the gravitational potential in Einstein
gravity
\be
	h^{massless}_{\mu\nu}(q^2) = - {8\pi\over M_P^2}{1\over q^2}
	\left(T_{\mu\nu} - {1\over 2}\eta_{\mu\nu}T_\alpha^\alpha\right)
\label{potential-massless}
\ee
it nevertheless has a distinct tensor structure.  In the
limit of vanishing mass, these five degrees of freedom may be
decomposed into a massless tensor (the graviton), a massless vector (a
graviphoton which decouples from any conserved matter source) and a
massless scalar.  This massless scalar persists as an extra degree of
freedom in all regimes of the theory.  Thus, a massive gravity theory
is distinct from Einstein gravity, {\em even in the limit where the graviton
mass vanishes} as one can see when comparing Eqs.~(\ref{potential-massive})
and (\ref{potential-massless}).  This discrepancy is a formulation of the
van~Dam--Veltman--Zakharov (VDVZ) discontinuity\cite{Iwasaki:1971uz,vanDam:1970vg,Z}.

The most accessible physical consequence of the VDVZ discontinuity is
the gravitational field of a star or other compact, spherically
symmetric source.  The ratio of the strength of the static (Newtonian)
potential to that of the gravitomagnetic potential is different for
Einstein gravity compared to massive gravity, even in the massless
limit.  Indeed the ratio is altered by a factor of order unity.  Thus,
such effects as the predicted light deflection by a star would be affected significantly if the
graviton had even an infinitesimal mass.

This discrepancy appears for the gravitational field of any compact
object.  An even more dramatic example of the VDVZ discontinuity
occurs for a cosmic string.  A cosmic string has no static potential
in Einstein gravity; however, the same does not hold for a cosmic
string in massive tensor gravity.  One can see why using the
potentials Eqs.~(\ref{potential-massive}) and (\ref{potential-massless}).
The potential between a cosmic string with
$T_{\mu\nu} = {\rm diag}(T,-T,0,0)$ and a test particle with
$\tilde{T}_{\mu\nu} = {\rm diag}(2\tilde{M}^2,0,0,0)$ is
\be
	V_{massless} = 0\ ,\ \ \ 
	V_{massive} \sim {T\tilde{M}\over M_P^2}\ln r \ ,
\ee
where the last expression is taken in the limit $m \rightarrow
0$.  Thus in a massive gravity theory, we expect a cosmic string to
attract a static test particle, whereas in general relativity, no such
attraction occurs.  The attraction in the massive case can be
attributed to the exchange of the remnant light scalar mode that comes
from the decomposition of the massive graviton modes in the massless
limit.

The gravitational potential in DGP gravity, Eq.~(\ref{prop}), has the same tensor
structure as that for a massive graviton and
perturbatively has the same VDVZ problem in the limit that the
graviton linewidth (effectively $r_0^{-1}$) vanishes.  Again, this tensor structure
is the result of an effective new scalar that may be associated with a brane fluctuation
mode, or more properly, the fluctuations of the extrinsic curvature of the brane.
Because, in this theory, the brane is tensionless, its fluctuations represent a very
light mode and one may seriously ask the question as to whether standard tests
of scalar-tensor theories, such as light-deflection by the sun, already rule out
DGP gravity by wide margins.

It is an important and relevant question to ask.  We are precisely interested in the
limit when $r_0^{-1} \rightarrow 0$ for all intents and purposes.  We want $r_0$ to
be the size of the visible Universe today, while all our reliable measurements of
gravity are on much smaller scales.  However, the answers to questions of observational
relevance are not straightforward.  Even in massive gravity, the presence of the VDVZ
discontinuity is more subtle
than just described.  The potential
Eq.~(\ref{potential-massive}) is only derived perturbatively to lowest order
in $h_{\mu\nu}$ or $T_{\mu\nu}$.   Vainshtein proposed that 
this discontinuity does not persist in the fully-nonlinear classical theory
\cite{Vainshtein:1972sx}.  However, doubts remain
\cite{Boulware:1973my} since no self-consistent, fully-nonlinear theory of massive
tensor gravity exists (see, for example, Ref.~\cite{Gabadadze:2003jq}).

If the corrections to Einstein gravity remain large even in limit $r_0 \rightarrow \infty$,
the phenomenology of DGP gravity is not viable.
The paradox in DGP gravity seems to be that while it is clear that a perturbative,
VDVZ--like discontinuity occurs in the potential somewhere (i.e., Einstein gravity
is not recovered at short distances), no such discontinuity appears in the cosmological
solutions;  at high Hubble scales, the theory on the brane appears safely like
general relativity \cite{Deffayet:2001uk}.  What does this mean?  What is clear is
that the cosmological solutions at high Hubble scales are extremely nonlinear, and
that perhaps, just as Vainshtein suggested for massive gravity, nonlinear effects
become important in resolving the DGP version of the VDVZ discontinuity.

\subsection{Case Study:  Cosmic Strings}

We may ask the question of how nonlinear, nonperturbative effects
change the potential Eq.~(\ref{prop}), per se.  Indeed, as a stark and straightforward exercise, we may
ask the question in DGP gravity, does a cosmic string attract a static test particle
or not in the limit?  We will see that corrections remain small and that the recovery of Einstein
gravity is subtle and directly analogous to Vainstein's proposal for massive gravity.
DGP cosmic strings provided the first understanding of how the recovery of Einstein
gravity occurs in noncosmological solutions \cite{Lue:2001gc}.  Cosmic strings
offer a conceptually clean environment and a geometrically appealing picture for
how nonperturbative effects drive the loss and recover of the Einstein limit in DGP
gravity.  Once it is understood how the VDVZ issue is resolved in this simpler system,
understanding it for the Schwarzschild-like solution becomes a straightforward affair.

\subsubsection{The Einstein Solution}

Before we attempt to solve the full five-dimensional problem for the
cosmic string in DGP gravity, it is useful to review the cosmic string
solution in four-dimensional Einstein gravity \cite{Vilenkin:1981zs,Gregory:1987gh}.
For a cosmic string with tension $T$, the exact metric
may be represented by the line element:
\be
	ds^2 = dt^2 - dx^2
	- \left(1 - 2GT\right)^{-2}dr^2 - r^2d\phi^2\ .
\label{metric-GRstring}
\ee
This represents a flat space with a deficit angle $4\pi GT$.  If one chooses,
one can imagine suppressing the $x$--coordinate and imagining that this
analysis is that for a particle in (2+1)--dimensional general relativity.
Equation~(\ref{metric-GRstring}) indicates that there is no Newtonian
potential (i.e., the potential between static sources arising from $g_{00}$)
between a cosmic string and a static test particle.  However, a test particle
(massive or massless) suffers an azimuthal deflection of $4\pi GT$ when
scattered around the cosmic string, resulting from the deficit angle cut from
spacetime.  Another way of interpreting this deflection effect may be
illuminated through a different coordinate choice.  The line element
Eq.~(\ref{metric-GRstring}) can be rewritten as
\be
	ds^2 = dt^2 - dx^2
	- (y^2+z^2)^{-2GT}[dy^2 + dz^2]\ .
\label{metric-GRstring2}
\ee
Again, there is no Newtonian gravitational potential between
a cosmic string and a static test particle.  There is no longer an explicit
deficit angle cut from spacetime;  however, in this coordinate choice, the
deflection of a moving test particle results rather from a gravitomagnetic
force generated by the cosmic string.

In the weak field limit, one may rewrite Eq.~(\ref{metric-GRstring2}) as
a perturbation around flat space, i.e., $g_{\mu\nu} = \eta_{\mu\nu}+h_{\mu\nu}$,
as a series in the small parameter $GT$ such that
\bea
	h_{00} = h_{xx} &=& 0	\\
	h_{yy} = h_{zz} &=& 4GT\ln r\ ,
\eea
where $r = \sqrt{y^2+z^2}$ is the radial distance from the cosmic string.
So, interestingly, one does recover the logarithmic potentials that are
expected for codimension--2 objects like cosmic strings in (3+1)--dimensions
or point particles in (2+1)--dimensions.  They appear, however, only in
the gravitomagnetic potentials in Einstein gravity, rather than in the
gravitoelectric (Newtonian) potential.

\subsubsection{DGP Cosmic Strings:  The Weak-Brane Limit}

We wish to fine the spacetime around a perfectly straight, infinitely
thin cosmic string with a tension $T$, located on the surface of our brane
Universe (see Fig.~\ref{fig:flat}).
Alternatively, we can again think of suppressing the coordinate along the
string so that we consider the spacetime of a point particle, located on a
two dimensional brane existing in a (3+1)--dimensional bulk.
As in the cosmological solution, we assume a mirror, ${\cal Z}_2$--symmetry,
across the brane surface at $z = {\pi\over 2}$.
The Einstein equations Eqs.~(\ref{Einstein}) may now be solved for this
system.

\begin{figure} \begin{center}\PSbox{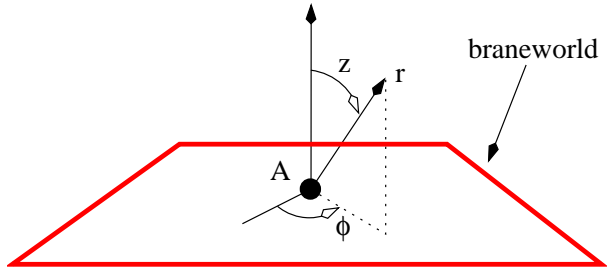
hscale=100 vscale=100 hoffset=-50 voffset=20}{2in}{1.5in}\end{center}
\caption{
A schematic representation of a spatial slice through a
cosmic string located at $A$.  The coordinate $x$ along the cosmic
string is suppressed.  The coordinate $r$ represents the 3-dimensional
distance from the cosmic string $A$, while the coordinate $z$ denotes
the polar angle from the vertical axis.  In the no-gravity limit,
the braneworld is the horizontal plane, $z = {\pi\over 2}$.  The
coordinate $\phi$ is the azimuthal coordinate.  Figure from Ref.~[47].
}
\label{fig:flat}
\end{figure}

There is certainly a regime where one may take a perturbative limit when
$GT$ is small and so that given $g_{AB} = \eta_{AB} + h_{AB}$, the
four-dimensional Fourier transform of the metric potential on the brane
is given by Eq.~(\ref{prop}).  For a cosmic string, this implies that when $r\gg r_0$,
\bea
	h_{00} = h_{xx} &=& -{1\over 3}{4r_0GT\over r}	\\
	h_{yy} = h_{zz} &=& -{2\over 3}{4r_0GT\over r}\ .
\eea
Graviton modes localized on the brane evaporate into the bulk
over distances comparable to $r_0$.  The presence of the brane becomes
increasingly irrelevant as $r/r_0 \rightarrow \infty$ and a cosmic
string on the brane acts as a codimension-three object in the full
bulk.  When $r\ll r_0$,
\bea
	h_{00} = h_{xx} &=& {1\over 3}4GT\ln r	\\
	h_{yy} = h_{zz} &=& {2\over 3}4GT\ln r\ .
\eea
The metric potentials when $r\ll r_0$ represent a
conical space with deficit angle ${2\over
3}4\pi GT$.  Thus in the weak
field limit, we expect not only an extra light scalar field generating
the Newtonian potential, but also a discrepancy in the
deficit angle with respect to the Einstein solution.

We can ask the domain of validity of the perturbative solution.  The
perturbative solution considered only terms in Eqs.~(\ref{Einstein})
linear in $h_{AB}$, or correspondingly, linear in $GT$.  When $GT \ll 1$,
this should be a perfectly valid approach to self-consistenly solving
Eqs.~(\ref{Einstein}).  However, there is an important catch.  While $GT$
is indeed a small parameter, DGP gravity introduces a large parameter
$r_0$ into the field equations.  Actually, since $r_0$ is dimensionful,
the large parameter is more properly $r_0/r$.  Thus, there are distances
for which nonlinear terms in Eqs.~(\ref{Einstein}) resulting from contributions
from the extrinsic curvature of the brane
\be
	\sim\ {r_0\over r}(GT)^2\ ,
\ee
cannot be ignored, even though they are clearly higher order in $GT$.
Nonlinear terms such
as these may only be ignored when \cite{Lue:2001gc}
\be
	r  \gg r_0\sqrt{4\pi GT}\ .
\label{limits-weak}
\ee
Thus, the perturbative solution given by the metric potential Eq.~(\ref{prop})
is not valid in all regions.  In particular, the perturbative solution is not valid in the limit where
everything is held fixed and $r_0\rightarrow \infty$, which is precisely the
limit of interest.

\subsubsection{The $r/r_0 \rightarrow 0$ Limit}

For values of $r$ violating Eq.~(\ref{limits-weak}), nonlinear contributions to the
Einstein tensor become important and the weak field approximation breaks down,
{\em even when the components $h_{\mu\nu} \ll 1$}.  What happens when
$r \ll r_0\sqrt{4\pi GT}$?  We need to find a new small expansion parameter in
order to find a new solution that applies for small $r$.
Actually, the full field equations Eqs.~(\ref{Einstein}) provide a clue \cite{Lue:2001gc}.
A solution that is five-dimensional Ricci flat in the bulk, sporting a brane surface
that is four-dimensional Ricci flat, is clearly a solution.
Figure~\ref{fig:space} is an example of such a solution (almost).  The bulk is
pure vanilla five-dimensional Minkowski space, clearly Ricci flat.  The brane is a
conical deficit space, a space whose intrinsic
curvature is strictly zero.  The field equations Eqs.~(\ref{Einstein}) should be
solved.

\begin{figure} \begin{center}\PSbox{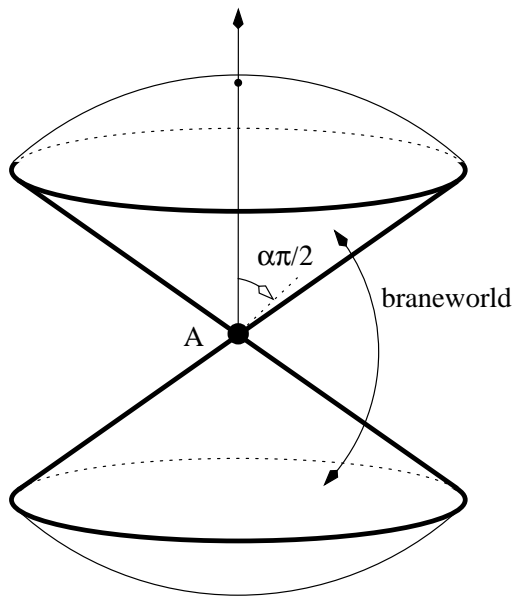
hscale=100 vscale=100 hoffset=-10 voffset=20}{2in}{3.25in}\end{center}
\caption{
A spatial slice through the cosmic string located at $A$.
As in Fig.~\ref{fig:flat} the coordinate $x$ along the cosmic string
is suppressed.  The solid angle wedge exterior to the cone is removed
from the space, and the upper and lower branches of the cone are
identified.  This conical surface is the braneworld ($z={\pi\alpha\over 2}$
or $\sin z = \beta$).  The bulk space now exhibits a deficit polar
angle (cf. Fig.~\ref{fig:flat}).  Note that this deficit in polar
angle translates into a conical deficit in the braneworld space.  Figure from Ref.~[47].
}
\label{fig:space}
\end{figure}

Why the space depicted in Fig.~\ref{fig:space} is not exactly a solution comes from
the ${\cal Z}_2$--symmetry of the bulk across the brane.  The brane surface has
nontrivial {\em extrinsic} curvature even though it has vanishing intrinsic curvature.
Thus a polar deficit angle space has a residual bulk curvature that is a delta-function
at the brane surface, and Eqs.~(\ref{Einstein}) are not exactly zero everywhere for that space.
Fortunately, the residual curvature is subleading in $r/r_0$, and one may perform
a new systematic perturbation in this new parameter, $r/r_0$, starting with the space
depicted in Fig.~\ref{fig:space} as the zeroth-order limit.

The new perturbative solution on the brane is given using the line element
\be
	ds^2 = N^2(r)|_{\sin z = \beta}\ (dt^2-dx^2)
	- A^2(r)|_{\sin z = \beta}\ dr^2 - \beta^2r^2d\phi^2\ ,
\label{brane-metric}
\ee
where the metric components on the brane are \cite{Lue:2001gc}
\bea
	N(r)|_{\sin z =\beta} &=& 1
	+ {\sqrt{1-\beta^2}\over 2\beta}{r\over r_0} \label{2d-braneN}	\\
	A(r)|_{\sin z =\beta} &=& 1 - {\sqrt{1-\beta^2}\over2\beta}{r\over r_0}
	\ , \label{2d-braneA}
\eea
and the deficit polar angle in the bulk is $\pi(1-\alpha)$ where
$\sin{\pi\alpha\over 2} = \beta$, while the deficit azimuthal angle
in the brane itself is $2\pi(1-\beta)$.  The deficit angle on the
brane is given by
\be
	\beta = 1 - 2GT\ ,
\label{alpha}
\ee
which is precisely equivalent to the Einstein result.  The perturbative
scheme is valid when
\be
	r\ \ll\ r_0{\sqrt{1-\beta^2}\over\beta}\
	\sim\ r_0\sqrt{{4\pi GT}} \ ,
\label{limit-space}
\ee
which is complementary to the regime of validity for the weak-brane
perturbation.  Moreover, Eq.~(\ref{limit-space}) is the regime
of interest when
taking $r_0\rightarrow\infty$ while holding everything else fixed, i.e.,
the one of interest in addressing the VDVZ discontinutiy.  What we see
is that, just like for the cosmological solutions, DGP cosmic strings
do no suffer a VDVZ--like discontinuity.  Einstein gravity is recovered
in the $r_0\rightarrow\infty$ limit, precisely because of effects nonlinear,
and indeed nonperturbative, in $GT$.

\subsubsection{The Picture}

\begin{figure} \begin{center}\PSbox{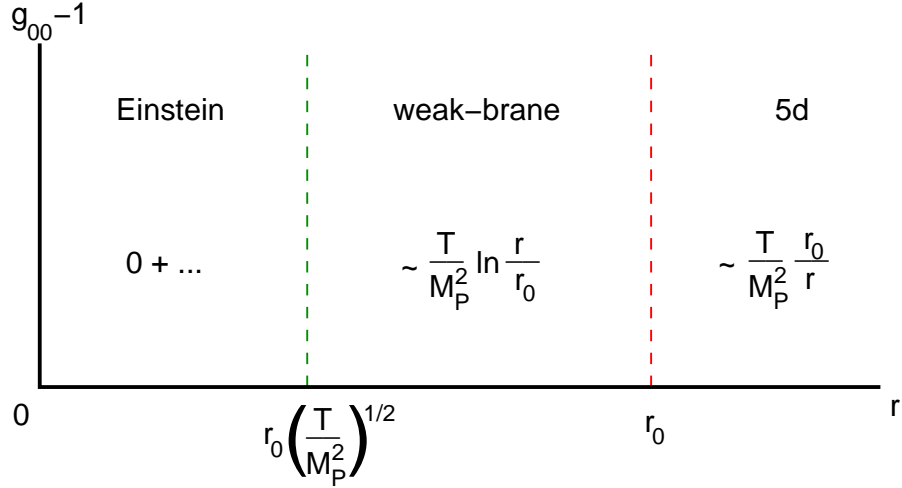
hscale=80 vscale=80 hoffset=0 voffset=0}{5in}{2.3in}\end{center}
\caption{
The Newtonian potential for a cosmic string has the following regimes:  outside $r_0$, the
cosmic string appears as a codimension--3 object, i.e., a Schwarzschild source, and so its
potential goes as $r^{-1}$;  inside $r_0$, the string appears as a codimension--2 object, i.e.
a true string source.  Outside $r_0(T/M_P^2)^{1/2}$, however, the theory appears Brans-Dicke
and one generates a logarithmic scalar potential associated with codimension--2 objects.
Inside the radius $r_0(T/M_P^2)^{1/2}$ from the string source, Einstein gravity is recovered
and there is no Newtonian potential.
}
\label{fig:string2}
\end{figure}

Figure~\ref{fig:string2} depicts how in different parametric regimes, we find different
qualitative behaviors for the brane metric around a cosmic string in DGP gravity.
Though we have not set out the details here, the different perturbative solutions
discussed are part of the same solution in the bulk and on the brane \cite{Lue:2001gc},
i.e., the trunk and the tail of the elephant, as it were.  For an
observer at a distance $r \gg r_0$ from the cosmic string, where $r_0^{-1}$
characterizes the graviton's effective linewidth, the cosmic
string appears as a codimension-three object in the full bulk.  The
metric is Schwarzschild-like in this regime.  When $r \ll r_0$, brane
effects become important, and the cosmic string appears as a
codimension-two object on the brane.  If the source is weak (i.e.,
$GT$ is small), the Einstein solution with a
deficit angle of ${4\pi GT}$ holds on the brane only when $r
\ll r_0\sqrt{4\pi GT}$.  In the region on the brane when $r
\gg r_0\sqrt{{4\pi GT}}$ (but still where $r \ll r_0$), the
weak field approximation prevails, the cosmic string exhibits a
nonvanishing Newtonian potential and space suffers a deficit angle
different from ${4\pi GT}$.

The solution presented here supports the Einstein solution near the
cosmic string in the limit that $r_0 \rightarrow \infty$ and recovery of
Einstein gravity proceeded precisely as Vainshtein suggested it would
in the case of massive gravity:  nonperturbative effects play a crucial
role in suppressing the coupling the extra scalar mode.  Far from the source, the
gravitational field is weak, and the geometry of the brane (i.e.,
its extrinsic curvature with respect to the bulk) is not substantially altered
by the presence of the cosmic string.  The solution is a perturbation in $GT$
around the trivial space depicted in Fig.~\ref{fig:flat}.  Propagation of the light scalar
mode is permitted and the solution does not correspond to that from
general relativity.  However near the source, the gravitational fields
induce a nonperturbative extrinsic curvature in the brane, in a manner
reminiscent of the popular science picture used to explain how matter
sources warp geometry.  Here, the picture is literally true.  The solution here
is a perturbation in $r/r_0$ around the space depicted in Fig.~\ref{fig:space}.
The brane's extrinsic curvature suppresses the coupling of the scalar mode to matter and only
the tensor mode remains, thus Einstein gravity is recovered.

\subsection{The Schwarzschild-like Solution}

So while four-dimensional Einstein gravity is recovered in a region
near a cosmic string source, it is recovered in a region much smaller
than the region where one naively expected the extra dimension to
be hidden, i.e., the larger radius $r_0$.  Einstein gravity is only recovered within
a region much smaller than $r_0$, a region whose size is dictated by the source strength.
Do the insights elucidated using cosmic string translate for a
star-like object?  If so, that would have fantastic observational
consequences.  We would have a strong handle for observing this
theory in a region that is accessible in principle, i.e., at distances much
smaller than today's Hubble radius.
Indeed, Gruzinov first showed how that recovery of Einstein
gravity in
the Schwarzschild-like solution is exactly analogous to what was found
for the cosmic string, and moreover, is also in
exactly the spirit of Vainshtein's resolution of the VDVZ discontinuity
for massive gravity \cite{Gruzinov:2001hp}.

\subsubsection{The Field Equations}

We are interested in finding the metric for a static, compact, spherical
source in a Minkowski background.   Under this circumstance, one can choose
a coordinate system in which the
metric is static (i.e., has a timelike Killing vector)
while still respecting the spherical symmetry of the matter source.
Let the line element be
\be
ds^{2} = N^2(r,z) dt^{2}
         - A^2(r,z)dr^2 - B^2(r,z)[d\theta^2 + \sin^2\theta d\phi^2]-dz^{2}\ .
\label{metric}
\ee
This is the most general static metric with spherical symmetry on the
brane.  The bulk Einstein tensor for this metric is:
\ \\
\bea
G_t^t &=& {1\over B^2}
-{1\over A^2}\left[{2B''\over B} - {2A'\over A}{B'\over B}
  + {B'^2\over B^2}\right]
-\left[{A_{zz}\over A} + {2B_{zz}\over B}
  + 2{A_z\over A}{B_z\over B} + {B_z^2\over B^2}\right]
\nonumber  \\
G_r^r &=& {1\over B^2}
-{1\over A^2}\left[2{N'\over N}{B'\over B} + {B'^2\over B^2}\right]
- \left[{N_{zz}\over N} + {2B_{zz}\over B}
  + 2{N_z\over N}{B_z\over B} + {B_z^2\over B^2}\right]
\nonumber  \\
G_\theta^\theta &=& G_\phi^\phi =
-{1\over A^2}\left[{N''\over N} + {B''\over B}
  - {N'\over N}{A'\over A}+{N'\over N}{B'\over B}-{A'\over A}{B'\over B}\right]
  \nonumber \\
&& \ \ \ \ \ \ \ \ \ \ \ \ \ \ \ \ \ \ \ \ \ \ - \left[{N_{zz}\over N} + {A_{zz}\over A} + {B_{zz}\over B}
  + {N_z\over N}{A_z\over A} + {N_z\over N}{B_z\over B}
  + {A_z\over A}{B_z\over B}\right]
\label{5dEinstein}\\
G_z^z &=& {1\over B^2}
-{1\over A^2}\left[{N''\over N} + {2B''\over B}
  - {N'\over N}{A'\over A}+2{N'\over N}{B'\over B}-2{A'\over A}{B'\over B}
  + {B'^2\over B^2}\right]     \nonumber \\
&& \ \ \ \ \ \ \ \ \ \ \ \ \ \ \ \ \ \ \ \ \ \ \ \ \ \ \ \ \ \ \ \ \ \ \ \ \ \
- \left[{N_z\over N}{A_z\over A} + 2{N_z\over N}{B_z\over B}
  + 2{A_z\over A}{B_z\over B} + {B_z^2\over B^2}\right]
 \nonumber  \\
G_{zr} &=& -\left[{N_z'\over N} + {2B_z'\over B}\right] +
{A_z\over A}\left({N'\over N} + {2B'\over B}\right)\ . \nonumber
\eea
\ \\
The prime denotes partial differentiation with respect to $r$,
whereas the subscript $z$ represents partial differentiation with respect to $z$.

We wish to solve the five-dimensional field equations,
Eq.~(\ref{Einstein}).  This implies that all components of the
Einstein tensor, Eqs.~(\ref{5dEinstein}), vanish in the bulk but
satisfy the following modified boundary relationships on the brane.
Fixing the residual gauge $B|_{z=0}=r$ and imposing
${\cal Z}_2$--symmetry across the brane
\bea
-\left({A_z\over A} + {2B_z\over B}\right)
&=& {r_0\over A^2}\left[-{2\over r}{A'\over A} + {1\over r^2}(1-A^2)\right]
+ {8\pi r_0\over M_P^2}\rho(r)
\nonumber \\
-\left({N_z\over N} + {2B_z\over B}\right)
&=&  {r_0\over A^2}\left[{2\over r}{N'\over N} + {1\over r^2}(1-A^2)\right]
- {8\pi r_0\over M_P^2}p(r)
\label{branebc}\\
-\left({N_z\over N} + {A_z\over A} + {B_z\over B}\right)
&=&  {r_0\over A^2}\left[{N''\over N} - {N'\over N}{A'\over A}
+ {1\over r}\left({N'\over N} - {A'\over A}\right)\right]
- {8\pi r_0\over M_P^2}p(r)\ ,
\nonumber
\eea
when $z=0$.  These brane boundary relations come from $G_{tt}$, $G_{rr}$
and $G_{\theta\theta}$, respectively.  We have chosen a gauge in which the
brane, while still dynamical, appears flat.  All the important extrinsic curvature
effects discussed in the last section will appear in the $z$--derivatives of the
metric components evaluated at the brane surface, rather than through any explicit
shape of the brane.

We are interested in a static matter distribution $\rho(r)$, and we may define an
effective radially-dependent Schwarzschild radius
\be
      R_g(r) = {8\pi\over \mpsq}\int_0^r r^2\rho(r) dr\ ,
\label{rg-mink}
\ee
where we will also use the true Schwarzschild radius, $r_g = R_g(r\rightarrow \infty)$.
We are interested only in weak matter sources, $\rho_g(r)$.
Moreover, we are most interested in those parts of spacetime where
deviations of the metric from Minkowski are small.  Then, it is
convenient to define the functions $\{n(r,z),a(r,z),b(r,z)\}$ such
that
\bea
N(r,z) &=& 1+n(r,z)     \\
A(r,z) &=& 1+a(r,z)
\label{linearize}      \\
B(r,z) &=& r~[1+b(r,z)]\ .
\eea
Since we are primarily concerned with the metric on the brane, we
can make a gauge choice such that $b(r,z=0) = 0$ identically so that
on the brane, the line element
\be
	ds^2 = \left[1+n(r)|_{z=0}\right]^2dt^2 - \left[1+a(r)|_{z=0}\right]^2dr^2 - r^2d\Omega\ ,
\label{metric-brane}
\ee
takes the standard form with two potentials, $n(r)|_{z=0}$ and $a(r)|_{z=0}$, the
Newtonian potential and a gravitomagnetic potential.
Here we use $d\Omega$ as shorthand for the usual differential solid angle.

We will be interested in small deviations from flat Minkowski space, or more
properly, we are only concerned when $n(r,z), a(r,z)$ and $b(r,z) \ll 1$.  We can
then rewrite our field equations, Eqs.~(\ref{5dEinstein}), and brane boundary
conditions, Eqs.~(\ref{branebc}), in terms of these quantities and keep only
leading orders.  The brane boundary conditions become
\bea
-(a_z+2b_z) &=& r_0\left[-{2a' \over r} - {2a\over r^2}\right]
			+ {r_0\over r^2}R'_g(r)
\nonumber    \\
-(n_z + 2b_z) &=& r_0\left[{2n' \over r} - {2a\over r^2}\right]
\label{branebc1}     \\
-(n_z + a_z+b_z) &=& r_0\left[n'' + {n'\over r} - {a'\over r} \right]\ .
\nonumber
\eea
Covariant conservation of the source on the brane allows one to
ascertain the source pressure, $p(r)$, given the source density
$\rho(r)$:
\be
	{p_g}' = - n'\rho_g\ .
\label{covariant}
\ee
The pressure terms were dropped from Eqs.~(\ref{branebc1})
because there are subleading here.

\subsubsection{The Weak-Brane Limit}

Just as for the cosmic string, there is again a regime where one may take a properly
take the perturbative limit when $r_g$ is small.  Again, given $g_{AB} = \eta_{AB} + h_{AB}$,
the four-dimensional Fourier transform of the metric potential on the brane is given by
Eq.~(\ref{prop})
\bea
	\tilde{h}_{\mu\nu}(p) = {8\pi\over M_P^2}{1\over p^2 + p/r_0}
		\left[\tilde{T}_{\mu\nu} - {1\over 3}\eta_{\mu\nu}\tilde{T}_\alpha^\alpha\right]\ .
\nonumber
\eea
For a Schwarzschild solution, this implies that when $r\gg r_0$
\bea
	h_{00} &=& -{4\over 3}{r_0r_g\over r^2}	\\
	h_{xx} = h_{yy} = h_{zz} &=& -{2\over 3}{r_0r_g\over r^2}\ ,
\eea
and $r\ll r_0$
\bea
	h_{00} &=& -{4\over 3}{r_g\over r^2}	\\
	h_{xx} = h_{yy} = h_{zz} &=& -{2\over 3}{r_g\over r^2}\ .
\eea
It is convenient to write the latter in terms of our new potentials for the line element,
Eq.~(\ref{metric-brane}),
\bea
	n(r)|_{z=0} &=& -{4\over 3}{r_g\over 2r}	\\
\label{weakmink1}
	a(r)|_{z=0} &=& +{2\over 3}{r_g\over 2r}\ .
\label{weakmink2}
\eea
This is actually the set of potentials one expects from Brans-Dicke scalar-tensor
gravity with Dicke parameter $\omega = 0$.  Einstein gravity would correspond to
potentials whose values are $-r_g/2r$ and $+r_G/2r$, respectively.  As discussed
earlier, there is an extra light scalar mode coupled to the matter source.  That
mode may be interpreted as the fluctuations of the free brane surface.

Again, just as in the cosmic string case, we see that significant deviations from
Einstein gravity yield nonzero contributions to the right-hand side of Eqs.~(\ref{branebc}).
Because these are mulitplied by $r_0$, this implies that the extrinsic curvatures
(as represented by the $z$--derivatives of the metric components at $z=0$) can be quite large.
Thus, while we neglected nonlinear contributions to the field equations, Eqs.~(\ref{5dEinstein}),
bilinear terms in those equations of the form ${A_z\over A}{B_z\over B}$, for example, are only
negligible when \cite{Gruzinov:2001hp}
\be
	r \gg r_* \equiv \left(r_gr_0^2\right)^{1/3}\ ,
\ee
even when $r_g$ is small and $n,a \ll 1$.  When $r\ll r_*$, we need to identify
a new perturbation scheme.

\subsubsection{The $r/r_0\rightarrow 0$ Limit}

\begin{figure} \begin{center}\PSbox{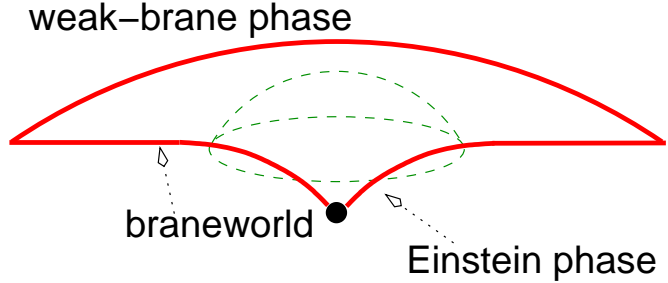
hscale=100 vscale=100 hoffset=-20 voffset=0}{3in}{1.3in}\end{center}
\caption{
Given a mass source located on the brane, inside the radius $r_*$ (the green hemisphere),
the brane is dimpled significantly generating a nonperturbative extrinsic curvature.  Brane
fluctuations are suppressed in this region (Einstein phase).  Outside this radius, the brane
is free to fluctuate.  The bulk in this picture is above the brane.  The mirror copy of the bulk
space below the brane is suppressed.
}
\label{fig:regime}
\end{figure}

The key to identifying a solution when $r\ll r_*$ is to recognize that one only needs to
keep certain nonlinear terms in Eqs.~(\ref{5dEinstein}).
So long as $n, a, b \ll 1$, or equivalently $r \gg r_g$, the only nonlinear
terms that need to be included are those terms bilinear in ${A_z\over A}$ and ${B_z\over B}$  \cite{Gruzinov:2001hp}.
Consider a point mass source such that $R_g(r) = r_g = {\rm constant}$.
Then, the following set of potentials on the brane are \cite{Gruzinov:2001hp}
\bea
	n &=& -{r_g\over 2r} + \sqrt{r_g r\over 2r_0^2}
\label{mink1}     \\
	a &=& +{r_g\over 2r} - \sqrt{r_g r\over 8r_0^2}\ .
\label{mink2}
\eea
The full bulk solution and how one arrives at that solution will be spelled out in
Sec.~\ref{sec:modforce} when we consider the more general case of the
Schwarzschild-like solution in the background of a general cosmology, a subset
of which is this Minkowski background solution.

\begin{figure} \begin{center}\PSbox{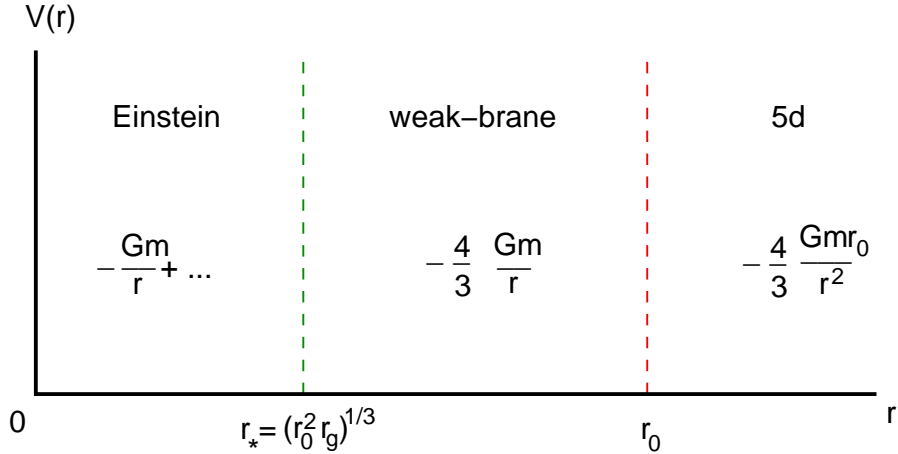
hscale=80 vscale=80 hoffset=0 voffset=0}{5in}{2.3in}\end{center}
\caption{
The Newtonian potential $V(r) = g_{00} - 1$ has the following regimes:  outside $r_0$, the potential
exhibits five-dimensional behavior (i.e., $1/r^2$);  inside $r_0$, the potential is indeed
four-dimensional (i.e., $1/r$) but with a coefficient that depends on $r$.  Outside $r_*$ we have
Brans-Dicke potential while inside $r_*$ we have a true four-dimensional Einstein potential.
}
\label{fig:schwarz}
\end{figure}

That the inclusion of terms only nonlinear in $a_z$ and $b_z$ was sufficient to find solutions
valid when $r\ll r_*$ is indicative that the nonlinear behavior arises from purely spatial
geometric factors \cite{Gruzinov:2001hp}.  In particular, inserting the potentials
Eqs.~(\ref{mink1}) and~(\ref{mink2})
into the expressions Eqs.~(\ref{branebc1}) indicates that the extrinsic curvatures of the
brane, i.e., $a_z|_{z=0}$ and $b_z|_{z=0}$, play a crucial role in the
nonlinear nature of this solution, indeed a solution inherently nonperturbative in the
source strength $r_g$.  This again is directly analogous to the cosmic string but rather than
exhibiting a conical distortion, the brane is now cuspy.  The picture of
what happens physically to the brane is depicted in Fig.~\ref{fig:regime}.  When a mass source is introduced in the brane, its gravitational effect includes
a nonperturbative dimpling of the brane surface (in direct analogy with the
popular physics picture of how general relativity works).  The brane is
dimpled significantly in a region within a radius $r_*$ of the matter source.
The extrinsic curvature suppresses the light brane bending mode associated
with the extra scalar field inside this region, whereas outside this region, the
brane bending mode is free to propagate.  Thus four-dimensional Einstein
gravity is recovered close the mass source, but at distances less than $r_*$,
not distances less than $r_0$.  Outside $r_*$, the theory appears like a
four-dimensional scalar-tensor theory, in particular, four-dimensional linearized
Brans--Dicke, with parameter $\omega = 0$.  A marked departure from Einstein
gravity persists down to distances much shorter than $r_0$.  Figure~\ref{fig:schwarz}
depicts the hierarchy of scales in this system.

\section{Modified Gravitational Forces}
\label{sec:modforce}

So, we expect a marked departure of the metric of a spherical, compact mass at distances
comparable to $r_*$ and greater.  The potentials Eqs.~(\ref{mink1}) and (\ref{mink2})
provide the form of the corrections to Einstein gravity as $r$ approaches $r_*$ while the
in the weak-brane phase, i.e., when $r\gg r_*$ but when $r$ is still much smaller than $r_0$,
the potentials are given by Eqs.~(\ref{weakmink1}) and (\ref{weakmink2}).  From our treatment
of the cosmic expansion history of DGP gravity, if we are to comprehend the contemporary
cosmic acceleration as a manifestation of extra-dimensional effects, wish to set $r_0\sim H_0^{-1}$.
The distance $r_*$ clearly plays an important role in DGP phenomenology.
Table~\ref{table:rstar} gives a few examples of what $r_*$ would be if given source masses
were isolated in an empty Universe when $r_0\sim H_0^{-1}$.

\begin{table}[t]
\caption{Example Values for $r_*$}
\begin{center}
Earth \ \ \ \ \ \ \ \ \ \ \ \ \ \ \ \ \ \ \ \ \ \ \ \ \ \ \ \ \ \ \ \ \ \ \ \ \ \ \ \ \ \ \ 1.2 pc\\
Sun \ \ \ \ \ \ \ \ \ \ \ \ \ \ \ \ \ \ \ \ \ \ \ \ \ \ \ \ \ \ \ \ \ \ \ \ \ \ \ \ \ \ \ \ \ 150 pc\\
Milky Way ($10^{12}M_{\sun}$) \ \ \ \ \ \ \ \ \ \ \ \ \ \ \ \ \ \ \ \ 1.2 Mpc\\
\end{center}
\label{table:rstar}
\end{table}

However, there is a complication when we wish to understand the new gravitational forces
implied by Eqs.~(\ref{weakmink1}) and (\ref{weakmink2}) the context of cosmology \cite{Lue:2002sw}.
The complication arises when we consider cosmologies whose Hubble radii, $H^{-1}$, are
comparable or even smaller than $r_0$.  Take $H^{-1} = {\rm constant} = r_0$ for example.  In
such an example, one may actually regard the Hubble flow as being as making an effective
contribution to the metric potentials
\be
	N_{\rm cosmo} = - {r^2\over r_0^2}\ .
\ee
Figure~\ref{fig:corr} depicts a representative situation.  The corrections computed in the
Minkowski case become important just at the length scales where the cosmology dominates
the gravitational potential.  I.e., the regime $r\gtrsim r_*$ is just that region where cosmology
is more important than the localized source.  Inside this radius, an observer is bound in the
gravity well of the central matter source.  Outside this radius, an observer is swept away
into the cosmological flow.  Thus, one cannot reliably apply the results from a Minkowski
background under the circumstance of a nontrivial cosmology, particularly, when we are
interested in DGP gravity because of its anomalous cosmological evolution.
We need to redo the computation to include the background cosmology, and indeed
we will find a cosmology dependent new gravitational force \cite{Lue:2002sw,Lue:2004rj,Lue:2004za}.

\begin{figure}
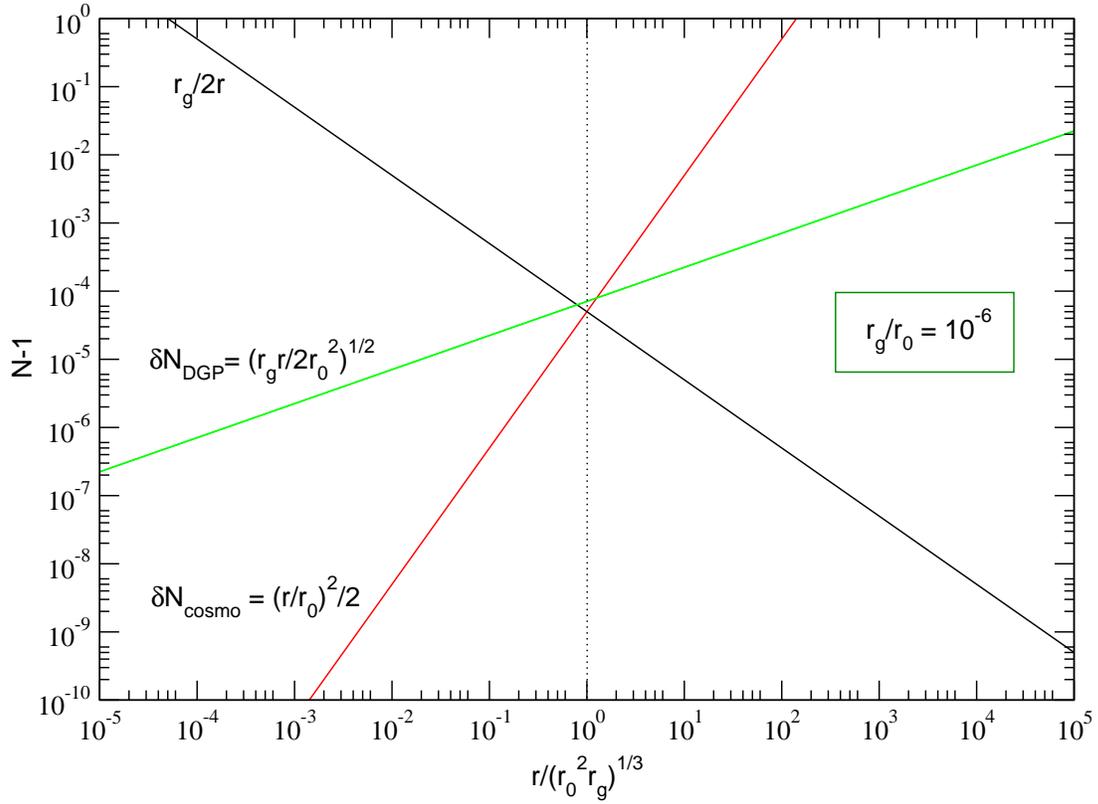
 \begin{center}\PSbox{corr.eps
hscale=60 vscale=60 hoffset=-20 voffset=-20}{6in}{4in}\end{center}
\caption{
The corrections to the Newtonian potential become as large as the potential itself
just as the contributions from cosmology become dominant.  It is for this reason
that we expect the background cosmology to has a significant effect on the
modified potential in the weak-brane regime.
}
\label{fig:corr}
\end{figure}

This computation is more nuanced than for a static matter source in a static Minkowski
background.   We are still interested in finding the metric for compact, spherically symmetric overdensities, but now there is time-evolution to contend with.  However, if we restrict our
attention to distance scales such that $rH \ll 1$ and to nonrelativistic matter sources,\footnote{
	There are a number of simplifications that result from these approximations.  See Ref.~\cite{Lue:2004rj} for an enumeration of these as well as the caveats concerning straying to far from the
	approximations.}
then to leading-order in $r^2H^2$ and $zH$, the solutions to the field equations Eqs.~(\ref{Einstein})
are also solutions to the static equations, i.e. the metric is quasistatic, 
where the only time dependence comes from the slow evolution of the extrinsic curvature
of the brane.  To be explicit, we are looking at the nonrelativistic limit, where the gravitational
potentials of a matter source depend only on the
instantaneous location of the matter's constituents, and not on the motion of those
constituents.  Incidentally, we left the details of the solution to the Minkowski problem
(which were first treated in Refs.~\cite{Gruzinov:2001hp,Porrati:2002cp}) to this section.
All the arguments to be employed may be used, a fortiori, as a
special case of the more general cosmological background.

\subsection{Background Cosmology}

One can choose a coordinate system in which the
cosmological metric respects the spherical symmetry of the matter source.
We are concerned with processes at distances, $r$, such that $rH \ll 1$.  
Under that circumstance it is useful to change coordinates to a frame 
that surrenders explicit brane spatial homogeneity but preserves isotropy
\bea
r(\tau,\lambda^i) &=& a(\tau)\lambda     \\
t(\tau,\lambda^i) &=& \tau + {\lambda^2\over 2} H(\tau)a^2(\tau)\ ,
\eea
for all $z$ and where $\lambda^2 = \delta_{ij}\lambda^i\lambda^j$.
The line element becomes
\be
ds^2 = \left[1 \mp 2(H+\dot{H}/H)|z| - (H^2+\dot{H})r^2\right]dt^2
-  \left[1 \mp 2H|z|\right]\left[(1 + H^2r^2)dr^2 + r^2d\Omega\right] - dz^2\ ,
\label{cosmo2}
\ee
where here dot repreresents differentiation with respect to the new time coordinate, $t$.  Moreover, $H = H(t)$ in this coordinate system.  
All terms of ${\cal O}(r^3H^3)$ or ${\cal O}(z^2H^2,zHrH)$ and higher have been neglected.
The key is that because we are interested primarily in phenomena 
whose size is much smaller than the cosmic horizon, 
the effect of cosmology is almost exclusively to control the extrinsic curvature, 
of the brane. 
{\em This can be  interpreted as a modulation of the brane's stiffness  or 
 the strength of the scalar gravitational mode.}
 
In the coordinate system described by Eq.~(\ref{cosmo2}), the bulk is like a Rindler space.  
This has a fairly natural
interpretation if one imagines the bulk picture~\cite{Deffayet,Lue:2002fe}.
One imagines riding a local patch of the brane, which appears as hyperspherical surface expanding into 
(or away from) a five-dimensional Minkowski bulk.  This surface either accelerates or decelerates 
in its motion with respect to the bulk, creating a Rindler-type potential.
Note that we are keeping the upper-and-lower-sign convention to represent
the two cosmological phases.  While we are
nominally focussed on self-acceleration, we will see that contributions from
the sign have important effects on the modified gravitational potentials.

\subsection{Metric Potentials}

We have chosen a coordinate system, Eq.~(\ref{metric}), in which a
compact spherical matter source may have a quasistatic metric, yet still
exist within a background cosmology that is nontrivial (i.e., deSitter
expansion).  Let us treat the matter distribution to be that required
for the background cosmology, Eq.~(\ref{cosmo2}), and
add to that a compact spherically symmetric matter source, located on
the brane around the origin ($r=0,z=0$)
\be
T^A_B|_{\rm brane}= ~\delta (z)\ {\rm diag}
\left(\delta\rho(r)+\rho_B,-\delta p(r)-p_B,
-\delta p(r)-p_B,-\delta p(r)-p_B,~0 \right)\ ,
\label{matter-EM}
\ee
where $\rho_B$ and $p_B$ are the density and pressure of the
background cosmology, and where $\delta\rho(r)$ is the overdensity
of interest and $\delta p(r)$ is chosen to ensure the matter distribution and
metric are quasistatic.  We may define an effective Schwarzschild radius
\be
      R_g(r,t) = {8\pi\over \mpsq}\int_0^r r^2\delta\rho(r,t) dr\ .
\label{rg}
\ee
We solve the perturbed Einstein equations in quasistatic approximation 
by generalizing the method used in~\cite{Lue:2002sw}, 
obtaining the metric of a spherical mass overdensity $\delta\rho(t,r)$ 
in the background of the cosmology described by Eqs.~(\ref{Fried}) and~(\ref{friedmann2}).
Because we are interested only in weak matter sources, $\rho_g(r)$, and
since we are interested in solutions well away from the cosmic horizon,
we can still exploit doing an expansion using Eqs.~(\ref{linearize}) and keeping
only leading orders.  Now, we just need to take care to include leading orders
in terms of $H$ as well as other parameters of interest.

We are particularly concerned with the evaluation of the metric on
the brane.  However, we need to take care that when such an evaluation
is performed, proper boundary conditions in the bulk are satisfied,
i.e., that there no singularities in the bulk, which is tantamount to having
spurious mass sources there.  In order to determine the metric on the brane,
we implement the following approximation \cite{Lue:2002sw,Lue:2004rj,Lue:2004za}:
\be
\left. n_z\right|_{z=0} = \mp \left(H + {\dot{H}\over H}\right)\ .
\label{assumption}
\ee
Note that this is just the value $n_z$ would take if there were only a
background cosmology, but we are making an assumption that the
presence of the mass source only contributes negligibly to this quantity
at the brane surface.
With this one specification, a complete set of equations,
represented by the brane boundary conditions Eqs.~(\ref{branebc}) and
$G_z^z=0$, exists on the brane so that the metric functions may be
solved on that surface without reference to the bulk.  At the end of this section,
we check that the assumption Eq.~(\ref{assumption}) indeed turns out the
be the correct one that ensures that no pathologies arise in the bulk.

The brane boundary conditions Eqs.~(\ref{branebc}) now take the form
\bea
-(a_z+2b_z) &=& r_0\left[-{2a' \over r} - {2a\over r^2}\right]
			+ {r_0\over r^2}R'_g(r) + 3H(r_0H \pm 1)
\nonumber    \\
-2b_z &=& r_0\left[{2n' \over r} - {2a\over r^2}\right]
	               	+ r_0(3H^2+2\dot{H}) \pm {2\over H}(H^2+\dot{H})
\label{branebc2}     \\
-(a_z+b_z) &=& r_0\left[n'' + {n'\over r} - {a'\over r} \right]
	                 	+ r_0(3H^2+2\dot{H}) \pm {2\over H}(H^2+\dot{H})\ ,
\nonumber
\eea
where we have substituted Eqs.~(\ref{Fried}) and~(\ref{friedmann2}) for
the background cosmological density and pressure and where we have
neglected second-order contributions (including
those from the pressure necessary to keep the compact matter
source quasistatic).

\begin{figure} \begin{center}\PSbox{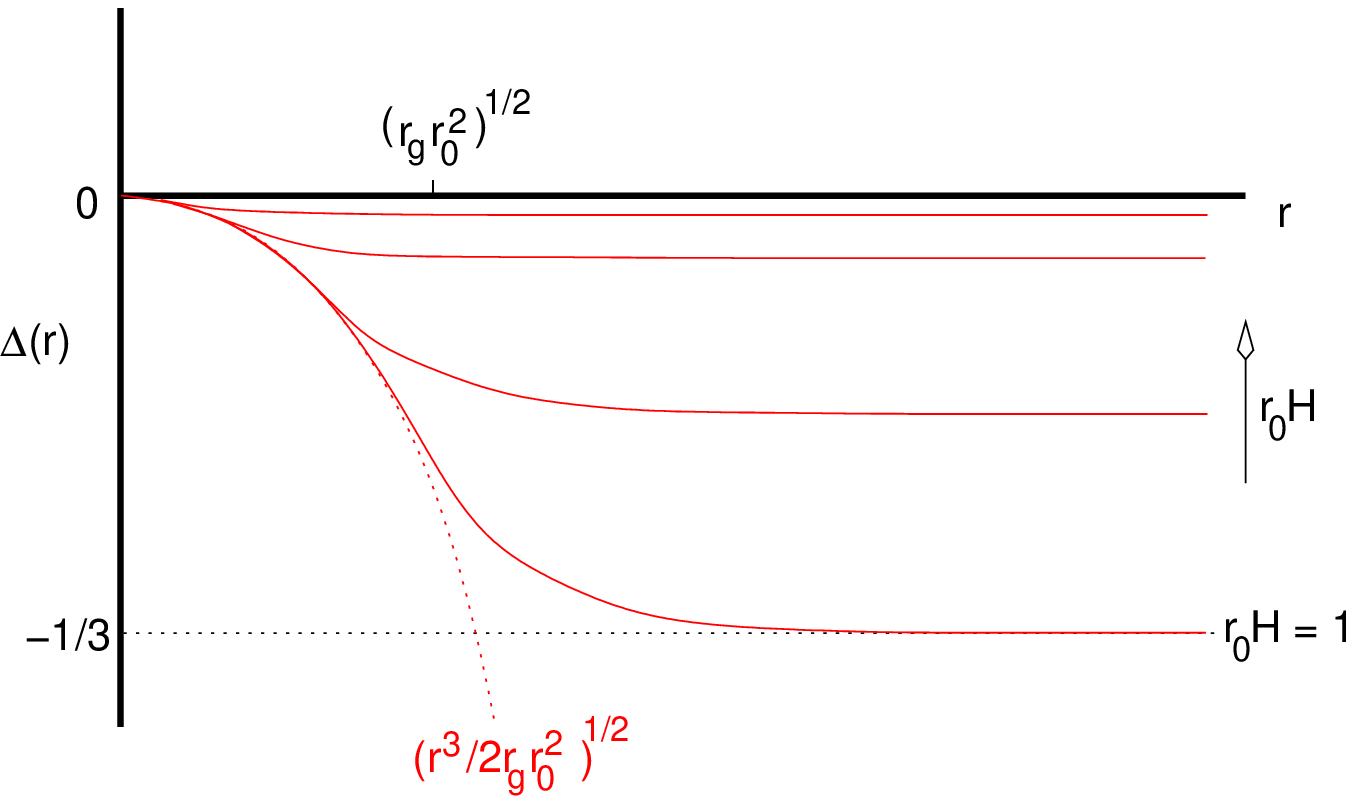
hscale=100 vscale=100 hoffset=0 voffset=0}{5.5in}{3in}\end{center}
\caption{
The function $\Delta(r)$ represents a normalized correction to Newton's constant, $G$, i.e.,
$G_{\rm eff} = G\left[1+\Delta(r)\right]$.  In the self-accelerating cosmological phase,
for small $r$, $\Delta(r)$ asymptotes to
$-(r^3/2r_gr_0^2)^{1/2}$, i.e., a correction independent of cosmology.  For large $r$ (but
also when $r \ll H^{-1}$), $\Delta(r)$ asymptotes the constant value $1/3\beta$.  This value
is $-{1\over 3}$ in the saturated limit $r_0H = 1$, and goes like ${\cal O}(1/r_0H)$ as
$r_0H \rightarrow \infty$.  The boundary between the two regimes is thus
$r_* = (r_gr_0^2/\beta^2)^{1/3}$.  For the FLRW phase, the graph is
just changed by a sign flip, with the exception that the most extreme curve occurs
not when $r_0H =1$, but rather when $H=0$.
}
\label{fig:Geff}
\end{figure}

There are now five equations on the brane with five unknowns.  The solution on the brane
is given by the following.
For a cosmological background with {\em arbitrary} evolution $H(\tau)$,
we find that \cite{Lue:2002sw,Lue:2004rj,Lue:2004za}
\bea
rn'(t,r)|_{\rm brane} &=& {R_g\over 2r}\left[1+\Delta(r)\right] - (H^2+\dot{H})r^2
\label{brane-n}\\
a(t,r)|_{\rm brane} &=& {R_g\over 2r}\left[1-\Delta(r)\right] + {1\over 2}H^2r^2\ .
\label{brane-a}
\eea
Note that the cosmological
background contribution is included in these metric components.  The
function $\Delta(r)$ is defined as
\be
     \Delta(r) = {3\beta r^3\over 4 r_0^2R_g}
     \left[\sqrt{1+{8r_0^2R_g\over 9\beta^2r^3}}-1\right]\ ;
\label{Delta}
\ee
and
\be
     \beta = 1\pm2r_0H\left(1 + {\dot{H}\over 3H^2}\right)\ .
\label{beta}
\ee
Just as for the modified Friedmann equation, Eq.~(\ref{Fried}), there is a sign degeneracy in Eq.~(\ref{beta}).  The lower sign corresponds to the self-accelerating cosmologies.
These expressions are valid on the brane when $r \ll r_0, H^{-1}$.  In both Eqs.~(\ref{brane-n})
and (\ref{brane-a}), the first term represent the usual Schwarzschild contribution
with a correction governed by $\Delta(r)$ resulting from brane dynamics (as
depicted in Fig.~\ref{fig:Geff}),
whereas the second term represents the leading cosmological contribution.
Let us try to understand the character of the corrections.

\subsection{Gravitational Regimes}

One may consolidate all our results and show from
Eqs.~(\ref{brane-n})--(\ref{Delta}) that there exists a scale \cite{Lue:2002sw,Lue:2004rj,Lue:2004za},
\be
	r_* = \left[{r_0^2R_g\over\beta^2}\right]^{1/3}\ ,
\label{radius2}
\ee
with
\bea
	     \beta = 1\pm2r_0H\left(1 + {\dot{H}\over 3H^2}\right)\ .
\nonumber
\eea
Inside a radius $r_*$ the metric is dominated by Einstein gravity but has
corrections which depend on the global cosmological phase.  Outside this radius (but at
distances much smaller than both the crossover scale, $r_0$, and
the cosmological horizon, $H^{-1}$) the metric is weak-brane and
resembles a scalar-tensor gravity in the background of a deSitter
expansion.
This scale is modulated both by the nonperturbative 
extrinsic curvature effects of the source itself as well as the extrinsic curvature of the brane
generated by its cosmology.  The qualitative picture described
in Fig.~\ref{fig:regime} is generalized to the picture shown in
Fig.~\ref{fig:bulk}.

\begin{figure} \begin{center}\PSbox{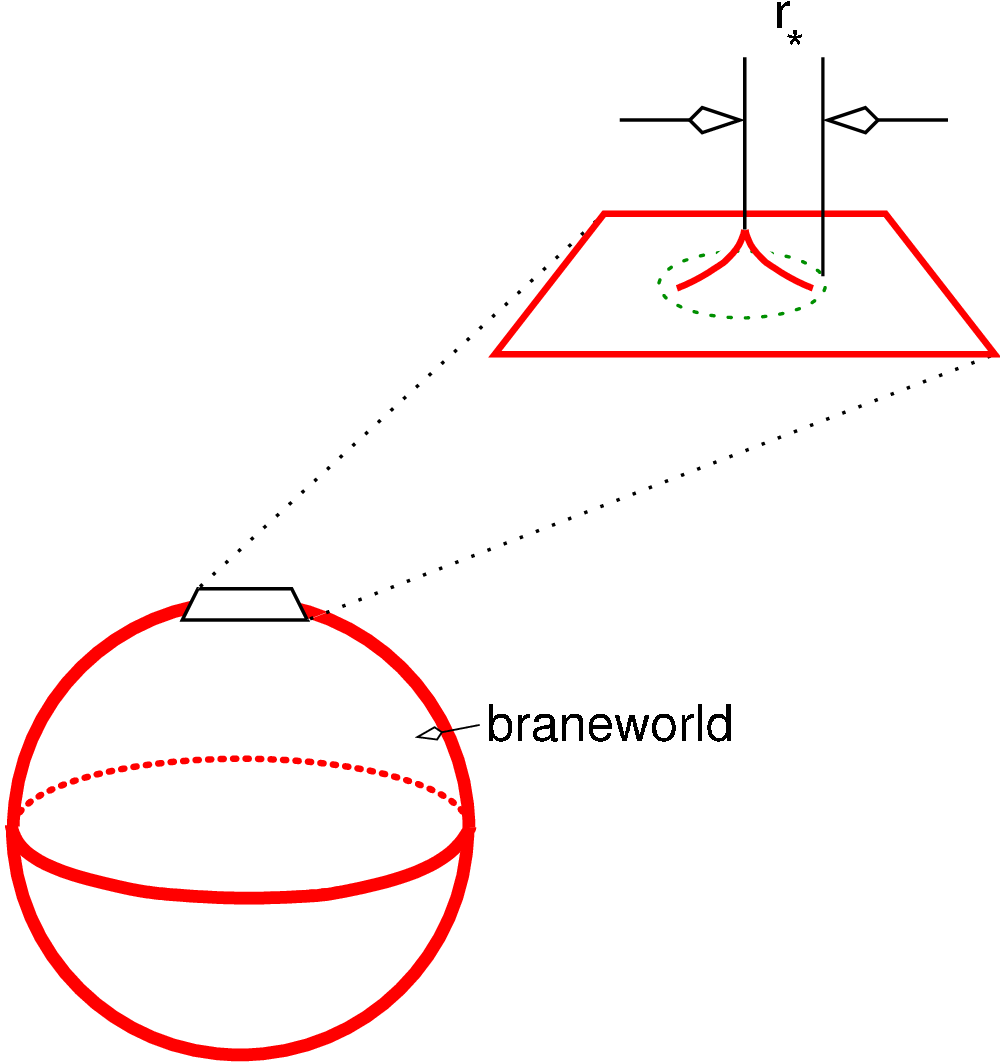
hscale=80 vscale=80 hoffset=0 voffset=0}{3in}{3in}\end{center}
\caption{
The four-dimensional universe where we live is denoted by the large spherical brane.  A local mass source located, for example, near its north pole dynamically dimples the brane, inducing a nonperturbative extrinsic curvature.  That extrinsic curvature suppress coupling of the mass source to the extra scalar mode and, within the region dictated by the radius $r_*$ given by Eq.~(\ref{radius2}), Einstein gravity is recovered.  Outside $r_*$, the gravitational field is still modulated by the effects of the extrinsic curvature of the brane generated by the background cosmology.
}
\label{fig:bulk}
\end{figure}

Keeping Fig.~\ref{fig:Geff} in mind, there are important asymptotic limits of
physical relevance for the metric on the brane, Eqs.~(\ref{brane-n}) and (\ref{brane-a}).
First, when $r\ll r_*$, the metric is close to the Schwarzschild solution of
four-dimensional general relativity.  Corrections to that solution are small:
\bea
	n &=& -{R_g\over 2r} \pm \sqrt{R_g r\over 2r_0^2}
\label{Einstein-n}     \\
	a &=& {R_g\over 2r} \mp \sqrt{R_g r\over 8r_0^2}\ .
\label{Einstein-a}
\eea
The background cosmological expansion becomes largely unimportant
and the corrections are dominated by effects already seen in the Minkowski
background.  Indeed, there is no explicit dependence on the parameter
governing cosmological expansion, $H$.  However, the sign of the
correction to the Schwarzschild solution is dependent on the global
properties of the cosmological phase.  Thus, we may ascertain
information about bulk, five-dimensional cosmological behavior from
testing details of the metric where naively one would not expect
cosmological contributions to be important.

Cosmological effects become important when $r \gg r_*$.  The metric
is dominated by the cosmological flow, but there is still an attractive
potential associated with the central mass source.  Within the cosmological
horizon, $r\ll H^{-1}$, this residual potential is
\bea
        \delta n &=& -{R_g\over 2r}\left[1 + {1\over 3\beta}\right]
\label{weakbrane-n}     \\
        \delta a &=& {R_g\over 2r}\left[1 - {1\over 3\beta}\right]\ .
\label{weakbrane-a}
\eea
This is the direct analog of the weak-brane phase one finds for
compact sources in Minkowski space.  The residual effect of the
matter source is a linearized scalar-tensor gravity with
Brans--Dicke parameter
\be
	\omega = {3\over 2}(\beta -1)\ .
\label{BD}
\ee
Notice  that as  $r_0H \rightarrow  \infty$, we  recover  the Einstein
solution, corroborating  results  found for  linearized  cosmological
perturbations  \cite{Deffayet:2002fn}.  At redshifts of order unity,
$r_0H \sim {\cal O}(1)$ and corrections to four-dimensional Einstein
gravity are substantial and $H(z)$--dependent.  These results found in this
analysis were also followed up in Ref.~\cite{Deffayet:2004ru}.

\subsection{Bulk Solutions}
\label{sec:bulk}

We should elaborate how the solution Eqs.~(\ref{brane-n}) and
(\ref{brane-a}) were ascertained as well as explicitly write the solutions of the potentials
$\{n, a, b\}$ in the bulk  \cite{Lue:2002sw,Lue:2004rj,Lue:2004za}.  The key to our solution is the
assumption Eq.~(\ref{assumption}).
It allows one to solve the brane equations independently of the rest of the
bulk and guarantees the asymptotic bulk boundary conditions.
In order to see why Eq.~(\ref{assumption}) is a reasonable approximation,
we need to explore the full solution to the bulk Einstein equations,
\be
	G_{AB}(r,z) = 0\ ,
\ee
satisfying the brane boundary conditions Eqs.~(\ref{branebc2}), as well as
specifying that the metric approach the cosmological background
Eq.~(\ref{cosmo2}) for large values of $r$ and $z$, i.e., far
away from the compact matter source.

First, it is convenient to consider not only the components of the Einstein
tensor Eqs.~(\ref{5dEinstein}), but also the following components of the
bulk Ricci tensor (which also vanishes in the bulk):
\bea
R_t^t &=& {1\over A^2}\left[{N''\over N} - {N'\over N}{A'\over A}
  + 2{N'\over N}{B'\over B}\right]
+ \left[{N_{zz}\over N} + {N_z\over N}{A_z\over A}
  + 2{N_z\over N}{B_z\over B}\right]
\label{5dRicci-tt}     \\
R_z^z &=& {N_{zz}\over N} + {A_{zz}\over A} + {2B_{zz}\over B}\ .
\label{5dRicci-zz}
\eea
We wish to take $G_{zr}=0$, $G_z^z=0$, and $R_z^z=0$ and derive
expressions for $A(r,z)$ and $B(r,z)$ in terms of $N(r,z)$.  Only two of
these three equations are independent, but it is useful to use all
three to ascertain the desired expressions.

\subsubsection{Weak-Field Analysis}

Since we are only interested in metric when $r,z \ll r_0, H^{-1}$ for
a weak matter source, we may rewrite the necessary field equations
using the expressions Eqs.~(\ref{linearize}).  Since the functions,
$\{n(r,z),a(r,z),b(r,z)\}$ are small, we need only keep nonlinear
terms that include $z$--derivatives.  The brane boundary conditions,
Eqs.~(\ref{branebc}), suggest that $a_z$ and $b_z$ terms may
be sufficiently large to warrant inclusion of their nonlinear
contributions.  It is these $z$--derivative nonlinear terms
that are crucial to the recover of Einstein gravity near the matter
source.  If one neglected these bilinear terms as well, one would
revert to the linearized, weak-brane solution
(cf. Ref.~\cite{Gruzinov:2001hp}).

Integrating Eq.~(\ref{5dRicci-zz}) twice with respect to the
$z$--coordinate, we get
\be
	n + a + 2b = zg_1(r) + g_2(r)\ ,
\label{app-relation1}
\ee
where $g_1(r)$ and $g_2(r)$ are to be specified by the brane
boundary conditions, Eqs.~(\ref{branebc2}), and the residual
gauge freedom $\delta b(r)|_{z=0} = 0$, respectively.
Integrating the $G_{zr}$--component of the bulk Einstein tensor
Eqs.~(\ref{5dEinstein}) with respect to the $z$--coordinate yields
\be
	r\left(n + 2b\right)' - 2\left(a-b\right) = g_3(r)\ .
\label{app-relation2}
\ee
The functions $g_1(r)$, $g_2(r)$, and $g_3(r)$ are not all
independent, and one can ascertain their relationship with one
another by substituting Eqs.~(\ref{app-relation1}) and
(\ref{app-relation2}) into the $G_z^z$ bulk equation.  If one can
approximate $n_z = \mp (H+\dot{H}/H)$ for all $z$, then one can see
that  $G_{zr}=0$, $G_z^z=0$, and $R_z^z=0$ are all consistently
satisfied by Eqs.~(\ref{app-relation1}) and (\ref{app-relation2}),
where the functions $g_1(r)$, $g_2(r)$, and $g_3(r)$ are
determined at the brane using Eqs.~(\ref{brane-n}) and
(\ref{brane-a}) and the residual gauge freedom
$b(r)|_{z=0} = 0$:
\bea
	g_1(r) &=& \mp \left(4H+{\dot{H}\over H}\right) - {r_0\over r^2}\left(R_g\Delta\right)'
\label{g1}	\\
	g_2(r) &=& -{1\over 2}r^2\dot{H} + {R_g\over 2r}(1-\Delta)
		+ \int_0^r dr~{R_g\over 2r^2}(1+\Delta)
\label{g2}	\\
	g_3(r) &=& {R_g\over 2r}(1-3\Delta) - \left(2H^2+\dot{H}\right)r^2\ ,
\label{g3}
\eea
where we have used the function $\Delta(r)$, defined in
Eq.~(\ref{Delta}).  Using Eqs.~(\ref{app-relation1})--(\ref{g3}), we
now have expressions for $a(r,z)$ and $b(r,z)$ completely in
terms of $n(r,z)$ for all $(r,z)$.

Now we wish to find $n(r,z)$ and to confirm that
$n_z = \mp (H+\dot{H}/H)$ is a good approximation everywhere of interest.
Equation~(\ref{5dRicci-tt}) becomes
\be
	n'' + {2n'\over r} + n_{zz} = \pm {H^2+\dot{H}\over H}\left[g_1(r)\pm {H^2+\dot{H}\over H}\right]\ ,
\ee
where again we have neglected contributions if we are only
concerned with $r,z \ll r_0, H^{-1}$.  Using the expression
Eq.~(\ref{g1}), we find
\be
	n'' + {2n'\over r} + n_{zz} = -3\left(H^2+\dot{H}\right)
		\mp {r_0\left(H^2+\dot{H}\right)\over r^2H}\left[R_g\Delta(r)\right]'\ .
\ee
Then, if we let
\be
	n = 1 \mp \left(H+{\dot{H}\over H}\right)z - {1\over 2}\left(H^2+\dot{H}\right)r^2
			\mp r_0{H^2+\dot{H}\over H}\int_0^r dr~{1\over r^2}R_g(r)\Delta(r)
			+ \delta n(r,z)\ ,
\label{app-relation3}
\ee
where $\delta n(r,z)$ satisfies the equation
\be
	\delta n'' + {2\delta n'\over r} + \delta n_{zz} = 0\ ,
\label{potential}
\ee
we can solve Eq.~(\ref{potential}) by requiring that $\delta n$
vanish as $r,z\rightarrow \infty$ and applying the condition
\be
	r~\delta n'|_{z=0} = {R_g\over 2r}
	\left[1 + \left(1\pm 2r_0{H^2+\dot{H}\over H}\right)\Delta(r)\right]\ ,
\label{brane-dn}
\ee
on the brane as an alternative to the appropriate brane boundary
condition for $\delta n(r,z)$ coming from a linear combination of
Eqs.~(\ref{branebc2}).  We can write the solution explicitly:
\be
        \delta n(r,z) = \int_0^{\infty}dk~c(k)e^{-kz}\sin kr\ ,
\ee
where
\be
        c(k) = {2\over \pi}\int_0^\infty dr~r\sin kr
	        \left.\delta n\right|_{z=0}(r)\ .
\ee
We can then compute $\left. \delta n_z\right|_{z=0}$, arriving at the bound
\be
	\left.\delta n_z\right|_{z=0} \lesssim
		{1\over r}\int_0^r dr~{R_g(r)\over r^2}\ ,
\ee
for all $r \ll r_0, H^{-1}$.  Then,
\be
	\left. n_z\right|_{z=0} = \mp \left(H+{\dot{H}\over H}\right) + \left.\delta n_z\right|_{z=0}\ .
\label{dotn}
\ee
When the first term in Eq.~(\ref{dotn}) is much larger than the
second, Eq.~(\ref{assumption}) is a good approximation.  When
the two terms in Eq.~(\ref{dotn}) are comparable or when the
second term is much larger than the first, neither term is
important in the determination of Eqs.~(\ref{brane-n}) and
(\ref{brane-a}).  Thus, Eq.~(\ref{assumption}) is still a safe
approximation.

One can confirm that all the components of the five-dimensional
Einstein tensor, Eqs.~(\ref{5dEinstein}), vanish in the bulk using
field variables satisfying the relationships
Eqs.~(\ref{app-relation1}), (\ref{app-relation2}), and
(\ref{app-relation3}).  The field variables $a(r,z)$ and $b(r,z)$ both
have terms that grow with $z$, stemming from the presence of the
matter source.  However, one can see that with the following
redefinition of coordinates:
\bea
	R &=& r - zr_0{R_g\Delta\over r^2}	\\
	Z &=& z + \int_0^r dr~ {R_g\Delta\over r^2}\ ,
\eea
that to leading order as $z \rightarrow H^{-1}$, the
desired $Z$--dependence is recovered for $a(R,Z)$ and $b(R,Z)$
(i.e., $\mp HZ$), and the Newtonian potential takes the form
\be
	n(R,Z) = \mp \left(H+{\dot{H}\over H}\right)Z -  {1\over 2}(H^2+\dot{H})R^2 + \cdots\ .
\ee
Thus, we recover the desired asymptotic form for the metric
of a static, compact matter source in the background of a
cosmological expansion.

\subsubsection{A Note on Bulk Boundary Conditions}

A number of studies have been performed for DGP gravity that have either arrived at or
used modified force laws different than those given by Eqs.~(\ref{brane-n})--(\ref{beta}) \cite{Kofinas:2001qd,Kofinas:2002gq,Song:2005gm,Knox:2005rg,Ishak:2005zs}.  How does one understand
or reconcile such a discrepancy?  Remember that one may ascertain the metric on the brane
without reference to the bulk because there are five unknown quantities, $\{N(r),A(r),N_z(r),A_z(r),B_z(r)\}$, and there are four independent equations (the three brane boundary conditions, Eqs.~(\ref{branebc2}),
and the $G_z^z$--component of the bulk Einstein equations) on the brane in terms of only these
quantities.  One need only choose an additional relationship between the five quantities in order to
form a closed, solvable system.  We chose Eq.~(\ref{assumption}) and showed here that it was
equivalent to choosing the bulk boundary condition that as $z$ became large, one recovers the background
cosmology.  The choice made in these other analyses is tantamount to a different choice of bulk
boundary conditions.  One must be very careful in the analysis of DGP gravitational fields that one is
properly treating the asymptotic bulk space, as it has a significant effect on the form of the metric
on the brane.

\section{Anomalous Orbit Precession}
\label{sec:solarsystem}

We have established that DGP gravity is capable of generating a contemporary cosmic
acceleration with dark energy.  However, it is of utmost interest to understand how one
may differentiate such a radical new theory from a more conventional dark energy model
concocted to mimic a cosmic expansion history identical to that of DGP gravity.  The results
of the previous sections involving the nontrivial recovery of Einstein gravity and large
deviations of this theory from Einstein gravity in observably accessible regimes is the
key to this observational differentiation.  Again, there are two
regimes where one can expect to test this theory and thus, there are two clear regimes in which to
observationally challenge the theory.  First deep within the gravity well, where $r\ll r_*$ and where
the corrections to general relativity are small, but the uncertainties are also
correspondingly well-controlled.
The second regime is out in the cosmological flow, where $r\gg r_*$ (but still $r\ll r_0\sim H_0^{-1}$)
and where corrections to general relativity are large, but
our observations are also not as precise.  We focus on the first possibility in this section and
will go to the latter in the next section.

\subsection{Nearly Circular Orbits}

Deep in the gravity well of a matter source, where the effects of cosmology are ostensibly
irrelevant, the correction to the gravitational potentials may be represented by effective
correction to Newton's constant
\be
	\Delta(r) = \pm\sqrt{{r^3\over 2r_0^2 R_g}}\ ,
\ee
that appears in the expressions Eqs.~(\ref{brane-n}) and (\ref{brane-a}).
Though there is no explicit dependence on the background $H(\tau)$ evolution, there
is a residual dependence on the cosmological phase through the overall sign.  In DGP gravity,
tests of a source's Newtonian force leads to discrepancies with general relativity \cite{Gruzinov:2001hp,Lue:2002sw,Dvali:2002vf}.

Imagine a body orbiting a mass source where $R_g(r) = r_g = {\rm
constant}$.  The perihelion precession per orbit may be determined
in the usual way given a metric of the form Eqs.~(\ref{linearize})
and (\ref{metric-brane})
\be
       \Delta\phi = \int dr~{J\over r^2}
		  {AN\over \sqrt{E^2 - N^2\left(1+{J^2\over r^2}\right)}}\ ,
\ee
where $E = N^2dt/ds$ and $J = r^2d\phi/ds$ are constants of
motion resulting from the isometries of the metric, and $ds$ is the differential
proper time of the orbiting body. With a nearly circular orbit deep within the
Einstein regime (i.e., when $r\ll r_*$ so that we may use
Eqs.~(\ref{Einstein-n}) and (\ref{Einstein-a})), the above expression yields
\be
       \Delta\phi = 2\pi + {3\pi r_g\over r}
       \mp {3\pi\over 2}\left(r^3\over 2r_0^2r_g\right)^{1/2}\ .
\ee
The second term is the famous precession correction from general
relativity.  The last term is the new anomalous precession due to DGP
brane effects.  This latter correction is a purely Newtonian effect.
Recall that any deviations of a Newtonian central potential from $1/r$
results in an orbit precession.  The DPG correction to the precession rate is
now \cite{Gruzinov:2001hp,Lue:2002sw,Dvali:2002vf}
\be
{d\over dt}{\Delta\phi}_{\rm DGP} = \mp {3\over 8r_0}
= \mp 5~\mu{\rm as/year}\ .
\label{corr-DGP}
\ee
Note that this result is independent of the source mass, implying that
this precession rate is a universal quantity dependent only on the
graviton's effective linewidth ($r_0^{-1}$) and the overall cosmological
phase.  Moreover, the final result depends on the sign of the cosmological
phase \cite{Lue:2002sw}.  Thus one can tell by the sign of the precession whether self-acceleration
is the correct driving force for today's cosmic acceleration.  It is extraordinary that a
local measurement, e.g. in the inner solar system, can have something definitive
to say about gravity on the largest observable scales.

\subsection{Solar System Tests}

Nordtvedt \cite{Nordtvedt:ts} quotes precision for perihelion precession at $430~\mu{\rm as/year}$ for Mercury and $10~\mu{\rm as/year}$ for Mars.  Improvements in lunar ranging measurements \cite{Williams:1995nq,Dvali:2002vf} suggest that the Moon will be sensitive to the DGP correction Eq.~(\ref{corr-DGP}) in the future lunar ranging studies.  Also, BepiColombo (an ESA satellite) and MESSENGER (NASA) are being sent to Mercury at the end of the decade, will also be sensitive to this correction \cite{Milani:2002hw}.  Counterintuitively, future {\em outer} solar system probes that possess precision ranging instrumentation, such as Cassini \cite{Bertotti:2003rm}, may also provide ideal tests of the anomalous precession.  Unlike post-Newtonian deviations arising from Einstein corrections, this anomaly does not attenuate with distance from the sun;  indeed, it amplifies.  This is not surprising
since we know that corrections to Einstein gravity grow as gravity weakens.
More work is needed to ascertain whether important inner or outer solar system systematics allow this anomalous precession effect to manifest itself effectively.  If so, we may enjoy precision tests of general relativity even in unanticipated regimes \cite{Adelberger:2003zx,Will:2004xi,Turyshev:2005aw,Turyshev:2005ux,Iorio:2005qn}.  The solar system seems to provide a most promising means to constrain this anomalous precession from DGP gravity.\footnote{
	One should also note that DGP gravity cannot explain the famous Pioneer anomaly \cite{Anderson:2001sg}.  First, the functional form of the anomalous acceleration in DGP is not correct:  it decays as a
	$r^{-1/2}$ power-law with distance from the sun.  Moreover, the magnitude of the effect is
	far too small in the outer solar system.}

it is also interesting to contrast solar system numbers with those for binary pulsars.  The rate of periastron advance for the object PSR~1913+16 is known to a precision of $4\times 10^4~\mu{\rm as/year}$ \cite{Will:2001mx}.  This precision is not as good as that for the inner solar system.
A tightly bound
system such as a binary pulsar is better at finding general relativity corrections to Newtonian gravity
because it is a stronger gravity system than the solar system.  It is for precisely the same reason that
binary pulsars prove to be a worse environment to test DGP effects, as these latter effects become less prominent as the
gravitational strength of the sources become stronger.

As a final note, one must be extremely careful about the application of Eqs.~(\ref{brane-n})
and (\ref{brane-a}).  Remember they were derived for a spherical source in isolation.  We
found that  the resulting metric, while in a weak-field regime, i.e., $r_g/r \ll 1$, was nevertheless
still very nonlinear.  Thus, superposibility of central sources is no longer possible.  This introduces
a major problem for the practical application of Eqs.~(\ref{brane-n}) and (\ref{brane-a}) to real
systems.  If we take the intuition developed in the last two sections, however, we can develop
a sensible picture of when these equations may be applicable.  We found that being in the
Einstein regime corresponds to being deep within the gravity well of a given matter
source.  Within that gravity well, the extrinsic curvature of the brane generated by the source
suppressed coupling of the extra scalar mode.  Outside a source's gravity well, one is
caught up in the cosmological flow, the brane is free to fluctuate, and the gravitational attraction
is augmented by an extra scalar field.

\begin{figure} \begin{center}\PSbox{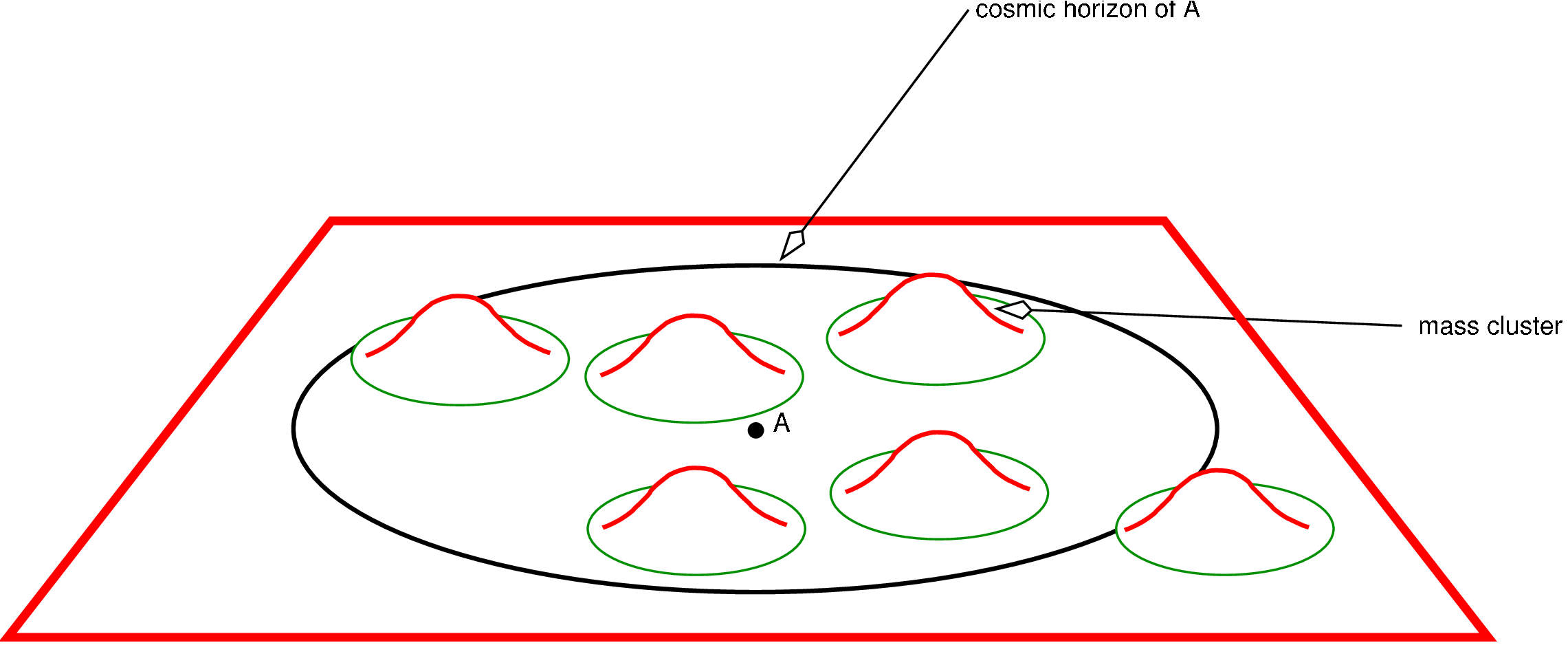
hscale=60 vscale=60 hoffset=-80 voffset=0}{3in}{1.8in}\end{center}
\caption{
The universe is populated by a variety of matter sources.  For sources massive enough to
deform the brane substantially, Einstein gravity is recovered within the gravity well
of each region (the green circles whose sizes
are governed by $r_* \sim ({\rm mass})^{1/3}$).  Outside
these gravity wells, extra scalar forces play a role.
}
\label{fig:comoving}
\end{figure}

If one takes a ``rubber sheet" picture of how this DGP effect works, we can imagine several
mass sources placed on the brane sheet, each deforming the brane in a region whose size
corresponds with the mass of the source (see Fig.~\ref{fig:comoving}).  These
deformed regions will in general overlap with each other in some nontrivial fashion.  However,
the DGP corrections for the orbit of a particular test object should be dominated by
the gravity well in which the test object is orbiting.  For example, within the solar system, we are clearly
in the gravity well of the sun, even though we also exist inside a larger galaxy, which in turn
is inside a cluster, etc.  Nevertheless, the spacetime curvature around us is dominated by the
sun, and for DGP that implies the extrinsic curvature of the brane in our solar system is also
dominated by the sun.
This picture also implies that if we are not rooted in the gravity well of a single body, then
the quantitative form of the correction given by Eqs.~(\ref{Einstein-n}) and (\ref{Einstein-a})
is simply invalid.  I.e., three body systems need to be completely reanalyzed.  This may
have relevant consequences for the moon which has substantial gravitational influences
from both the earth and the sun.

\section{Large-Scale Structure}
\label{sec:lss}

Future solar system tests have the possibility of probing 
the residual deviation from four-dimensional Einstein gravity at distances well 
below $r_*$.
Nevertheless, it would be ideal to test gravitational physics where dramatic differences 
from Einstein gravity are anticipated.  Again, this is the crucial element to the program
of differentiating a modified-gravity scenario such as DGP from a dark-energy explanation
of today's cosmic acceleration that has an identical expansion history.  A detailed study of
large scale structure in the Universe can provide a test of gravitational physics at large distance
scales where we expect anomalous effects from DGP corrections to be large.
In the last section, we have seen that the modified force law is sensitive to the background cosmological expansion, 
since this expansion is intimately tied to the extrinsic curvature of the brane~\cite{Deffayet,Lue:2002fe}, 
and this curvature controls the effective Newtonian potential \cite{Lue:2002sw,Lue:2004rj,Lue:2004za}.
This gives us some measure of
sensitivity to how cosmology affects the changes in the growth of structure through the
modulation of a cosmology-dependent Newtonian potential.  We may then proceed and compare those results 
to the standard cosmology, as well as to a cosmology that exactly mimics the DGP expansion history 
using dark energy.  The description of the work presented in this section first appeared in Ref.~\cite{Lue:2004rj}.

\subsection{Linear Growth}

The modified gravitational potentials given by Eqs.~(\ref{brane-n}) and (\ref{brane-a}) indicate that
substantial deviations from Einstein gravity occur when $r \gtrsim r_*$.  More generally, from our
discussion of the VDVZ discontinuity, we expect large deviations from general relativity any time
an analysis perturbative in the source strength is valid.  This is true when we want to understand
the linear growth of structure in our universe.  So long as we consider
early cosmological times, when perturbations around the homogeneous cosmology are small, or
at later times but also on scales much larger than the clustering scale, a linear analysis should be
safe.

The linear regime is precisely what we have analyzed with the potential given by Eq.~(\ref{prop}),
or, more generally, Eqs.~(\ref{weakbrane-n}) and (\ref{weakbrane-a})
\bea
        \delta n &=& -{R_g\over 2r}\left[1 + {1\over 3\beta}\right]
\nonumber     \\
        \delta a &=& {R_g\over 2r}\left[1 - {1\over 3\beta}\right]\ ,
\nonumber
\eea
in a background cosmology
(remember, we are still only considering nonrelativistic sources at scales smaller than the horizon
size).  Because we are in the linear regime, we may think of these results as Green's function
solutions to the general physical problem.  Thus, these potentials are applicable beyond just spherical sources and we deduce that
\bea
	\nabla^2 \delta n(r,t) &=& 4\pi G\left[1 + {1\over 3\beta}\right] \delta\rho({\bf x},t)
\label{poisson-n} \\
	\nabla^2 \delta a(r,t) &=& 4\pi G\left[1 - {1\over 3\beta}\right]\delta\rho({\bf x},t)\ ,\
\label{poisson-a}
\eea
for {\rm general} matter distributions, so long as they are nonrelativistic.  What results is a
linearized scalar-tensor theory with cosmology-dependent Brans--Dicke parameter Eq.~(\ref{BD})
\bea
	\omega = {3\over 2}(\beta - 1)\ .
\nonumber
\eea
Note this is only true for weak, linear perturbations around the cosmological background, and that
once overdensities becomes massive enough for self-binding, these results cease to be valid
and one needs to go to the nonlinear treatment.

Nonrelativistic matter perturbations with $\delta(\tau) = \delta\rho/\rho(t)$ evolve via
\be
	\ddot{\delta} + 2H\dot{\delta} = 4\pi G\rho\left(1+{1\over 3\beta}\right)\delta\ ,
\label{growth}
\ee
where $\rho(t)$ is the background cosmological energy density, 
implying that self-attraction of overdensities is governed by an evolving $G_{{\rm eff}}$:
\be
	G_{{\rm eff}} = G\left[1 + {1\over 3\beta}\right]\ .
\label{Geff}
\ee
Here the modification manifests itself  as a time-dependent effective Newton's constant, $G_{\rm eff}$.
Again, as we are focused on the self-accelerating phase, then from Eq.~(\ref{beta})
\bea
     \beta = 1 - 2r_0H\left(1 + {\dot{H}\over 3H^2}\right)\ .
\nonumber
\eea
As time evolves the effective gravitational constant decreases.  For example, if $\omt=0.3$, $G_{\rm eff}/G=0.72,0.86,0.92$ at $z=0,1,2$. 

One may best observe this anomalous repulsion (compared to general relativity) through the growth of large-scale structure in the early universe.  That growth is governed not only by the expansion of the universe, but also by the gravitational self-attraction of small overdensities.  Figure~\ref{fig:growth} depicts how growth of large-scale structure is altered by DGP gravity.  The results make two important points:  1) growth is suppressed compared to the standard cosmological model since the expansion history is equivalent to a $w(z)$ dark-energy model with an effective equation of
state given by Eq.~(\ref{weff})
\bea
w_{\rm eff}(z) = -\frac{1}{1+\om}\ ,
\nonumber
\eea
and 2) growth is suppressed even compared to a dark energy model that has identical expansion history from the alteration of the self attraction from the modified Newton's constant, Eq.~(\ref{Geff}).  The latter point reiterates the crucial feature that one can differentiate between this modified-gravity model and dark energy.

\begin{figure}[t]
\centerline{\epsfxsize=10cm\epsffile{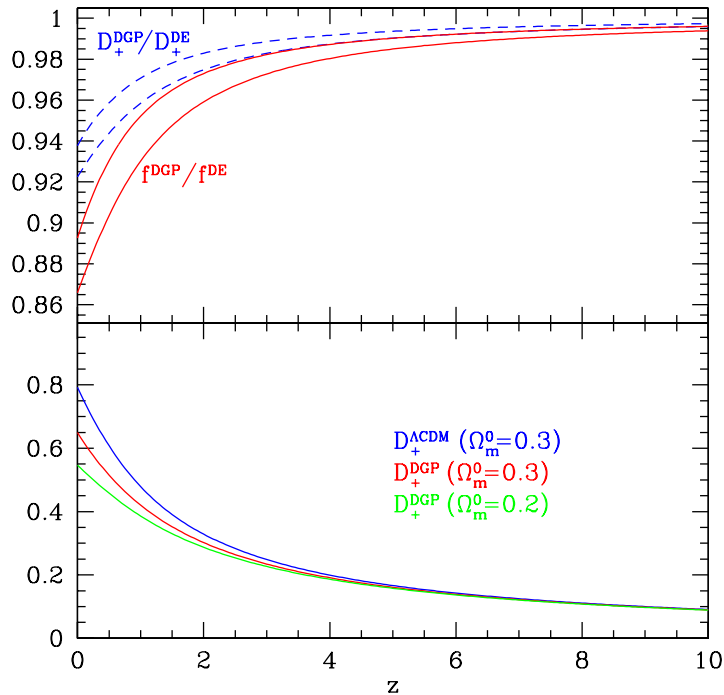}}
\caption{The top panel shows the ratio of the growth factors $D_+$ (dashed lines) in DGP gravity [Eq.~(\ref{growth})] and a model of dark energy (DE) with an equation of state such that it gives rise to the same expansion history (i.e. given by Eq.~(\ref{Fried}), but where the force law is still given by general relativity). The upper line corresponds to $\omt=0.3$, the lower one to $\omt=0.2$. The solid lines show the analogous result for velocity perturbations factors $f\equiv d\ln D_+/d\ln a$. The bottom panel shows the growth factors as a function of redshift for models with {\em different} expansion histories, corresponding to (from top to bottom) $\Lambda$CDM ($\omt=0.3$), and DGP  gravity with $\omt=0.3,0.2$ respectively.
Figure from Ref.~[56].}
\label{fig:growth}
\end{figure}

\subsection{Nonlinear Growth}
\label{sec:nlgrowth}

We certainly want to understand large scale structure beyond the linear analysis.
Unlike the standard cosmological scenario where the self-gravitation of
overdensities may be treated with a linear Newtonian gravitational field, in DGP gravity
the gravitational field is highly nonlinear, even though the field is weak.  This
nonlinearity, particularly the inability to convert the gravitational field into a superposition of
point-to-point forces, poses an enormous challenge to understanding growth of structure
in DGP gravity.

Nevertheless, we already have the tools to offer at least a primitive preliminary analysis.
We do understand the full nonlinear gravitational field around spherical, nonrelativistic
sources, Eqs.~(\ref{brane-n})--(\ref{beta}).  Consider the evolution of a spherical top-hat
perturbation $\delta(t,r)$ of top-hat radius $R_t$.  At subhorizon scales ($Hr \ll 1$),
the contribution from the Newtonian potential, $n(t,r)$, dominates the
geodesic evolution of the overdensity.  The equation of motion for the perturbation is
\cite{Lue:2004rj}
\be
\ddot{\delta} - \frac{4}{3} \frac{\dot{\delta}^2}{1+\delta}+2H\dot{\delta} = 4\pi G \rho\, \delta (1+\delta) \left[ 1 + \frac{2}{3\beta} \frac{1}{\epsilon} \left( \sqrt{1+\epsilon}-1\right)\right]\ ,
\label{sphc}
\ee
where $\epsilon\equiv8r_0^2R_g/9\beta^2R_t^3$.  Note that for large $\delta$,
Eq.~(\ref{sphc}) reduces to the standard evolution of spherical perturbations in
general relativity.

\begin{figure}[t]
\centerline{\epsfxsize=10cm\epsffile{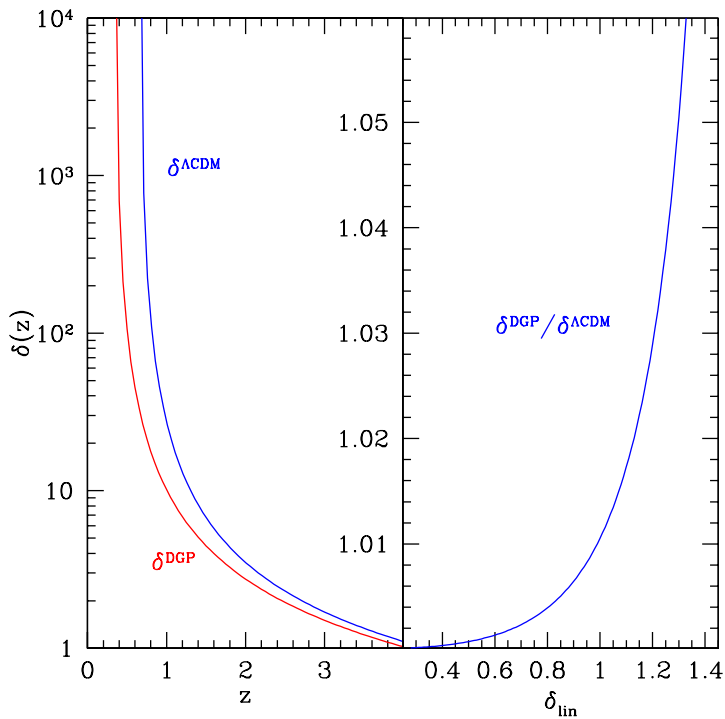}}
\caption{Numerical solution of the spherical collapse. 
The left panel shows the evolution for a spherical perturbation with $\delta_i=3\times 10^{-3}$ 
at $z_i=1000$ for $\omt=0.3$ in DGP gravity and in $\Lambda$CDM. 
The right panel shows the ratio of the solutions once they are both expressed 
as a function of their linear density contrasts.  Figure from Ref.~[56].}
\label{fig:SphColl}
\end{figure}

Figure~\ref{fig:SphColl} shows an example of a full solution of Eq.~(\ref{sphc})
and the corresponding solution in the cosmological constant case. 
Whereas such a perturbation collapses in the $\Lambda$CDM case at $z=0.66$ 
when its linearly extrapolated density contrast is $\delta_c=1.689$, 
for the DGP case the collapse happens much later at $z=0.35$ when its $\delta_c=1.656$.  
In terms of the linearly extrapolated density contrasts things do not look very different, 
in fact, when the full solutions are expressed as a function of the linearly extrapolated density contrasts, 
$\delta_{\rm lin} = D_+ \delta_i/(D_+)_i$ they are very similar to within a few percent.
This implies that all the higher-order moments of the density field 
are very close to those for $\Lambda$CDM models, for example, the skewness is less than a $1\%$
difference from $\Lambda$CDM. 

This close correspondence of the higher-order moments can be useful by allowing
the use of non-linear growth to constrain 
the bias between galaxies and dark matter in the same way as it is done in standard case, thus inferring 
the linear growth factor from the normalization of the power spectrum in the linear regime. 
Although the result in the right panel in Fig.~\ref{fig:SphColl} may seem a coincidence at first sight, 
Eq.~(\ref{sphc}) says that the nontrivial correction from DGP gravity in square brackets is maximum 
when $\delta=0$ (which gives the renormalization of Newton's constant). 
As $\delta$ increases the correction disappears (since DGP becomes Einstein at high-densities), 
so most of the difference between the two evolutions happens in the linear regime, 
which is encoded in the linear growth factor.

\subsection{Observational Consequences}
\label{sec:obs}

What are the implications of these results for testing DGP gravity using large-scale structure?  
A clear signature of DGP gravity is the suppressed (compared to $\Lambda$CDM) growth of perturbations in the linear regime due to the different expansion history and the addition of a repulsive contribution to the force law.  However, in order to predict the present normalization of the power spectrum at large scales, we need to know the normalization of the power spectrum at early times from the CMB. A fit of the to pre-WMAP CMB data was performed in Ref.~\cite{Deffayet:2002sp} using the angular diameter distance for DGP gravity, finding a best fit (flat) model with $\omt\simeq 0.3$, with a very similar CMB power spectrum to the standard cosmological constant model (with $\omt\simeq 0.3$ and $\Omega_\Lambda^0=0.7$) and other parameters kept fixed at the same value. Here we use this fact, plus the normalization obtained from the best-fit cosmological constant  power-law model from WMAP~\cite{Spergel} which has basically the same (relevant for large-scale structure) parameters as in Ref.~\cite{Deffayet:2002sp}, except for the normalization of the primordial fluctuations which has increased compared to pre-WMAP data (see e.g. Fig.~11 in Ref.~\cite{Hinshaw}). The normalization for the cosmological constant scale-invariant model corresponds to present {\em rms} fluctuations in spheres of 8 Mpc/$h$,  $\sigma_8=0.9\pm 0.1$ (see Table 2 in~\cite{Spergel}).

\begin{figure}[t!]
\centerline{\epsfxsize=10cm\epsffile{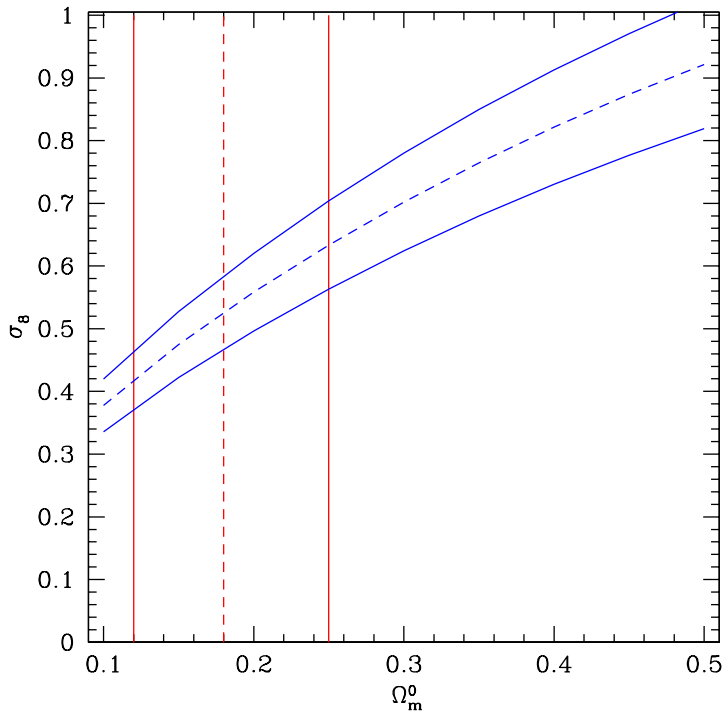}}
\caption{The linear power spectrum normalization, $\sigma_8$, for DGP gravity as a function of $\omt$. The vertical lines denote the best fit value and $68\%$ confidence level error bars from fitting to type-IA supernovae data from~\protect\cite{Deffayet:2002sp}, $\omt=0.18^{+0.07}_{-0.06}$. The other lines correspond to $\sigma_8$ as a function of $\omt$ obtained by evolving the primordial spectrum as determined by WMAP by the DGP growth factor. Figure from Ref.~[56].}
\label{fig:sigma8}
\end{figure}

Figure~\ref{fig:sigma8} shows the present value of $\sigma_8$ as a function of $\omt$ for DGP gravity, where we assume that the best-fit normalization of the {\em primordial} fluctuations stays constant as we change $\omt$, and recompute the transfer function and growth factor as we move away from $\omt=0.3$. Since most of the contribution to $\sigma_8$ comes from scales $r<100 h$/Mpc, we can calculate the transfer function using Einstein gravity, since these modes entered the Hubble radius at redshifts high enough that they evolve in the standard fashion. The value of $\sigma_8$ at $\omt=0.3$ is then given by $0.9$ times the ratio of the DGP to $\Lambda$CDM growth factors shown in the bottom panel of Fig.~\ref{fig:growth}. The error bars in $\sigma_8$ reflect the uncertainty in the normalization of primordial fluctuations, and we keep them a constant fraction as we vary $\omt$ away from $0.3$. We see in Fig.~\ref{fig:sigma8} that for the lower values of $\omt$ preferred by fitting the acceleration of the universe, the additional suppression of growth plus the change in the shape of the density power spectrum drive $\sigma_8$ to a rather small value. This could in part be ameliorated by increasing the Hubble constant, but not to the extent needed to keep $\sigma_8$ at reasonable values. The vertical lines show the best-fit and 1$\sigma$ error bars from fitting DGP gravity to the supernova data from Ref.~\cite{Deffayet:2002sp}. This shows that fitting the acceleration of the universe requires approximately $\sigma_8\leq0.7$ to 1$\sigma$ and $\sigma_8\leq0.8$ to 2$\sigma$. 

In order to compare this prediction of $\sigma_8$ to observations one must be careful since most determinations of $\sigma_8$ have built in the assumption of Einstein gravity or $\Lambda$CDM models. We use galaxy clustering, which in view of the results in Sect.~\ref{sec:nlgrowth} for higher-order moments, should provide a test of galaxy biasing independent of gravity being DGP or Einstein. 
Recent determinations of $\sigma_8$ from galaxy clustering in the SDSS survey~\cite{Tegmark03a} give $\sigma_8^*=0.89\pm 0.02$ for $L^*$ galaxies at an effective redshift of the survey $z_s=0.1$. We can convert this value to $\sigma_8$ for dark matter at $z=0$ as follows. We evolve to $z=0$ using a conservative growth factor, that of DGP for $\omt=0.2$. In order to convert from $L^*$ galaxies to dark matter, we use the results of the bispectrum analysis of the 2dF survey~\cite{Verde02} where $b=1.04\pm0.11$ for luminosity $L\simeq 1.9L^*$. We then scale to $L^*$ galaxies using the empirical relative bias relation obtained in~\cite{Norberg01} that $b/b^*=0.85+0.15(L/L^*)$, which is in very good agreement with SDSS (see Fig.~30 in Ref.~\cite{Tegmark03a}). This implies $\sigma_8=1.00\pm0.11$. Even if we allow for another $10\%$ systematic uncertainty in this procedure, the preferred value of $\omt$ in DGP gravity  that fits the supernovae data is about 2$\sigma$ away from that required by the growth of structure at $z=0$. 

Nevertheless, the main difficulty for DGP gravity to simultaneously explain cosmic acceleration and the growth of structure is easy to understand: the expansion history is already significantly different from a cosmological constant, corresponding to an effective equation of state with $w_{\rm eff}=-(1+\om)^{-1}$. This larger value of $w$ suppresses the growth somewhat due to earlier epoch of the onset of acceleration. In addition, the new repulsive contribution to the force law suppreses the growth even more, driving $\sigma_8$ to a rather low value, in contrast with observations. If as error bars shrink the supernovae results continue to be consistent with $w_{\rm eff}=-1$, this will drive the DGP fit to a yet lower value of $\omt$ and thus a smaller value of $\sigma_8$. For these reasons we expect the tension between explaining acceleration and the growth of structure to be robust to a more complete treatment of the comparison of DGP gravity against observations.
Already more ideas for how to approach testing Newton's constant on ultralarge scale and the
self-consistency of the DGP paradigm for explaining cosmic accleration have been taken \cite{Linder:2004ng,Sealfon:2004gz,Bernardeau:2004ar,Ishak:2005zs,Linder:2005in}.

\section{Gravitational Lensing}

\subsection{Localized Sources}

One clear path to differentiating DGP gravity from conventional Einstein
gravity is through an anomalous mismatch between the mass of a compact
object as computed by lensing measurements, versus the mass of an
object as computed using some measure of the Newtonian potential,
such as using the orbit of some satellite object, or other means such as the
source's x-ray temperature or through the SZ--effect.

The lensing of light by a compact matter source with metric
Eq.~(\ref{brane-n}) and (\ref{brane-a}) may be computed in the usual way.
The angle of deflection of a massless test particle is given by
\be
       \Delta\phi = \int dr~{J\over r^2}
		  {A\over \sqrt{{E^2\over N^2} - {J^2\over r^2}}}\ ,
\ee
where $E = N^2dt/d\lambda$ and $J = r^2d\phi/d\lambda$ are constants
of motion resulting from the isometries, and $d\lambda$ is the
differential affine parameter.  Removing the effect of the background
cosmology and just focussing on the deflection generated by passing
close to a matter source, the angle of deflection is \cite{Lue:2002sw}
\be
       \Delta\phi = \pi + 2b\int_b^{r_{\rm max}} 
       dr~{R_g(r) \over r^2\sqrt{r^2-b^2}}\ ,
\ee
where $b$ is the impact parameter.  This result is equivalent to
the Einstein result, implying light deflection is unaltered by
DGP corrections, even when those corrections are large.  This
result jibes with the picture that DGP corrections come solely
from a light scalar mode associated with brane fluctuations.
Since scalars do not couple to photons, the trajectory of light in
a DGP gravitational field should be identical to that in a Einstein
gravitational field generated by the same mass distribution.

\begin{figure}
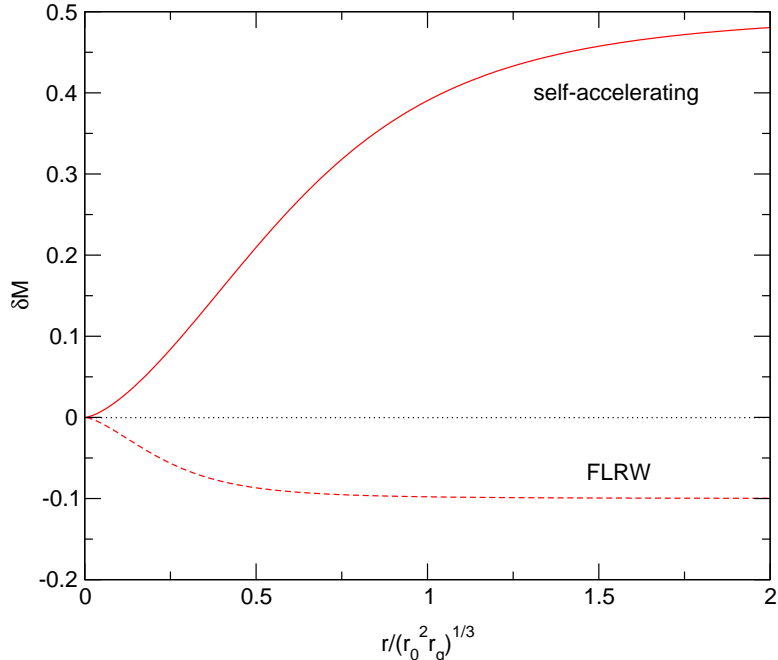
 \begin{center}\PSbox{mass.eps
hscale=50 vscale=50 hoffset=-60 voffset=-25}{3in}{3.1in}\end{center}
\caption{
Mass discrepancy, $\delta M$, for a static point source whose
Schwarzschild radius is $r_g$.  The solid curve is for a
self-accelerating background with $H = r_0^{-1}$.  The dashed curve is
for a FLRW background with $H = r_0^{-1}$.  Figure from Ref.~[55].
}
\label{fig:mass}
\end{figure}

Light deflection measurements thus probe the ``true" mass of a given
matter distribution.  Contrast this with a Newtonian measurement of
a matter distribution using the gravitational potential Eq.~(\ref{brane-n})
while incorrectly assuming general relativity holds.
The mass discrepancy between the lensing mass (the actual mass) and that
determined from the Newtonian force may be read directly from Eq.~(\ref{brane-n}),
\be
       \delta M = {M_{\rm lens}\over M_{\rm Newt}}-1
       = {1\over 1+\Delta(r)} - 1\ .
\ee
This ratio is depicted in Fig.~\ref{fig:mass} for both cosmological
phases, with an arbitrary background cosmology chosen.  When the mass is
measured deep within the Einstein regime, the mass discrepancy simplifies to
\be
       \delta M = \mp \left(r^3 \over 2r_0^2 R_g\right)^{1/3}\ .
\ee
Solar system measurements are too coarse to be able to resolve the DGP
discrepancy between lensing mass of the sun and its Newtonian mass.
The discrepancy $\delta M$ for the sun at ${\cal O}({\rm AU})$ scale
distances is approximately $10^{-11}$.  Limits on this discrepancy for
the solar system as characterized by the post-Newtonian parameter,
$\gamma-1$, are only constrained to be $< 3\times 10^{-4}$.

In the solar system, this is much too small an effect to be a serious
means of testing this theory, even with the most recent data from Cassini
\cite{Bertotti:2003rm}.  Indeed most state-of-the-art constraints on the post-Newtonian
parameter $\gamma$ will not do, because most of the tests of general relativity
focus on the strong-gravity regime near $r \sim r_g$.  However, in
this theory, this is where the anomalous corrections to general relativity are the
weakest.

The best place to test this theory is as close to the weak-brane regime
as we can get.  
A possibly more promising regime may be found in galaxy clusters.  For
$10^{14} \rightarrow 10^{15}~M_\odot$ clusters, the scale
$(r_0^2R_g)^{1/3}$ has the range $6\rightarrow 14~{\rm Mpc}$.  For
masses measured at the cluster virial radii of roughly $1\rightarrow
3~{\rm Mpc}$, this implies mass discrepancies of roughly $5\rightarrow
8\%$.  X-ray or Sunyaev--Zeldovich (SZ) measurements are poised to map
the Newtonian potential of the galaxy clusters, whereas weak lensing
measurements can directly measure the cluster mass profile.
Unfortunately, these measurements are far from achieving the desired
precisions.  If one can extend mass measurements to distances on the
order of $r_0$, Fig.~\ref{fig:mass} suggests discrepancies can be as
large as $-10\%$ for the FLRW phase or even $50\%$ for the
self-accelerating phase;  however, remember these asymptotic limits
get smaller as $(r_0H)^{-1}$ as a function of the redshift of the lensing mass.

It is a nontrivial result that
light deflection by a compact spherical source is identical to that in
four-dimensional Einstein gravity (even with potentials Eqs.~(\ref{brane-n})--(\ref{beta}) substantially differing from those of Einstein gravity) through the nonlinear transition between the Einstein phase and the weak-brane phase.  As such, there remains the possibility that for {\em aspherical} lenses that this
surprising null result does not persist through that transition and that DGP may manifest itself through some anomalous lensing feature.  However, given the intuition that the entirety of
the anomalous DGP effect comes from an extra gravitational scalar suggests that
even in a more general situation, photons should still probe the ``true" matter
distribution.  Were photon geodesics in DGP gravity to act just as in general relativity for
any distribution of lenses, this would provide a powerful tool for proceeding with the analysis
of weak lensing in DGP.  Weak lensing would then provide a clean window into the true
evolution of structure as an important method of differentiating it from general relativity.
Tanaka's analysis of gravitational fields in more general matter distributions in DGP gravity
provides a first step to answering the nature of photon geodesics \cite{Tanaka:2003zb}.

\subsection{The Late-Time ISW}

The late-time integrated Sachs--Wolfe (ISW) effect on the cosmic microwave background (CMB)
may be viewed as another possible ``lensing" signature.  The Sachs--Wolfe effect is intrinsically
connected to how photons move through and are altered by gravitational potentials (in this case,
time-dependent ones).  This effect is a direct probe of any possible alteration of the gravitational
potentials over time;  moreover, it is a probe of potential perturbations on the largest of distance scales,
safely in the linear regime.

Late-time ISW would then seem to be a promising candidate for modified-gravity theories of
the sort that anticipate cosmic acceleration, as we see that the potentials are altered by substantial
corrections.  Indeed, for a particular set of modified-gravity theories, this assertion is indeed the case
\cite{Lue:2003ky}.  However, for DGP gravity, there is an unfortunate catch.

Recall in Sec.~\ref{sec:lss}, we argued that when considering linear potentials satisfied
Eqs.~(\ref{poisson-n}) and (\ref{poisson-a}), our results generalized beyond just spherical
perturbations.  In order to bring our linear potentials Eqs.~(\ref{poisson-n}) and (\ref{poisson-a})
into line with how the late-time ISW is normally treated, we need to identify the gravitational
potentials as perturbations around a homogeneous cosmological background with the line
element
\be
     ds^2 = \left[1+2\Phi(\tau,\lambda)\right]d\tau^2
     - a^2(\tau)\left[1+2\Psi(\tau,\lambda)\right]
     \left[d\lambda^2+\lambda^2d\Omega\right]\ .
\label{potentials}
\ee
Here $\Phi(\tau,\lambda)$  and $\Psi(t,\lambda)$ are the relevant gravitational potentials and $\lambda$ is a comoving radial coordinate. 
In effect we want to determine $\Phi$ and $\Psi$ given $\delta n$ and $\delta a$. Unlike the case of Einstein's gravity, $\Phi \neq -\Psi$.   One may perform a coordinate transformation to determine that relationship.  
We find that, assigning $r = a(\tau)\lambda$, and
\bea
	\Phi &=& \delta n(\tau, r)	\\
	\Psi &=& -\int {dr\over r}\delta a(\tau, r)\ ,
\eea
keeping only the important terms when $rH \ll 1$.

The quantity of interest for the ISW effect is the time derivative of  $\Phi-\Psi$, the above
the above analysis implies
\be
	\nabla^2(\Phi-\Psi) = {8\pi \over M_P^2}a^2\rho\delta\ ,
	\label{poisson}
\ee
where $\nabla$ is the gradient in comoving spatial coordiantes, $\lambda^i$.
Just as we found for light deflection, this result is identical to the four-dimensional
Einstein result, the contributions from the brane effects exactly cancelling.
Again, the intuition of the anomalous DGP effects coming from a light gravitational
scalar is correct in suggesting the microwave background photons probe the
``true" matter fluctuations on the largest of scales.

Thus, the late-time ISW effect for DGP gravity will be identical to that of a dark
energy cosmology that mimics the DGP cosmic expansion history, Eq.~(\ref{Fried}),
at least at scales small compared to the horizon. Our approximation does not allow
us to address the ISW effect at the largest scales (relevant for the CMB at low multipoles),
but it is applicable to the cross-correlation of the CMB with galaxy surveys \cite{Fosalba:2003ge,Scranton:2003in}.  At larger scales, one
expects to encounter difficulties associated with leakage of gravity off
the brane (for order-unity redshifts) and other bulk effects
\cite{Lue:2002fe,Deffayet:2002fn,Deffayet:2004xg} that we were successfully able to
ignore at subhorizon scales.

\subsection{Leakage and Depletion of Anisotropic Power}

There is an important effect we have completely ignored up until now.  At
scales comparable to $r_0$, when the Hubble expansion rate is comparable
to $r_0^{-1}$, i.e., ${\cal O}(1)$ redshifts, gravitational perturbations can
substantially leak off the brane.  This was the original effect discussed from
the introduction of DGP braneworlds.  At the same time, perturbations that
exist in the bulk have an equal likelihood of impinging substantial perturbation
amplitude onto the brane from the outside bulk world.

This leads to a whole new arena of possibilities and headaches.  There is
little that can be done to control what exists outside in the free bulk.  However,
there are possible reasonable avenues one can take to simplify the situation.
In Sec.~\ref{sec:global} we saw how null worldlines through the bulk in
the FLRW phase could connect different events on the brane.  This
observation was a consequence of the convexity of the brane worldsheet
and the choice of bulk.  Conversely, if one chooses the bulk
corresponding to the self-accelerating phase, one may conclude that no
null lightray through the bulk connects two different events on the
brane.

Gravity in the bulk is just empty space five-dimensional Einstein gravity,
and thus, perturbations in the bulk must follow null geodesics.
Consider again Fig.~\ref{fig:brane}.  If we live in the self-accelerating
cosmological phase, then the bulk exists only exterior to the brane
worldsheet in this picture.  One can see that, unlike in the interior phase,
null geodesics can only intersect our brane Universe once.  I.e., once
a perturbation has left our brane, it can never get back.
Therefore, if we assume that the bulk is completely empty and all
perturbations originate from the brane Universe, that means that
perturbations can {\em only leak} from the brane, the net result of
which is a systematic loss of power with regard to the gravitational
fluctuations.

Let us attempt to quantify this depletion.  As a crude estimate, we can
take the propagator from the linear theory given by Eqs.~(\ref{Einstein}).
and treat only fluctuations in the Newtonian potential,
$\phi(x^A)$ where $g_{00} = 1 + 2\phi$.  For modes inside the
horizon, we may approximate evolution assuming a Minkowski
background.  For a mode whose spatial wavenumber is $k$, the equations of
motion are
\be
	{\d^2\phi\over\d\tau^2} - {\d^2\phi\over\d z^2} + k^2\phi = 0\ ,
\ee
subject to the boundary condition at the brane ($z=0$)
\be
	\left.{\d\phi\over \d z}\right|_{z=0}
		= -r_0\left({\d^2\phi\over\d \tau^2} -k^2\phi\right)\ .
\ee
Then for a mode initially localized near the brane, the amplitude on
the brane obeys the following form \cite{Lue:2002fe}: 
\be
	|\phi| = |\phi|_0e^{-{1\over 2}\sqrt{\tau\over kr_0^2}}\ ,
\ee
when $\tau \lesssim kr_0^2$.  Imagine the late universe as each mode
reenters the horizon.  Modes, being frozen outside the horizon, are
now free to evolve.  For a given mode, $k$, the time spent inside the
horizon is
\be
	\tau = r_0\left[1 - \left({1\over kr_0}\right)^3\right]\ ,
\ee
in a late-time, matter-dominated universe and where we have
approximated today's cosmic time to be $r_0$.  Then, the anomalous depletion resulting in DGP gravity is
\be
	\left.{\delta|\phi|\over |\phi|}\right|_{\rm DGP}
	= \exp\left[-{1\over 2}\sqrt{1\over kr_0}
		\left(1 - {1\over (kr_0)^3}\right)\right] - 1\ .
\ee
This depletion is concentrated at scales where $kr_0 \sim 1$.  It should
appear as an {\em enhancement} of the late-time integrated Sachs--Wolfe
effect at the largest of angular scales.

A more complete analysis of the perturbations of the full metric is
needed to get a better handle on leakage on sizes comparable to the
crossover scale, $r_0$.  Moreover, complicated global structure in the
bulk renders the situation even more baffling.  For example, the inclusion
of an initial inflationary stage completely alters the bulk boundary
conditions.  Other subtleties of bulk initial and boundary condition also
need to be taken into account for a proper treatment of leakage in a
cosmological setting \cite{Lue:2002fe,Deffayet:2002fn,Deffayet:2004xg}.

\section{Prospects and Complications}

We have presented a preliminary assessment of the observational viability of DGP gravity to simultaneously explain the acceleration of the universe and offer a promising set of
observables for the near future.
The theory is poised on the threshold of a more mature analysis that requires a new level of
computational sophistication.  In order to improve the comparison against observations a number
of issues need to be resolved.  The applicability of coarse graining is a pertinent issue that needs
to be addressed and tackled.  Understanding growth of structure into clustered objects is essential
in order to apply promising observing techniques in the near future.  To do a full comparison of the CMB power spectrum against data, it remains to properly treat the ISW effect at scales comparable to the
horizon.  Understanding how
primordial fluctuations were born in this theory likewise requires a much more detailed treatment
of technical issues.

\subsection{Spherical Symmetry and Birkhoff's Law}

An important outstanding question in DGP cosmology is whether a universe driven
by a uniform distribution of compact sources, such as galaxies or
galaxy clusters, actually drives the same overall expansion history
as a truly uniform distribution of pressureless matter.
The problem is very close to the same problem as in general relativity
(although in DGP, the system is more nonlinear) of how one recovers the
expansion history of a uniform distribution of matter with a statistically
homogeneous and isotropic distribution of discrete sources.  What
protects this coarse-graining in general relativity is that the theory possesses
a Birkhoff's law property (Fig.~\ref{fig:expansion}),
even in the fully-nonlinear case.

\begin{figure} \begin{center}\PSbox{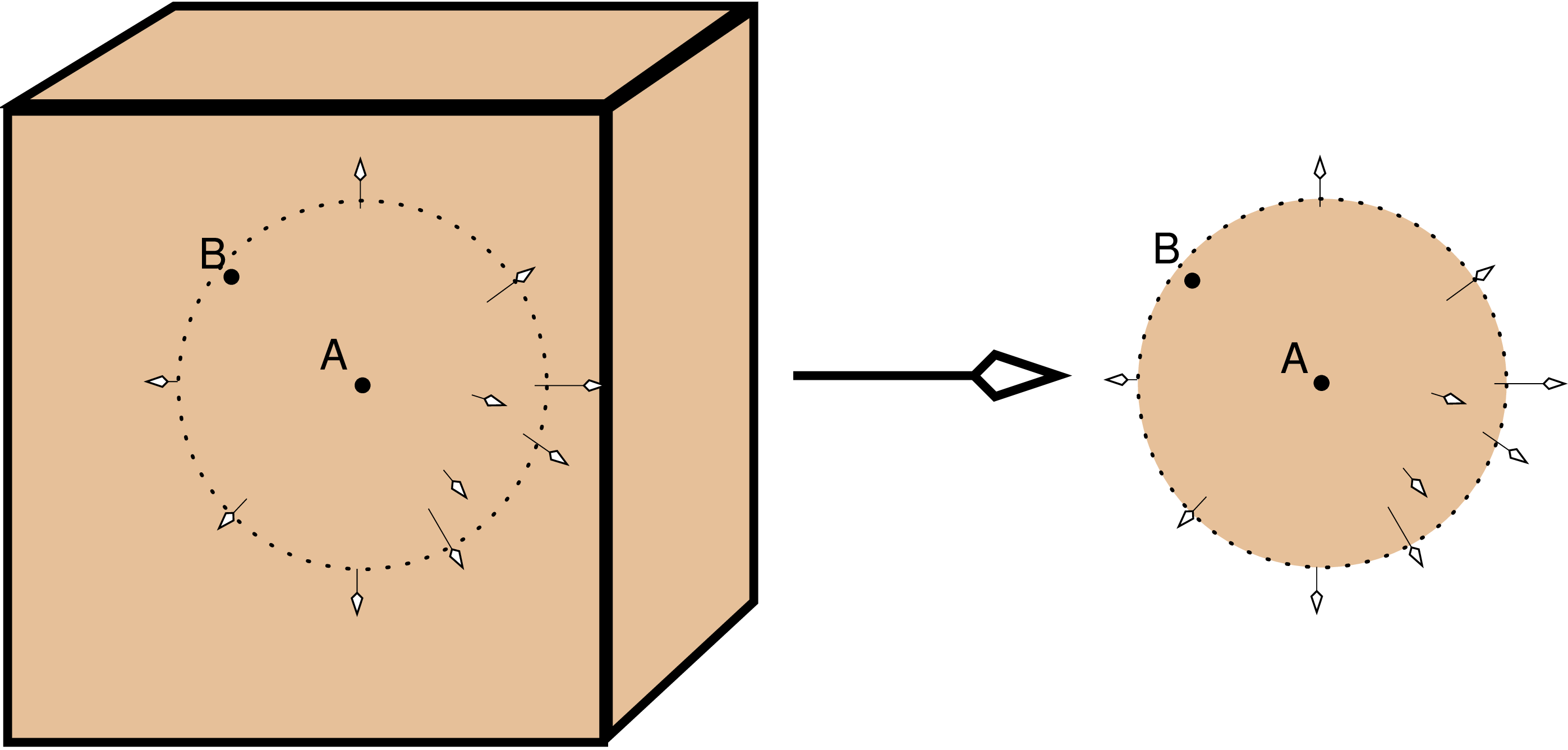
hscale=50 vscale=50 hoffset=0 voffset=0}{5.5in}{2.5in}\end{center}
\caption{
It is the Birkhoff's law property of general relativity that allows one take an arbitrary, though
spherically symmetric matter distribution (left) but describe the evolution of observer $B$ relatively to the origin of symmetry $A$ by excising all matter outside a spherical ball of mass around
the $A$ and condensing the matter within the ball to a single point.  DGP gravity, in a limited
fashion, obeys the same property.
}
\label{fig:expansion}
\end{figure}

It is clear from the force law generated from the metric components
Eqs.~(\ref{brane-n}--\ref{beta}) that Birkhoff's law does not strictly
apply in DGP gravity, even with a pressureless background
energy-momentum component.  The expressions in these equations
make a clear distinction between the role of matter that is part of
an overdensity and matter that is part of the background.  However,
there is a limited sense in which Birkhoff's law does apply.  We do see that the
evolution of overdensities is only dependent on the quantity $R_g(r)$.
In this sense, because for spherically-symmetric matter configurations,
only the integrated mass of the overdensity matters (as opposed to
more complicated details of that overdensity), Birkhoff's law does apply
in a limited fashion.  So, if all matter perturbations were
spherically-symmetric, then coarse graining would apply in DGP gravity.

Of course, matter perturbations are not spherically symmetric, not even
when considering general relativity.  We extrapolate that because we
are concerned with perturbations that are at least statistical isotropic, that
like in general relativity, coarse graining may be applied in this more
general circumstance.  It is widely believed that coarse
graining may be applied with impunity in general relativity.  Based on the arguments used
here, we assume that the same holds true for DGP.  However, it is
worthwhile to confirm this in some more systematic and thorough way, as
there are subtleties in DGP gravity that do not appear in general relativity.

\subsection{Beyond Isolated Spherical Perturbations}

Understanding the growth of structure in DGP gravity is an essential requirement
if one is to treat it as a realistic or sophisticated counterpart to the standard cosmological
model.  We need to understand large scale structure beyond a linear analysis and
beyond the primitive spherical collapse picture.  Without that understanding, analysis
of anything involving clustered structures are suspect, particularly those that have formed during the
epoch of ${\cal O}(1)$ redshift where the nonlinear effects are prominent and
irreducible.  This includes galaxy cluster counting, gravitational lens
distributions, and several favored methods for determining $\sigma_8$ observationally.
For analyses of DGP phenomenology such as Refs.~\cite{Jain:2002di,Lima:2003dd,Seo:2003pu,Zhu:2004vj,Uzan:2004my,Capozziello:2004jy,Alcaniz:2004ei,Song:2005gm,Knox:2005rg}} to
carry weight, they need to include such effects.

What is needed is the moral equivalent of the N--body simulation for DGP gravity.
But the native nonlinearity coursing throughout this model essentially forbids a simple
N--body approach in DGP gravity.
The ``rubber sheet" picture described at the end of Sec.~\ref{sec:solarsystem} and in
Fig.~{\ref{fig:comoving} must be
developed in a more precise and formal manner.  The truly daunting aspect of this
approach is that it represents a full five-dimensional boundary
value problem for a complex distribution of matter sources, even if one could use the
quasistatic approximation.  One possible simplification is the equivalent
of employing the assumption Eq.~(\ref{assumption}) but in a more general context of
an arbitrary matter distribution.  This allows one to reduce the five-dimensional boundary
value problem to a much simpler (though still terribly intractable) four-dimensional
boundary value problem.  Tanaka has provided a framework where one can begin
this program of extending our understanding of gravitational fields of interesting
matter distributions beyond just spherically-symmetric ones \cite{Tanaka:2003zb}.

If this program can be carried out properly, whole new vistas of approaches can be taken
for constraining DGP gravity and then this modified-gravity theory may take a place as a
legitimate alternative to the standard cosmological model.

\subsection{Exotic Phenomenology}

There are a number of intriguing exotic possibilities for the future in DGP gravity.  The model has
many rich features that invite us to explore them.  One important avenue that begs to be developed
is the proper inclusion of inflation in DGP gravity.  Key issues involve the proper treatment of the
bulk geometry, how (and at what scale) inflationary perturbation in this theory develop and what
role do brane boundary conditions play in the treatment of this system.  An intriguing possibility
in DGP gravity may come into play as the fundamental five-dimensional Planck scale is so low
(corresponding to $M \sim 100\ {\rm MeV}$, suggesting transplanckian effects may be important
at quite low energies.  Initial work investigating
inflation in DGP gravity has begun \cite{Papantonopoulos:2004bm,Zhang:2004in,Bouhmadi-Lopez:2004ax}.  Other work that has shown just a sample of the richness of DGP gravity and its observational
possibilities include the possibility of nonperturbatively dressed or screened starlike solutions \cite{Gabadadze:2004iy,Gabadadze:2005qy} and shock solutions \cite{Kaloper:2005az,Kaloper:2005wa}.

Indeed, a possible glimpse of how strange DGP phenomenology may be is exhibited by the
starlike solution posed in Refs. \cite{Gabadadze:2004iy,Gabadadze:2005qy}.  In these papers,
the solution appears as four-dimensional Einstein when $r\ll r_*$, just as the one described in
Sec.~\ref{sec:einstein}.  However, far from the matter source when $r\gg r_0$, rather than the
brane having no influence on the metric, in this new solution, there is a strong backreaction from
the brane curvature that screens the native mass, $M$, of the central object and the effective ADM
mass appears to be
\be
	M_{\rm eff} \sim M\left({r_g\over r_0}\right)^{1/3}\ ,
\ee
where $r_g$ is the Schwarzschild radius corresponding to the mass $M$.  Given the strongly
nonlinear nature of this system, that two completely disconnected metric solutions exist for the
same matter source is an unavoidable logical possibility.\footnote{There is a potential subtlety
	that may shed light on the nature of this system.  A coordinate transformation may take
	the solution given in Refs. \cite{Gabadadze:2004iy,Gabadadze:2005qy} into a metric of
	the form given by Eq.~(\ref{metric}).  From this point of view, it is clear that the form of the
	induced metric on the brane is constrained.  It would seem then that the imbedding of the
	brane in the bulk is not fully dynamical, and would thus imply that the solution given is
	not the result of purely gravitational effects.  I.e., that there must be some extra-gravitational
	force constraining the brane to hold a particular form.  A consensus has not yet been achieved
	on this last conculsion.}
The consequences of such an exotic phenomenology are still to be fully revealed.

\subsection{Ghosts and Instability}

There has been a suggestion that the scalar, brane-fluctuation mode is a
ghost in the self-accelerating phase \cite{Luty:2003vm,Nicolis:2004qq}.
Ghosts are troublesome because, having a negative kinetic term, high
momentum modes are extremely unstable leading to the rapid
devolution of the vacuum into increasingly short wavelength fluctuations.
This would imply that the vacuum is unstable to unacceptable fluctuations,
rendering moot all of the work done in DGP phenomenology.
The status of this situation is still not resolved.  Koyama suggests that, in
fact, around the self-accelerating empty space solution itself, the brane
fluctuation mode is not a ghost \cite{Koyama:2005tx}, implying that there
may be a subtlety involving the self-consistent treatment of perturbations
around the background.  Nevertheless, the situation may be even more
complicated, particularly in a system more complex than a quasistatic
deSitter background.

However, recall that the coupling of the scalar mode to matter is increasingly
suppressed at sort scales (or correspondingly, high momentum).  Even if
the extra scalar mode were a ghost, it is not clear that the normal mechanism for the
instability of the vacuum is as rapid here, possibly leaving a phenomenological
loophole:  with a normal ghost coupled to matter, the vacuum would almost instantaneously
dissolve into high momentum matter and ghosts, here because of the suppressed
coupling at high momentum, that devolution may proceed much more slowly.
A more quantitative analysis is necessary to see if this is a viable scenario.

\section{Gravity's Future}

What is gravity?  Our understanding of this fundamental, universal interaction has been a pillar of
modern science for centuries.  Physics has primarily focused on gravity as a theory incomplete in the ultraviolet regime, at incomprehensibly large energies above the Planck scale.  However, there still remains a tantalizing regime where we may challenge Einstein's theory of gravitation on the largest of scales, a new infrared frontier.  This new century's golden age of cosmology offers an unprecedented opportunity to understand new infrared fundamental physics.  And while particle cosmology is the celebration of the intimate connection between the very great and the very small, phenomena on immense scales {\em in of themselves} may be an indispensable route to understanding the true nature of our Universe.

The braneworld theory of Dvali, Gabadadze and Porrati (DGP) has pioneered this new line of investigation, offering an novel explanation for today's cosmic acceleration as resulting from the unveiling of an unseen extra dimension at distances comparable today's Hubble radius.  While the theory offers a specific prediction for the detailed expansion history of the universe, which may be tested
observationally, it offers a paradigm for nature truly distinct from dark energy explanations of
cosmic acceleration, even those that perfectly mimic the same expansion history.  DGP braneworld
theory alters the gravitational interaction itself, yielding unexpected phenomenological handles
beyond just expansion history.  Tests from the solar system, large scale structure, lensing all offer
a window into understanding the perplexing nature of the cosmic acceleration and, perhaps,
of gravity itself.

Understanding the complete nature of the gravitational interaction is the final frontier of theoretical physics and cosmology, and provides an indispensable and tantalizing opportunity to peel back the curtain of the unknown.  Regardless of the ultimate explanation, revealing the structure of physics at the largest of scales allows us to peer into its most fundamental structure.  What is gravity?  It is not a new question, but it is a good time to ask.

\acknowledgements

The author wishes to thank C.~Deffayet, G.~Dvali, G.~Gabadadze, M.~Porrati, A.~Gruzinov, G.~Starkman and R.~Scoccimarro for their crucial interactions, for their deep
insights and for their seminal and otherwise indispensable scientific contribution to this material.

\end{document}